\documentclass[12pt, a4paper]{article}

\usepackage[a4paper]{geometry}
\usepackage{authblk}
\usepackage{fullpage}
\usepackage[T1]{fontenc}
\usepackage[version=3]{mhchem}
\usepackage{lscape}

\usepackage{siunitx}
\usepackage{mathrsfs}
\usepackage{amsmath}
\usepackage{amsthm}
\usepackage{amssymb}
\theoremstyle{theorem}
\newtheorem{theorem}{Theorem}

\usepackage[ruled,linesnumbered,noline]{algorithm2e}

\usepackage[dvipsnames, svgnames, x11names]{xcolor}

\newcommand{\mvec}[1]{\boldsymbol{#1}} 		% math-mode vector: bold and italic
\newcommand{\mmat}[1]{\mathbf{#1}}   		% math-mode matrix: bold
\newcommand{\transpose}{\ensuremath{^{\scriptscriptstyle\top}}} % math and inline mode Transpose operator

\usepackage[utf8]{inputenc}
\usepackage[sort&compress,numbers,square]{natbib}
\usepackage{hyperref}
\hypersetup{
	unicode, 
	colorlinks=true, 
	breaklinks=true, 
	urlcolor=cyan, 
	linkcolor=blue,
	citecolor=cyan,
}
\usepackage[noabbrev,capitalize,nameinlink]{cleveref}

\usepackage[left]{lineno}
\usepackage{setspace}
\usepackage{booktabs}
\usepackage{multirow}
\usepackage{makecell} % fo wraping text in table

\newcommand{\mreplaced}[2]{\replaced{#2}{#1}}
\usepackage[final]{changes}				% final version
\makeatletter
\@namedef{Changes@AuthorColor}{red}
\colorlet{Changes@Color}{red}
\makeatother

\usepackage{graphicx, color}
\usepackage{caption} 	% required for subfigures
\usepackage{subcaption} % required for subfigures
\graphicspath{{figures/}}
\title{\textbf{\mreplaced{A novel spatial filtering framework for feature extraction in motor imagery brain computer interfaces}{\textcolor{black}{Scatter-based common spatial patterns - a unified spatial filtering framework}}}}

\author[1]{Jinlong Dong}
\author[1]{Milana Komosar}
\author[2]{Johannes Vorwerk}
\author[2, \footnote{Corresponding author$\colon$ Daniel Baumgarten (Daniel.Baumgarten@umit-tirol.at)}]{Daniel Baumgarten}%\footnote{Corresponding author： Daniel Baumgarten (Daniel.Baumgarten@umit-tirol.at)}}
\author[1]{Jens Haueisen}
\affil[1]{Institute of Biomedical Engineering and Informatics, Technische Universit\"{a}t Ilmenau, Ilmenau, Germany}
\affil[2]{Institute of Electrical and Biomedical Engineering, UMIT \added{TIROL} -- Private University for Health Sciences\mreplaced{, Medical Informatics and}{ and Health} Technology, Hall in Tirol, Austria}

\date{}
\begin{document}

\begin{singlespace}
	\maketitle
\end{singlespace}
%%==============================================================================
%%

%%==============================================================================
%%
\begin{abstract}
The common spatial pattern (CSP) approach is known as one of the most popular spatial filtering techniques for EEG classification in motor imagery (MI) based brain-computer interfaces (BCIs). However, it still suffers some drawbacks such as sensitivity to noise, non-stationarity, and limitation to binary classification.Therefore, we propose a novel spatial filtering framework called scaCSP based on the scatter matrices of spatial covariances of EEG signals, which works generally in both binary and multi-class problems whereas CSP can be cast into our framework as a special case when only the range space of the between-class scatter matrix is used in binary cases.We further propose subspace enhanced scaCSP algorithms which easily permit incorporating more discriminative information contained in other range spaces and null spaces of the between-class and within-class scatter matrices in two scenarios: a nullspace components reduction scenario and an additional spatial filter learning scenario.The proposed algorithms are evaluated on two data sets including 4 MI tasks. The classification performance is compared against state-of-the-art competing algorithms: CSP, Tikhonov regularized CSP (TRCSP), stationary CSP (sCSP) and stationary TRCSP (sTRCSP) in the binary problems whilst multi-class extensions of CSP based on pair-wise and one-versus-rest techniques in the multi-class problems. The results show that the proposed framework outperforms all the competing algorithms in terms of average classification accuracy and computational efficiency in both binary and multi-class problems.The proposed scsCSP works as a unified framework for general multi-class problems and is promising for improving the performance of MI-BCIs.
\end{abstract}
%%==============================================================================
%%

%%==============================================================================
%%
\section{Introduction}

\mreplaced{B}{A b}rain computer interface (BCI) aims to establish a direct communication pathway between humans and external devices while bypassing \added{the} brain's normal output pathways of peripheral nerves and muscles \citep{wolpaw2012BraincomputerInterfacesPrinciples, dornhege2007BraincomputerInterfacing}. It extracts specific features of the ongoing neural activities through electrophysiological or other brain signals and convert\added{s} them into artificial control commands that drive an output. During the past decades, BCI systems have been attracting great interest in a wide range of application areas \citep{brunner2015BNCIHorizon2020, chaudhary2016BrainComputerInterfaces, lebedev2017BrainMachineInterfacesBasic}. In terms of brain activity monitoring, electroencephalography (EEG) is among the most widely used techniques in BCI systems due to its low cost, high temporal resolution, ease of use, and noninvasive and portable characteristics compared to other monitoring modalities, such as magnetoencephalography (MEG), functional magnetic resonance imaging (fMRI), functional near-infrared spectroscopy (fNIRS), etc. At present, various types of EEG signals that might serve as control signals have been studied, among which P300 evoked potentials, slow cortical potentials (SCPs), steady-state visual evoked potentials (SSVEPs)\added{,} and sensorimotor rhythms (SMRs) are \added{some of } the most popular ones in EEG-based BCI systems \citep{ramadan2017BrainComputerInterface, he2015NoninvasiveBrainComputerInterfaces}.

One typical type of BCI paradigm\deleted{s} is based on the voluntary modulation of SMRs during motor imagery (MI) \highlight{\citep{arpaia2022HowSuccessfullyClassify}}. When a subject is imaging a motor activity, the power attenuation or power increase of SMRs \mreplaced{could}{can} be detected from the EEG recordings. Such changes in power are termed \deleted{as} event-related desynchronization (ERD) and event-related synchronization (ERS), respectively \citep{pfurtscheller2006MuRhythmSynchronization}. A motor imagery BCI (MI-BCI) is based on decoding the SMR ERD/ERS patterns. However, it is still a very challenging task to decode MI-related patterns correctly and put them into practical use due to the low signal-to-noise ratio (SNR) of EEG recordings, the non-stationary nature of brain signals, and variabilities of the occurrence of ERD/ERS patterns in frequency bands, spatial locations\added{,} and time intervals.

In order to maximize the classification accuracy and efficiency of MI-BCIs, it is of particular importance to extract the most effective features from the raw brain signals. The common spatial pattern (CSP) algorithm \citep{blankertz2008OptimizingSpatialFilters} is the most commonly used and successful approach for feature extraction in MI-BCIs. It aims to find optimal projections \added{onto which} \deleted{that maximize\deleted{s} }the variance ratio of the \mreplaced{transformed}{projected} signals between two different MI tasks\added{ is maximized}. In practice, the signal is \deleted{previously }band-pass filtered \mreplaced{within a specific frequency band of interest}{prior to the application of the CSP to attenuate the frequency band of interest}, e.g., \(\mu\) (7--13Hz) or $\beta$ (14--30 Hz) band. In this context, CSP\deleted{ is to} extract\added{s} the \mreplaced{bandpower}{band power} components that differ maximally. Therefore, it is well suitable to decode brain states that are characterized by ERD/ERS patterns. After CSP filtering, the extracted feature vector set is finally fed into a classifier for classification \citep{lotte2018ReviewClassificationAlgorithms}. To date, impressive improvements in MI-BCI performance have been achieved by using CSP \highlight{\citep{blankertz2004BCICompetition2003, blankertz2006BCICompetitionIII, tangermann2012ReviewBCICompetition, bennett2021NeurophysiologicalApproachSpatial, liu2021OnlineDetectionClassimbalanced, vidaurre2021ImprovingMotorImagery, zuo2021ClusterDecomposingMultiobjective}}.

Despite its popularity and efficiency, CSP is still suffering some drawbacks leading to suboptimal performance in terms of reliability and generalizability, which are mainly attributed to the non-stationarity, outlier sensitivity, overfitting issue, and limitation to binary classifications \citep{samek2014DivergenceBasedFrameworkCommon, lotte2011RegularizingCommonSpatial}. First\deleted{ of all}, the conventional CSP works on manually or heuristically selected frequency bands and time intervals, whereas the BCI performance could be potentially enhanced by extracting features adapting to the subject-specific characteristics \citep{blankertz2007NoninvasiveBerlinBrain, blankertz2008OptimizingSpatialFilters}. To this end, one direction of CSP-variants is to automatically learn the optimal subject-specific frequency bands and time intervals in conjunction with the CSP filters by incorporating temporal filters into CSP \citep{lemm2005SpatioSpectralFiltersImproving, dornhege2006CombinedOptimizationSpatial, wu2008ClassifyingSingleTrialEEG, higashi2013SimultaneousDesignFIR} or combining\deleted{ with} CSP\added{ with} a filter bank framework \highlight{\citep{park2018FilterBankRegularized, ang2012FilterBankCommon, liu2021DistinguishableSpatialspectralFeaturea}} and sliding window techniques \citep{gaur2021SlidingWindowCommon, feng2018CorrelationbasedTimeWindow}. Another important direction is mainly focused on the regularization or reformulation of the CSP objective function to make it more robust to outliers and non-stationarity \citep{samek2014DivergenceBasedFrameworkCommon, lotte2011RegularizingCommonSpatial, roijendijk2016ClassifyingRegularizedSensor, wang2021CommonSpatialPattern}. Alternatively, the regularization techniques can also be applied during the \deleted{the }covariance matrix estimation phase \highlight{\citep{olias2019EEGSignalProcessing, balzi2015ImportanceweightedCovarianceEstimation, azab2020DynamicTimeWarpingbased}}. More specifically, a large group of regularized CSP algorithms has been shown to outperform the classical CSP\deleted{,} by incorporating into the objective function of CSP \textit{a priori} information in terms of $L_2$-norm penalties \highlight{\citep{ghanbar2021CorrelationbasedCommonSpatial, lotte2011RegularizingCommonSpatial}}, sparsity in terms of $L_1$-norm penalties \citep{jin2021InternalFeatureSelection, wang2016RegularizedFiltersL1NormBased, wang2012L1NormBasedCommonSpatial} or more general $L_p$-norms \citep{cai2021SingleTrialEEGClassification, gu2021CommonSpatialPattern}, and other information such as the phase value of signals \citep{chakraborty2020DesigningPhaseSensitiveCommon, li2019CommonSpatialPatterns}. Accounting for the non-stationarity, many variants of CSP have been proposed \mreplaced{towards to}{for} extracting stationary features such as stationary CSP (sCSP) \citep{samek2012StationaryCommonSpatial} and Kullback-Leibler CSP (KL-CSP) \citep{arvaneh2013OptimizingSpatialFilters} by regularizing the CSP using within-class scatters information and a more generalized divergence-based CSP (divCSP) by reformulating CSP as a divergence maximization framework \citep{samek2014DivergenceBasedFrameworkCommon}. Finally, since the CSP algorithm has initially \added{been} proposed for two-class paradigms, multi-class extensions of CSP \mreplaced{could be}{were} realized by using pair-wise (PW) or one-versus-rest (OVR) techniques which separate the original multi-class problem into several binary problems \citep{ang2012FilterBankCommon}, the joint approximate diagonalization (JAD) approach which is equivalent to ICA \citep{grosse-wentrup2008MulticlassCommonSpatial, gouy-pailler2010NonstationaryBrainSource}, and KL divergence approach \citep{wang2012HarmonicMeanKullback}.

In this work, we propose\deleted{d} a novel scatter-based spatial filtering framework (scaCSP) for robust and effective feature extraction in multichannel EEG based MI-BCIs. The scaCSP algorithms are formulated using the range spaces and null spaces of scatter matrices of spatial covariances including the between-class scatter matrix, the within-class scatter matrix and the total scatter matrix, by converting covariance matrices into higher dimensional vectors with a linear transformation. \deleted{However, }\mreplaced{t}{T}he proposed scaCSP works as a unified framework for more general multi-class problems (both binary and multi-class classification problems). \added{Specifically, we prove that the conventional CSP algorithms can be subsumed within the proposed framework in that they can be derived from scaCSP as a special case in the binary classification scenario where only the range space of the between-class scatter matrix is taken into consideration.} Moreover, the scatter-based framework paves a way for enhancing the classification performance by incorporating more discriminative information contained in other subspaces of \added{the} scatter matrices in two different scenarios: the subspace component reduction scenario for improving the task-related SNR and the extra spatial filter learning scenario for seeking more robust and stationary filters.

\mreplaced{At}{In} the end\added{,} we would like to stress that the proposed spatial filtering framework is \textit{by no means} limited to MI-BCI applications. On the contrary\added{,} it is a completely generic new signal processing technique that can be \mreplaced{applicable}{applied} for all general single-trial settings for brain signal analysis that require discrimination \added{of} brain states based on modulations of brain rhythms.

\noindent \textit{Notations}: The following terminology, notations, and mathematical operations are used throughout this paper. All vectors are column vectors denoted by bold and italic symbols in lower case and matrices by bold symbols in upper case. The superscript $\transpose$ represents transpose. For example, $\mvec{a} = [a_1, \cdots, a_n]\transpose$ is an $n$-dimensional vector with entries $a_i$ while $\mmat{A} = [\mvec{a}_1, \cdots, \mvec{a}_n]$ is an $n \times n$ matrix with $n$-dimensional column vectors $\mvec{a}_i$. The vectorization operator vec($\cdot$) creates a column vector from a matrix $\mmat{A}$ by stacking the columns of $\mmat{A}$ below one another, for example, $\text{vec}(\mmat{A}) = [\mvec{a}_1\transpose, \cdots, \mvec{a}_n\transpose]\transpose$. \mreplaced{O}{The o}perator $\otimes$ gives the Kronecker product of matrices or vectors and the identity $\text{vec}(\mvec{a} \mvec{a}\transpose) = \mvec{a} \otimes \mvec{a}$ holds for all column vectors. 
%%==============================================================================
%%

%%==============================================================================
%%
\section{Materials and methods}

\subsection{CSP algorithms} 

In this section, we briefly introduce the conventional CSP algorithm and its variants in terms of regularization and multi-class extensions considered in this work. 

\subsubsection{CSP} \label{sec:csp}

Given the spatial covariance matrices of two different MI tasks (e.g., left hand and right hand MI), the CSP algorithm computes optimal linear spatial filters in a supervised manner such that the variance of the spatially filtered signals is maximized for one task while minimized for the other. Let \(\mmat{X}_i = \left[\mvec{x}_{i1}, \cdots, \mvec{x}_{iN_t}\right] \in \mathbb{R}^{N_c \times N_t}\) be the data matrix of an EEG signal trial measured from \(N_c\) channels and $N_t$ time samples, where $i \in \Omega = \Omega_1 \cup \Omega_2$ and $\Omega_1, \Omega_2$ are the set of trial labels corresponding to class 1 and class 2, respectively. The number of EEG trials for class $k$, $k = 1, 2$ is denoted by $|\Omega_k|$ and the total number of trials $|\Omega| = |\Omega_1| + |\Omega_2|$. The signals are assumed to have been bandpass filtered in a specific frequency band or centered, i.e., $\mmat{X}_i$ has zero mean for each row. The sample covariance matrix $\mmat{C}_i \in \mathbb{R}^{N_c \times N_c}$ for $i$-th trial is computed by $\mmat{C}_i = \frac{1}{N_t - 1} \mmat{X}_i \mmat{X}_i\transpose$. The class-mean and composite covariance matrices are then given by $\tilde{\mmat{C}}_k = \frac{1}{|\Omega_k|} \sum_{i \in \Omega_k} \mmat{C}_i, k = 1, 2$ and $\tilde{\mmat{C}} = \tilde{\mmat{C}}_1 + \tilde{\mmat{C}}_2$, respectively.

The CSP approach is to find the spatial filter  $\mvec{w} \in \mathbb{R}^{N_c}$ that \mreplaced{extremizes}{optimizes} (maximizes or minimizes) the objective function $J(\mvec{w})$ formulated as the Rayleigh quotient:
\begin{equation} \label{eq:J1-csp}
	J(\mvec{w}) = \dfrac{\mvec{w}\transpose \tilde{\mmat{C}}_1 \mvec{w}}{\mvec{w}\transpose (\tilde{\mmat{C}}_1 + \tilde{\mmat{C}}_2) \mvec{w}} = \dfrac{\mvec{w}\transpose \tilde{\mmat{C}}_1 \mvec{w}}{\mvec{w}\transpose \tilde{\mmat{C}} \mvec{w}}
\end{equation}
whose solution can be given by the eigenvectors of the following generalized eigenvalue problem: $\tilde{\mmat{C}}_1 \mvec{w} = J(\mvec{w}) \tilde{\mmat{C}} \mvec{w}$. Alternatively, the CSP optimization problem can also be solved by the simultaneous diagonalization technique \citep{fukunaga1990IntroductionStatisticalPattern}, including a whitening process and a principal component analysis (PCA) process. More precisely, noting that $J(a\mvec{w}) = J(\mvec{w}), \forall a \in \mathbb{R}\backslash 0$, it is only the direction of vector $\mvec{w}$ that matters whatever its magnitude. Thus\added{,} it is possible to rescale $\mvec{w}$ such that $\mvec{w}\transpose \tilde{\mmat{C}} \mvec{w} = 1$ with $J(\mvec{w})$ unchanged. Since $\tilde{\mmat{C}}$ is symmetric, it has the eigendecomposition $\tilde{\mmat{C}} = \tilde{\mmat{U}}_c \tilde{\mmat{\Lambda}}_c \tilde{\mmat{U}}_c\transpose$. Then the whitening transform with $\mmat{P}_c = \tilde{\mmat{U}}_c \tilde{\mmat{\Lambda}}_c^{(-1/2)}$ leads to the whitened covariance matrices for each EEG trial $\mmat{R}_i = \mmat{P}_c\transpose \mmat{C}_i \mmat{P}_c$, where the whitened class-mean covariance matrices $\tilde{\mmat{R}}_k = \mmat{P}_c\transpose \tilde{\mmat{C}}_k \mmat{P}_c, k = 1, 2$ for $k$-th class and the whitened composite covariance matrix $\tilde{\mmat{R}} = \tilde{\mmat{R}}_1 + \tilde{\mmat{R}}_2 = \mmat{P}_c\transpose \tilde{\mmat{C}} \mmat{P}_c = \mmat{I}$. The initial maximization/minimization of \eqref{eq:J1-csp} is then equivalent to extremizing the function 
\begin{equation} \label{eq:J2-csp-whitened}
	J^*(\mvec{w}) = \dfrac{\mvec{w}\transpose \tilde{\mmat{R}}_1 \mvec{w}}{\mvec{w}\transpose (\tilde{\mmat{R}}_1 + \tilde{\mmat{R}}_2) \mvec{w}} = \dfrac{\mvec{w}\transpose \tilde{\mmat{R}}_1 \mvec{w}}{\mvec{w}\transpose \mvec{w}}
\end{equation}
in addition to a whitening transform with $\mmat{P}_c$.

% \eqref{eq:J1-csp}

Given the eigenvalue decomposition of $\tilde{\mmat{R}}_1$:
\begin{equation} \label{eq:eigen-tilde-R1}
	\tilde{\mmat{R}}_1  = \tilde{\mmat{U}}_1 \tilde{\mmat{\Lambda}}_1 \tilde{\mmat{U}}_1\transpose %= \sum_{j=1}^{N_c} \tilde{\lambda}_j \tilde{\mvec{u}}_j \otimes \tilde{\mvec{u}}_j
\end{equation}
we have
\begin{equation} \label{eq:eigen-tilde-R2}
	\tilde{\mmat{R}}_2  = \tilde{\mmat{U}}_1 (\mmat{I}^{N_c} - \tilde{\mmat{\Lambda}}_1) \tilde{\mmat{U}}_1\transpose 
\end{equation}
which indicates that $\tilde{\mmat{R}}_1$ and $\tilde{\mmat{R}}_2$ share the same eigenvectors but with reversely ordered corresponding eigenvalues, i.e., the eigenvector with the largest eigenvalue for $\tilde{\mmat{R}}_1$ is associated with the smallest eigenvalue for $\tilde{\mmat{R}}_2$, and vice versa. In \added{the} above equations, $\tilde{\mmat{U}}_1 = \left[\tilde{\mvec{u}}_1, \cdots, \tilde{\mvec{u}}_{N_c}\right]$ is an orthogonal matrix whose columns are the eigenvectors of $\tilde{\mmat{R}}_1$, $\tilde{\mmat{\Lambda}}_1 = \text{diag}(\tilde{\lambda}_1, \cdots, \tilde{\lambda}_{N_c})$ is a diagonal matrix whose diagonal entries are the corresponding eigenvalues, which are sorted in non-increasing order, and $\mmat{I}^{N_c}$ is the $N_c \times N_c$ identity matrix.

Finally, we can construct the projection matrix $\mmat{W} = \mmat{P}_c \tilde{\mmat{U}}_1 = \left[\mvec{w}_1, \cdots, \mvec{w}_{N_c} \right]$ whose columns $\mvec{w}_i = \mmat{P}_c \tilde{\mvec{u}}_i, i = 1, \cdots, N_c,$ are the obtained $N_c$ spatial filters by CSP, leading to the spatially filtered signals \(\mmat{Y}_i \in \mathbb{R}^{N_c \times N_t}\) by projecting EEG signal samples $\mmat{X}_i$ onto $\mmat{W}$, i.e., $\mmat{Y}_i = \mmat{W} \transpose \mmat{X}_i = [\mvec{w}_1 \transpose \mmat{X}_i, \cdots, \mvec{w}_{N_c} \transpose \mmat{X}_i]\transpose$. The extracted features $\mvec{f}_i \in \mathbb{R}^{N_c}$ are the variances of the projected EEG signals for each channel:
\begin{equation} \label{eq:features}
	\mvec{f}_i = \text{var} (\mmat{Y}_i) = \begin{bmatrix}
		\mvec{w}_1 \transpose \mmat{C}_i \mvec{w}_1 \\
		\vdots \\
		\mvec{w}_{N_c} \transpose \mmat{C}_i \mvec{w}_{N_c}
	\end{bmatrix} = \begin{bmatrix}
	\tilde{\mmat{u}}_1 \transpose \mmat{R}_i \tilde{\mmat{u}}_1 \\
	\vdots \\
	\tilde{\mmat{u}}_{N_c} \transpose \mmat{R}_i \tilde{\mmat{u}}_{N_c}
	\end{bmatrix}
\end{equation}
or their logarithm values in order to normalize the features such that they are much closer to the Gaussian distribution.

The eigenvalues $\tilde{\lambda}_i$ measure the variance ratio between class 1 and \added{the} composite of both classes. Note that the composite variance of both classes is kept to be one by the whitening process using $\mmat{P}_c$. Then a large $\tilde{\lambda}_i$ indicates high variance of class 1 and low variance of class 2, whilst a small $\tilde{\lambda}_i$ indicates the opposite.  In \deleted{an}other word\added{s}, $\mvec{w}_1$ maximizes the variance ratio ${\mvec{w}\transpose \tilde{\mmat{C}}_1 \mvec{w}}/{\mvec{w}\transpose \tilde{\mmat{C}} \mvec{w}}$ in \eqref{eq:J1-csp} for class 1 while $\mvec{w}_{N_c}$ maximizes the variance ratio ${\mvec{w}\transpose \tilde{\mmat{C}}_2 \mvec{w}}/{\mvec{w}\transpose \tilde{\mmat{C}} \mvec{w}}$ for class 2. The spatial filter $\mvec{w}_1$ (respectively, $\mvec{w}_{N_c}$) is, in the sense of separating the \mreplaced{bandpower}{band power} maximally, the best filter for class 1 (resp. class 2). Practically, we choose $m$ ($2m \leq N_c$) spatial filters corresponding to the $m$ largest eigenvalues and $m$ filters corresponding to the $m$ smallest eigenvalues in order to improve the classification accuracy. The optimal selection of $m$ is a difficult and subject-specific problem. As suggested by \cite{blankertz2008OptimizingSpatialFilters}, $m=$ 2 or 3 would be a good option for general settings. The selected features are finally fed into a classifier for classification.

\subsubsection{Regularizations of CSP} \label{sec:rcsp}

The regularized CSP (RCSP) can be performed by adding a penalty term $P(\mvec{w})$ to the denominator of the CSP objective function in \eqref{eq:J1-csp} in order to regularize solutions that do not satisfy \textit{\mreplaced{a prior}{a priori}} information. The RCSP algorithms have been shown to be able to reduce the sensitivity of CSP to noise and artifacts and overcome the nonstationary and overfitting issues by incorporating variant penalty terms \citep{lotte2011RegularizingCommonSpatial, samek2012StationaryCommonSpatial, ghanbar2021CorrelationbasedCommonSpatial, onaran2013SparseSpatialFilter}.

Note that in this case, the best spatial filter for class 1 should also minimize the penalty $P(\mvec{w})$ which does not ensure a minimum variance ratio for class 2. Therefore, the optimized filters should be computed separately for each class by maximizing the modified objective function of CSP in \eqref{eq:J1-csp} as follows:
\begin{equation} \label{eq:J-rcsp}
	J_{P_k}(\mvec{w}) = \dfrac{\mvec{w}\transpose \tilde{\mmat{C}}_k \mvec{w}}{\mvec{w}\transpose \tilde{\mmat{C}} \mvec{w} + \alpha P(\mvec{w})}, k = 1, 2
\end{equation}
where $\alpha$ is the user-defined regularization parameter ($\alpha \geq 0$) that modulates the effect of the penalty. The higher $\alpha$ is, the more satisfied prior information is enforced for the filters. The penalty term $P(\mvec{w})$ measures how much the spatial filter $\mvec{w}$ satisfies a given prior. The more $\mvec{w}$ satisfies it, the smaller $P(\mvec{w})$ is. Therefore, to maximize $J_{P_k}$, $P(\mvec{w})$ must be minimized, leading to spatial filters fulfill\added{ing} the specified prior. Thereby the regularization guides the filters toward robustness and stationar\mreplaced{y}{ity}. Generally, $P(\mvec{w})$ can be formulated in terms of quadratic or non-quadratic forms, leading to different CSP variants. In this work, the following regularizations are considered.

\paragraph*{TRCSP} \label{sec:trcsp}

For \added{the} Tikhonov regularized CSP (TRCSP) algorithm \citep{lotte2011RegularizingCommonSpatial}, a quadratic function $P_{TR}(\mvec{w}) = \mvec{w}\transpose \mvec{w}$ is used as the penalty term\deleted{, where \added{the} matrix $\mmat{K}$ that encodes the prior information is simply the identity matrix, i.e., $\mmat{K} = \mmat{I}$}. Thus the penalty term is then the squared $L_2$-norm of $\mvec{w}$ noting $P_{TR}(\mvec{w}) = \mvec{w}\transpose \mmat{I} \mvec{w} = \|\mvec{w}\|_2^2$. The objective function of TRCSP is then  
\begin{equation} \label{eq:J-trcsp}
	J_{TR_k}(\mvec{w}) = \dfrac{\mvec{w}\transpose \tilde{\mmat{C}}_k \mvec{w}}{\mvec{w}\transpose (\tilde{\mmat{C}} + \alpha \mmat{I}) \mvec{w}}, k = 1, 2
\end{equation}
Since the penalty enforces the filters with a small $L_2$-norm, solutions $\mvec{w}$ with large weights are avoided. Thus\added{,} optimal spatial filters \mreplaced{could}{can} be obtained especially with small and noisy training datasets. Compared to CSP, TRCSP \mreplaced{could}{can} reduce the sensitivity to artifacts and the tendency to overfitting.

\paragraph*{sCSP} \label{sec:scsp}

The stationary CSP (sCSP) algorithm \citep{samek2012StationaryCommonSpatial} aims to extract discriminative and stationary features. To this end, the algorithm introduces a penalty term that measures the within-class stationarity, which is the sum of absolute differences between the projected class-mean  variance and the projected variance in each trial
\begin{equation} 
	P_s(\mvec{w}) = \sum_{k=1}^2 \sum_{i \in \Omega_k} |\mvec{w}\transpose \mmat{C}_i \mvec{w} - \mvec{w}\transpose \tilde{\mmat{C}}_k \mvec{w}|
\end{equation}
However, $P_s(\mvec{w})$ is a non-quadratic form and cannot be incorporated directly into \added{the} Rayleigh quotient in \eqref{eq:J-rcsp}. Maximizing the objective function \eqref{eq:J-rcsp} is difficult and a non-convex problem which has no closed-form solution. \cite{samek2012StationaryCommonSpatial} proposed an approximation of $P_s(\mvec{w})$ by flipping the signs of all the negative eigenvalues of the resultant difference matrices $\mmat{C}_i - \tilde{\mmat{C}}_k$. Thus, the estimated difference matrices become positive definite, yielding a quadratic form penalty. Thereby the sCSP optimization problem can be directly solved as a generalized eigenvalue problem.

\paragraph*{sTRCSP} \label{sec:strcsp}
sCSP aims at extracting stationary features, but it is not able to handle rank deficient matrices and does not reduce the sensitivity to noise and overfitting as TRCSP \mreplaced{dose}{does}. Therefore, \added{the} stationary TRCSP (sTRCSP) algorithm \citep{samek2012StationaryCommonSpatial}, by combining both sCSP and TRCSP, is to maximize the following objective function: 
\begin{equation} \label{eq:J-strcsp}
	J_{sTR_k}(\mvec{w}) = \dfrac{\mvec{w}\transpose \tilde{\mmat{C}}_k \mvec{w}}{\mvec{w}\transpose \tilde{\mmat{C}} \mvec{w} + \alpha P_s(\mvec{w}) + \beta P_{TR}(\mvec{w})}, k = 1, 2
\end{equation}
where $P_s(\mvec{w})$ and $P_{TR}(\mvec{w})$ are the penalty terms of sCSP and TRCSP, respectively, and $\alpha$ and $\beta$ are the corresponding regularization parameters to be determined.

\subsubsection{Multi-class extension of CSP}

The conventional CSP algorithms are designed for binary classification. In this work, the widely used pair-wise (PW) and one-versus-rest (OVR) based multi-class extensions of CSP \citep{ang2012FilterBankCommon} are considered (CSP-PW and CSP-OVR for short). Both \deleted{CSP-PW and CSP-OVR }approaches are based \added{on} separating the original multi-class problem into several binary problems. For instance, given that $\Omega = \{L, R, T, F\}$ representing the set of class labels of left, right, tongue, and foot MI tasks, the PW approach computes the CSP features for all the binary pair\added{s} of MI tasks. For each pair, a binary classifier is trained using the extracted features. For the 4 classes of MI tasks in DS1 and DS2 used in this work, $4 \times (4 - 1)/2 = 6$ binary pairs are required, yielding 6 binary classification results (predicted class labels). Therefore, 6 CSP procedures and 6 binary classifiers are needed. Finally a majority voting scheme based on the binary classification is used to realize the final 4-class classification. On the other hand, the OVR method computes the CSP features that discriminate\deleted{s} each class from the rest of the MI tasks. For the 4 classes problem, totally 4 CSP procedures and 4 binary classifiers are required. The final classification result is achieved based on the value of classifiers' outputs (the sample is classified as the class that the corresponding binary classifier gives the largest output).

\subsection{Proposed scatter-based framework}

\subsubsection{Overview}

\mreplaced{From the previous}{As explained in} \cref{sec:csp}, the conventional CSP algorithm includes two steps: whitening the class-mean covariance matrices with the composite covariance and then applying the orthogonal projection which maximizes $J^*(\mvec{w})$ in \eqref{eq:J2-csp-whitened}. It could be noted that maximizing $J^*(\mvec{w})$ is in fact a principal component analysis (PCA) process to the whitened covariance \citep{ghojogh2019eigenvalue}. Moreover, both the filter computation and the feature extraction processes are operated on the $N_c \times N_c$ covariance matrices in terms of quadratic forms. In this work, the covariance matrices are transform\added{ed} into $N_c^2$-dimensional vectors in the vectorized covariance space (a $N_c^2$-dimensional vector space). Then with the help of vectorization of matrices, the above mentioned quadratic forms in terms of class-mean covariance matrices can be transformed into linear projections in terms of $N_c^2$-dimensional class-mean covariance vectors. This section introduces the proposed scatter-based framework (scaCSP) for computing spatial filters by using scatter matrices defined in the $N_c^2$-dimensional vectorized covariance space.

\subsubsection{scaCSP in \added{the} binary case\deleted{: an equivalent to CSP}}

Given the whitened covariance matrices $\mmat{R}_i = \mmat{P}_c\transpose \mmat{C}_i \mmat{P}_c, i \in \Omega$, their vectorizations are denoted by $\mathbb{R}^{N_c^2} \ni \mvec{r}_i = \text{vec}(\mmat{R}_i)$, where vec($\cdot$) is the vectorization operator which concentrates the columns of a matrix into a single column vector, $\Omega = \bigcup_{k=1}^{N_\Omega} \Omega_k$ is the set of trial labels, and $N_\Omega$ is the number of classes (in the binary case, $N_\Omega = 2$). The vectorizations of class-mean covariance matrices are then 
\begin{equation} \label{eq:tilde-rk}
	\tilde{\mvec{r}}_k = \text{vec}(\tilde{\mmat{R}}_k) = \frac{1}{|\Omega_k|} \sum_{i \in \Omega_k} \mvec{r}_i
\end{equation}
Let $\tilde{\mvec{r}} \in \mathbb{R}^{N_c^2}$ be the mean vector of vectorizations of all the whitened covariance matrices given by 
\begin{equation} \label{eq:tilde-r}
	\tilde{\mvec{r}} = \frac{1}{|\Omega|} \sum_{i \in \Omega} \mvec{r}_i 
	= \dfrac{1}{|\Omega|} \sum_{k=1}^{N_\Omega} |\Omega_k| \tilde{\mvec{r}}_k
\end{equation}
Note that $\tilde{\mvec{r}}$ is not the vectorization of the composite covariance matrix $\tilde{\mmat{R}}$. Then the within-class scatter matrix $\mmat{S}_w \in \mathbb{R}^{N_c^2 \times N_c^2}$, between-class scatter matrix $\mmat{S}_b \in \mathbb{R}^{N_c^2 \times N_c^2}$, and the total scatter matrix $\mmat{S}_t \in \mathbb{R}^{N_c^2 \times N_c^2}$ are defined as \citep{tharwat2017LinearDiscriminantAnalysis, bishop2006PatternRecognitionMachine}
\begin{subequations} \label{eq:scatter-of-covariance}
	\begin{align}
		&\mmat{S}_w = \sum_{k=1}^{N_\Omega} \sum_{i \in \Omega_k} \left( \mvec{r}_i - \tilde{\mvec{r}}_k \right) \left( \mvec{r}_i - \tilde{\mvec{r}}_k \right)\transpose \label{eq:subeq-Sw} \\
		&\mmat{S}_b = \sum_{k=1}^{N_\Omega} |\Omega_k| \left( \tilde{\mvec{r}}_k - \tilde{\mvec{r}} \right) \left( \tilde{\mvec{r}}_k - \tilde{\mvec{r}} \right)\transpose \label{eq:subeq-Sb} \\
		&\mmat{S}_t = \sum_{i \in \Omega} (\mvec{r}_i - \tilde{\mvec{r}}) (\mvec{r}_i - \tilde{\mvec{r}})\transpose \label{eq:subeq-St}
	\end{align}
\end{subequations}
with $\mmat{S}_t = \mmat{S}_w + \mmat{S}_b$, whose ranks are given by 
\begin{subequations} \label{eq:rank-of-scatter-matrix}
	\begin{align}
		\text{rank}(\mmat{S}_w) &= \text{min}\{|\Omega| - N_\Omega, \frac{N_c(N_c + 1)}{2}\} \label{eq:subeq-rank-Sw} \\
		\text{rank}(\mmat{S}_b) &= \text{min}\{N_\Omega - 1, \frac{N_c(N_c + 1)}{2}\} \label{eq:subeq-rank-Sb} \\
		\text{rank}(\mmat{S}_t) &= \text{min}\{|\Omega| - 1, \frac{N_c(N_c + 1)}{2}\} \label{eq:subeq-rank-St}
	\end{align}
\end{subequations}
The scatter matrices can never be full rank since all the covariance matrices are symmetric, leading the ranks no larger than $\frac{N_c(N_c + 1)}{2}$. Practically\added{,} in BCI applications the class number is much smaller than the number of EEG channels, i.e., $N_\Omega - 1 < \frac{N_c(N_c + 1)}{2}$. In this work, we assume that $\text{rank}(\mmat{S}_b) = N_\Omega - 1$.

In the vectorized covariance space, the proposed framework aims to find a projection vector $\mvec{v} \in \mathbb{R}^{N_c^2}$ that maximizes the between-class scatters of the projected samples (here the samples are the vectorized covariance matrices $\mvec{r}_i$ for the EEG trials), i.e. maximizing the Rayleigh quotient: 
\begin{equation} \label{eq:Qv-Sb}
	Q(\mvec{v}) = \dfrac{\mvec{v}\transpose \mmat{S}_b \mvec{v}}{\mvec{v}\transpose \mvec{v}}
\end{equation}
whose solution is given by the eigenvectors of the eigendecomposition problem: 
\begin{equation} \label{eq:eigen-Sb}
	\mmat{S}_b \mmat{V} = \mmat{V} \mmat{\Lambda}
\end{equation}
where diagonal entries of $\mmat{\Lambda} = \text{diag}(\lambda_1, \cdots, \lambda_{N_c^2})$ are the eigenvalues and  the columns of $\mmat{V} = [\mvec{v}_1, \cdots, \mvec{v}_{N_c^2}]$ are the corresponding eigenvectors.

In case of the binary classification problem ($N_\Omega = 2$), only one eigenvalue is nonzero. Let $\mvec{v}$ be the eigenvector corresponding to the nonzero eigenvalue and $\mvec{v} = \text{vec}(\mmat{A})$ with $\mmat{A} = \mmat{A}\transpose$. The eigendecomposition of $\mmat{A}$ is given by:
\begin{equation} \label{eq:eigen-A}
	\mmat{A} = \mmat{U}_a \Lambda_a \mmat{U}_a\transpose
\end{equation}
where the eigenvalues in $\mmat{\Lambda}_a = \text{diag}(\lambda_{a1}, \cdots, \lambda_{aN_c})$ are sorted in non-increasing order and columns of $\mmat{U}_a = [\mvec{u}_{a1}, \cdots, \mvec{u}_{aN_c}]$ are the corresponding eigenvectors. Finally we can construct $\mathbb{R}^{N_c \times N_c} \ni \mmat{W}_a = \mmat{P}_c \mmat{U}_a$ whose columns $\mvec{w}_{ai} = \mmat{P}_c \mvec{u}_{ai}$ are the spatial filters computed by scaCSP.

\begin{theorem} \label{theorem1}
	Let $\mathbb{R}^{N_c \times N_c} \ni \mmat{W} = \mmat{P}_c \tilde{\mmat{U}}_1$ be the spatial filters (columns that are sorted by $\tilde{\lambda}_i$) computed by the CSP algorithm in \cref{sec:csp} and $\mathbb{R}^{N_c \times N_c} \ni \mmat{W}_a = \mmat{P}_c \mmat{U}_a$ the spatial filters (columns that are sorted by $\lambda_{ai}$) by the scatter-based framework in this section, both of which can be decomposed into a whitening projection $\mmat{P}_c \in \mathbb{R}^{N_c \times N_c}$ with $\mmat{P}_c\transpose \tilde{\mmat{C}} \mmat{P}_c = \mmat{I}$ and an orthogonal projection $\tilde{\mmat{U}}_1 \in \mathbb{R}^{N_c \times N_c}$ from \eqref{eq:eigen-tilde-R1} or $\mmat{U}_a \in \mathbb{R}^{N_c \times N_c}$ from \eqref{eq:eigen-A}, then 
	\begin{subequations} \label{eq:theorem1}
		\begin{align}
			\mmat{W}_a &= \mmat{W} \\
			\mmat{U}_a &= \tilde{\mmat{U}}_1 \\
			\mvec{d}_a &= \dfrac{2\tilde{\mvec{d}}_1 - \mvec{1}^{N_c}}{\| 2\tilde{\mvec{d}}_1 - \mvec{1}^{N_c} \|_2}
		\end{align}
	\end{subequations}
	Here \deleted{\added{are} vectors} $\mathbb{R}^{N_c} \ni \mvec{d}_a = [\lambda_{a1}, \cdots, \lambda_{aN_c}]\transpose$ and $\mathbb{R}^{N_c} \ni \tilde{\mvec{d}}_1 = [\tilde{\lambda}_1, \cdots, \tilde{\lambda}_{N_c}] \transpose$ \mreplaced{by}{are vectors} collecting the eigenvalues from \eqref{eq:eigen-A} and \eqref{eq:eigen-tilde-R1} into a single vector respectively.
\end{theorem}

\noindent\textit{Proof.} The proof is given in \cref{sec:proof-theom1}. 
%\medskip
\newline

%\begin{proof}
%	The proof is given in \cref{sec:proof-theom1}. 
%\end{proof}

\cref{theorem1} says that the eigendecomposition \eqref{eq:eigen-A} for the linear projection $\mvec{v}$, which maximizes $Q(\mvec{v})$ in \eqref{eq:Qv-Sb}, yields the equivalent spatial filters to that computed by CSP. To keep the proposed scatter-based framework consistent with CSP in terms of formulation raising from un-whitened covariance matrices $\mmat{C}_i$, the modified objective function of \eqref{eq:Qv-Sb} and its solution are detailed in \cref{sec:scacsp-unwhitening}. It should be noted that scaCSP formulated with $\mmat{C}_i$ still provides equivalent CSP filters for binary problems. In that case, the whitening processing would be incorporated into the denominator of the Rayleigh quotient.

The quadratic forms for computing feature vectors in \eqref{eq:features} can be converted into a linear projection:
\begin{equation} \label{eq:feature-vectorized}
	\begin{split}
		\mvec{f}_i &= \mmat{V}_a\transpose \mvec{r}_i, \\
		\text{with } \mmat{V}_a &= [\mvec{u}_{a1} \otimes \mvec{u}_{a1}, \cdots, \mvec{u}_{aN_c} \otimes \mvec{u}_{aN_c}]
	\end{split}
\end{equation}
Moreover, the scatter-based framework provides an alternative criterion for feature selection. The eigenvalues in $\mvec{d}_a$ \mreplaced{is}{are} a measure of the normalized \mreplaced{bandpower}{band power} difference between class 1 and class 2. More precisely, noting that the eigenvalues $\tilde{\lambda}_i$ (respectively $1 - \tilde{\lambda}_i$) are the \mreplaced{bandpower}{band power} ratio between class 1 (resp. class 2) with the composite of both classes, the \mreplaced{bandpower}{band power} difference between class 1 and 2 are then given by $\tilde{\lambda}_i - (1 - \tilde{\lambda}_i) = 2\tilde{\lambda}_i - 1$. Collecting them into a single vector, which is then normalized by its $L_2$--norm, we finally obtain $\mvec{d}_a$. Therefore, a larger magnitude of $|\lambda_{ai}|$ indicates a more discriminative feature from that corresponding filter in terms of \mreplaced{bandpower}{band power}. 
%Therefore, in this work we choose features according to the absolute value of $\lambda_{ai}$, i.e., by $|\mvec{d}_a|$.

\subsubsection{scaCSP in the general multi-class case}

\begin{algorithm}[ht] 
	\caption{scaCSP in general multi-class settings.}
	\label{alg:multi-scaCSP}
	\BlankLine
	\KwIn{EEG data matrices $\mmat{X}_i$ from $N_\Omega$ classes $i \in \Omega = \bigcup_{k=1}^{N_\Omega} \Omega_k$; Number of CSP filters: $m$.}
	\KwOut{Feature vectors $\mvec{f}_i \in \mathbb{R}^{m(N_\Omega - 1)}$ for each EEG trial.}
	\BlankLine
	Compute covariance matrices for each trial $\mmat{C}_i = \frac{1}{N_t - 1} \mmat{X}_i \mmat{X}_i\transpose$ \;
	Calculate class-mean covariance matrix $\tilde{\mmat{C}}_k = \frac{1}{|\Omega_k|} \sum_{i \in \Omega_k} \mmat{C}_i$ and composite covariance matrix $\tilde{\mmat{C}} = \sum_{k=1}^{N_\Omega} \tilde{\mmat{C}}_k$ \;
	Compute whitening transform matrix $\mmat{P}_c = \tilde{\mmat{U}}_c \tilde{\mmat{\Lambda}}_c^{(-1/2)}$ from eigendecomposition $\tilde{\mmat{C}} = \tilde{\mmat{U}}_c \tilde{\mmat{\Lambda}}_c \tilde{\mmat{U}}_c\transpose$ \;
	Compute the whitened covariance matrices $\mmat{R}_i = \mmat{P}_c\transpose \mmat{C}_i \mmat{P}_c$ \;
	Compute the between-class scatter matrix $\mmat{S}_b$ according to \eqref{eq:scatter-of-covariance} using vectorized covariance in \eqref{eq:tilde-rk} and \eqref{eq:tilde-r} \;
	Find $N_\Omega - 1$ basis vectors as in \eqref{eq:range-Sb} from the eigendecomposition of $\mmat{S}_b$ in \eqref{eq:eigen-Sb} \;
	\For{$i = 1: N_\Omega - 1$}{
		Compute eigenvalues $\lambda_{aj}^i$ and eigenvectors $\mvec{u}_{aj}^i, j = 1, \cdots, N_c,$ using \eqref{eq:eigen-Ai} \;
		Choose $m$ eigenvectors $\mvec{u}_{aj}^i, j = 1, \cdots, m,$ according to $m$ largest $|\lambda_{aj}^i|$ \;
		$\mmat{V}_a^i \gets [\mvec{u}_{a1}^i \otimes \mvec{u}_{a1}^i, \cdots, \mvec{u}_{am}^i \otimes \mvec{u}_{am}^i]$ \;
	}
	$\mmat{V}_a \gets [\mmat{V}_a^1, \cdots, \mmat{V}_a^{N_\Omega-1}]$ \;
	Return the feature vectors computed by \eqref{eq:feature-vectorized}.
\end{algorithm}

The proposed scatter-based framework paves the way for us to revisit the conventional CSP algorithm in a new scenario. In the binary case ($N_\Omega = 2$), the rank of $\mmat{S}_b$ is one, leading to a one-dimensional range space $\mmat{S}_b^{range}$. Therefore, only one projection vector could be found in $\mmat{S}_b^{range}$, whose eigendecomposition yields the equivalent CSP filters. Following the basic idea of multi-class linear discriminant analysis (LDA) \citep{bishop2006PatternRecognitionMachine}, we can find $N_\Omega - 1$ independent projection vectors in $\mmat{S}_b^{range}$ which leads to a straightforward extension of scaCSP to the more general multi-class problem.

Considering the $N_\Omega$-class classification problem, the dimension of $\mmat{S}_b^{range}$ is $N_\Omega - 1$. Suppose $\mmat{S}_b^{range}$ is spanned by
\begin{equation} \label{eq:range-Sb}
	\begin{split}
		\mmat{S}_b^{range} &= \text{span}(\mvec{v}_1, \cdots, \mvec{v}_{N_\Omega - 1})
	\end{split}
\end{equation}
where basis vectors $\mvec{v}_i$ can be given by the eigenvectors corresponding to nonzero eigenvalues computed with \eqref{eq:eigen-Sb}. Let $\mvec{v}_i = \text{vec}(\mmat{A}_i)$ with $\mmat{A}_i = \mmat{A}_i\transpose$. Then $\mmat{A}_i$ is factorized into its eigendecomposition as 
\begin{equation} \label{eq:eigen-Ai}
	\begin{split}
		\mmat{A}_i &= \mmat{U}_a^i \mmat{\Lambda}_a^i {\mmat{U}_a^i}\transpose, i = 1, \cdots, N_\Omega - 1.
	\end{split}	
\end{equation}
where \added{the} diagonal matrix $\mmat{\Lambda}_a^i = \text{diag}(\lambda_{a1}^i, \cdots, \lambda_{aN_c}^i)$ consists of eigenvalues $\lambda_{aj}^i$ sorted in non-increasing order and columns of $\mmat{U}_a^i$ are the corresponding eigenvectors $\mvec{u}_{aj}^i$. From each basis orientation $\mvec{v}_i$ of $\mmat{S}_b^{range}$, we can construct $N_c$ spatial filters collected into columns of $\mmat{W}^i = [\mvec{w}_1^i, \cdots, \mvec{w}_{N_c}^i]$ with $\mvec{w}_j^i = \mmat{P}_c\transpose \mvec{u}_{aj}^i$. Finally, $\mmat{W}^i$ are concentrated into the multi-class spatial filters $\mmat{W} = [\mmat{W}^1, \cdots, \mmat{W}^{N_\Omega - 1}]$. We collect \added{the} eigenvalues $\lambda_{aj}^i$ into a column vector $\mvec{d}_a^i$ and vectors $\mvec{u}_{aj}^i \otimes \mvec{u}_{aj}^i$ into columns of $\mmat{V}_a^i$. Then the multi-class feature vectors can be computed by linearly projecting samples $\mvec{r}_i$ onto $\mmat{V}_a = [\mmat{V}_a^1, \cdots, \mmat{V}_a^{N_\Omega-1}]$ as in \eqref{eq:feature-vectorized}. For feature selection, we can choose those according to $|\mvec{d}_{a}^i|$ from each basis orientation individually. The procedure to calculate feature vectors using multi-scaCSP is described in \cref{alg:multi-scaCSP}.

\subsubsection{Subspace enhancement of scaCSP}

By introducing scatter matrices in the vectorized covariance space, there are generally six informative subspaces containing significant discriminant information that could be useful for classification (see \cref{sec:empirical} for an empirical demonstration): null space of $\mmat{S}_b$ ($\mmat{S}_b^{null}$), range space of $\mmat{S}_b$ ($\mmat{S}_b^{range}$), null space of $\mmat{S}_w$ ($\mmat{S}_w^{null}$), range space of $\mmat{S}_w$ ($\mmat{S}_w^{range}$), null space of $\mmat{S}_t$ ($\mmat{S}_t^{null}$), range space of $\mmat{S}_t$ ($\mmat{S}_t^{range}$). Note that the range spaces (resp., null spaces) are spanned by the eigenvectors corresponding to the nonzero eigenvalues (resp., zero eigenvalues) of the scatter matrices. Therefore, from \cref{theorem1} we see that only $\mmat{S}_b^{range}$ is used in the conventional CSP approach for the \added{computation of} spatial filters\deleted{ computation}, i.e., only discriminant information contained in $\mmat{S}_b^{range}$ \mreplaced{are}{is} taken into account whereas that in other subspaces \mreplaced{are}{is} ignored by CSP.

In this work, we introduced several extensions of the scaCSP by additionally using subspaces of $\mmat{S}_b$, $\mmat{S}_w$\added{,} and $\mmat{S}_t$ in order to incorporat\deleted{ing}\added{e} more discriminative information contained in these subspaces for \deleted{the }classification performance improvement. Furthermore, $\mmat{S}_w$ measures the within-class variations. Therefore, combining the objective function of scaCSP with subspaces of $\mmat{S}_w$ would be potentially helpful \mreplaced{to}{in} extracting stationary features. More precisely, extensions of scaCSP by incorporating range spaces and null spaces of the scatter matrices can be realized in two scenarios. On one hand, instead of extracting spatial filters from only $\mmat{S}_b^{range}$, we are seeking more extra spatial filters from other spaces. On the other hand, data components in some spaces may contain less effective discriminant information which should be removed in order to improve the signal-to-noise ratio. A more detailed empirical analysis of these two scenarios can be found in \cref{sec:empirical}. Here, the following possible extensions are highlighted:

1) \textit{\added{scaCSP enhanced by} \mreplaced{E}{e}xtra spatial filters from subspaces} (scaCSP\mreplaced{-extraSub}{$_{(extraSub)}$}): Following the basic idea of the multi-class extension of scaCSP, the scatter-based framework also provides another scenario to find more spatial filters from subspaces of the scatter matrices that could be possibly helpful for improving the classification performance. In this work, we considered \added{the} following subspaces: $\mmat{S}_b^{null}$, $\mmat{S}_w^{range}$, $\mmat{S}_w^{null}$, and $\mmat{S}_t^{range}$ as well as different combinations of them. The detailed steps for computing extra spatial filters from the considered subspace or subspace combinations are give\added{n} in \cref{alg:scaCSP-extraSub}. \added{The extra spatial filters are then concentrated into the scaCSP filters for the sequential feature computation. We denote the extra spatial filters enhanced scaCSP by scaCSP$_{(extraSub)}$, where the subscript "extraSub" indicates the used subspaces. For example, scaCSP$_{(\mmat{S}_b^{null})}$ (resp., scaCSP$_{(\mmat{S}_b^{null} + \mmat{S}_w^{range})}$) indicates that scaCSP filters are used additionally with extra spatial filters computed from subspace $\mmat{S}_b^{null}$ (resp., subspaces $\mmat{S}_b^{null}$ and $\mmat{S}_w^{range}$).}

\begin{algorithm}[ht] 
	\caption{Extra spatial filters from subspaces\deleted{: scaCSP-extraSub}.}
	\label{alg:scaCSP-extraSub}
	\BlankLine
	\KwIn{Scatter matrices \added{of covariance samples of} EEG data\deleted{ matrices} $\mmat{S}_b$, $\mmat{S}_w$, and $\mmat{S}_t$; Number of extra filters: $m$.}
	\KwOut{$m$ extra spatial filters.}
	\BlankLine
	Find the basis vectors of $\mmat{S}_b^{null}$, $\mmat{S}_w^{range}$, $\mmat{S}_w^{null}$, and $\mmat{S}_t^{range}$ by eigendecomposition \;
	Let $\mathbb{R}^{N_c^2 \times n} \ni \mmat{V}^{extra} = [\mvec{v}_1, \cdots, \mvec{v}_n]$ be the matrix whose columns span one or any combination of $\mmat{S}_b^{null}$, $\mmat{S}_w^{range}$, $\mmat{S}_w^{null}$, and $\mmat{S}_t^{range}$ \;
	Initialize $\mmat{W} = \mmat{0}$ and $\mvec{d} = \mvec{0}$ \;
	\For{$i = 1: n$}{
		$\mmat{A}$ $\gets$ reshape($\mvec{v}_i, N_c, N_c$) \;
		$\mmat{A}$ $\gets$ $\frac{1}{2} (\mmat{A} +\mmat{A}\transpose)$ \;
		Compute the \added{eigendecomposition} $\mmat{A} = \sum_{j=1}^{N_c} \lambda_{j} \mvec{u}_{j} \mvec{u}_{j}\transpose$ \;
		Concentrate $\mvec{u}_{j}$ into columns of matrix $\mmat{W}$: $\mmat{W}$ $\gets$ $[\mmat{W}, \mvec{u}_{1}, \cdots, \mvec{u}_{N_c}]$ \;
		Concentrate $\lambda_{j}$ into vector $\mvec{d}$: $\mvec{d}\transpose$ $\gets$ $[\mvec{d}\transpose, \lambda_{1}, \cdots, \lambda_{N_c}]$ \;
	}
	Return $m$ columns of $\mmat{W}$ corresponding to $m$ largest $|\lambda_{j}|$ in $|\mvec{d}|$.
\end{algorithm}

2) \textit{Null\mreplaced{ S}{s}pace\mreplaced{ }{-}Reduced scaCSP} (scaCSP-NSR): Note that scatter-based framework extracts features $\mvec{f}_i$ by finding linear projections $\mmat{V}$. The between-class and within-class scatters of $\mvec{f}_i$ are then given by $\mmat{V}\transpose \mmat{S}_b \mmat{V}$ and $\mmat{V}\transpose \mmat{S}_w \mmat{V}$, respectively. As for improving the classification performance, the optimal linear filters should give large between-class scatters whilst small within-class scatters, or in \deleted{an}other word\added{s} \deleted{the }Fisher's criterion $|\mmat{V}\transpose \mmat{S}_b \mmat{V}|/ |\mmat{V}\transpose \mmat{S}_w \mmat{V}|$ should be as large as possible \citep{fukunaga1990IntroductionStatisticalPattern, bishop2006PatternRecognitionMachine}. In terms of maximizing \deleted{the }Fisher's criterion, it is well known that $\mmat{S}_t^{null}$ does not contain any discriminant information since $\mmat{S}_t^{null}$ is the common null space of both $\mmat{S}_b$ and $\mmat{S}_w$ (i.e., $\mmat{S}_t^{null} = \mmat{S}_b^{null} \cap \mmat{S}_w^{null}$) \highlight{\citep{huang2002SolvingSmallSample}}. Therefore, we can remove the signal components in $\mmat{S}_t^{null}$ from the EEG data in order to suppress task-irrelevant signals. This process can be employed with the help of the eigendecomposition of $\mmat{S}_t$. For example, let $\mmat{S}_t^{null}$ be spanned by the columns of a matrix $\mmat{Q}$, which can be obtained by collecting the eigenvectors of $\mmat{S}_t$ corresponding to zero eigenvalues. Then the data components in $\mmat{S}_t^{null}$ can be computed by projecting the vectorized covariance samples $\mvec{r}_i$ onto $\mmat{Q} \mmat{Q}\transpose$, which are then removed\added{,} namely, by subtracting $\mmat{Q} \mmat{Q}\transpose \mvec{r}_i$ from $\mvec{r}_i$. Finally, we can compute the feature vectors using the reduced covariance samples. More precisely, let columns of $\mmat{V}$ be the spatial filters computed using the binary-scaCSP or multi-scaCSP or additionally scaCSP\mreplaced{-extraSub}{$_{(extraSub)}$}. Then the final common null space reduced (scaCSP-CNSR) feature vectors are given by $\mvec{f}_i = \mmat{V}\transpose (\mmat{I} - \mmat{Q} \mmat{Q}\transpose) \mvec{r}_i$. Moreover, the data components in $\mmat{S}_b^{null}$ lead\deleted{s} to a zero between-class scatter of the features, which makes no contribution to \deleted{the }Fisher's criterion. Therefore, removing the $\mmat{S}_b^{null}$ components from the data with the introduced process (termed as scaCSP-BNSR) may also have the potential to improve the classification performance.

3) \textit{Null Space Reduced scaCSP\mreplaced{-extraSub}{$_{(extraSub)}$}} (scaCSP-NSR\mreplaced{-extraSub}{$_{(extraSub)}$}): It can be noted that the above introduced extensions scaCSP\mreplaced{-extraSub}{$_{(extraSub)}$} and scaCSP-NSR use information from subspaces in two individual scenarios (spatial filter learning scenario and subspace component projecting scenario, respectively, see \cref{sec:empirical}). Then these two extensions can be combined together, i.e., on one hand, extract extra spatial filters from subspaces and\added{,} on the other hand, remove data components in $\mmat{S}_t^{null}$ (scaCSP-CNSR\mreplaced{-extraSub}{$_{(extraSub)}$}) or $\mmat{S}_b^{null}$ (scaCSP-BNSR\mreplaced{-extraSub}{$_{(extraSub)}$}).

\subsection{\added{Dataset description}} 

The proposed approach was evaluated on two datasets: a public dataset from BCI competition IV Dataset-2a with wet electrodes (DS1 for short) \citep{tangermann2012ReviewBCICompetition} and our in-house motor imagery dataset collected with dry electrodes (DS2).

\begin{figure}[ht!]
	\centering
	\begin{minipage}[ht!]{\textwidth}
		%\subfloat[][Timeline of the MI paradigm]{\includegraphics[width = 0.74\textwidth]{Timing_scheme_DS1_output.pdf}\label{subfig:timeline-ds1}}
		%\hfill
		%\subfloat[][Electrode layout in DS1\label{subfig:layout-ds1}]{\includegraphics[width = 0.25\textwidth]{layout_ds1.pdf}}
		\subfloat[][]{\includegraphics[width = 0.74\textwidth]{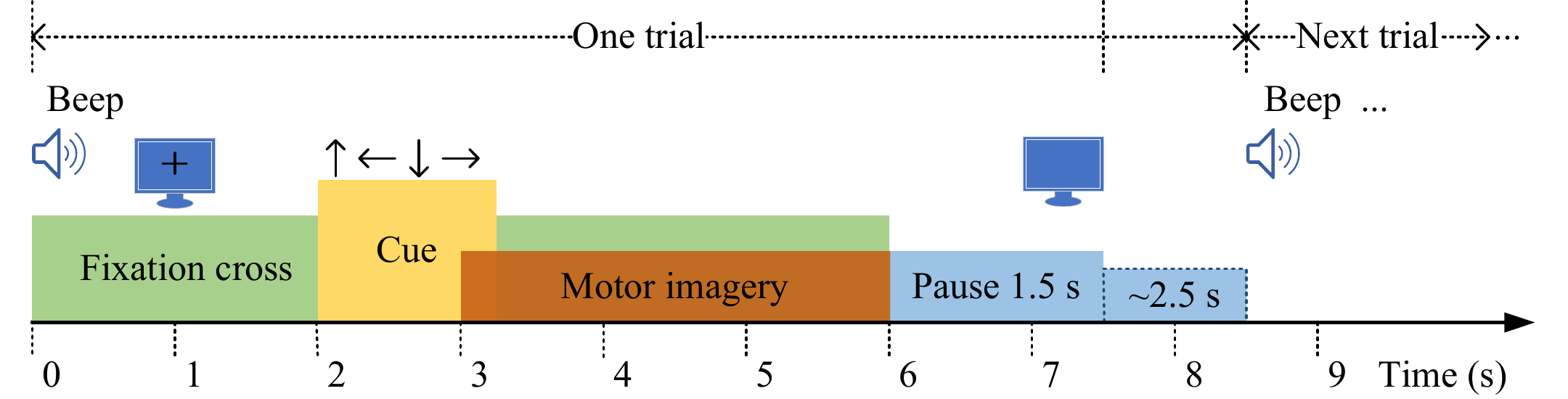}\label{subfig:timeline-ds1}}
		\hfill
		\subfloat[][\label{subfig:layout-ds1}]{\includegraphics[width = 0.25\textwidth]{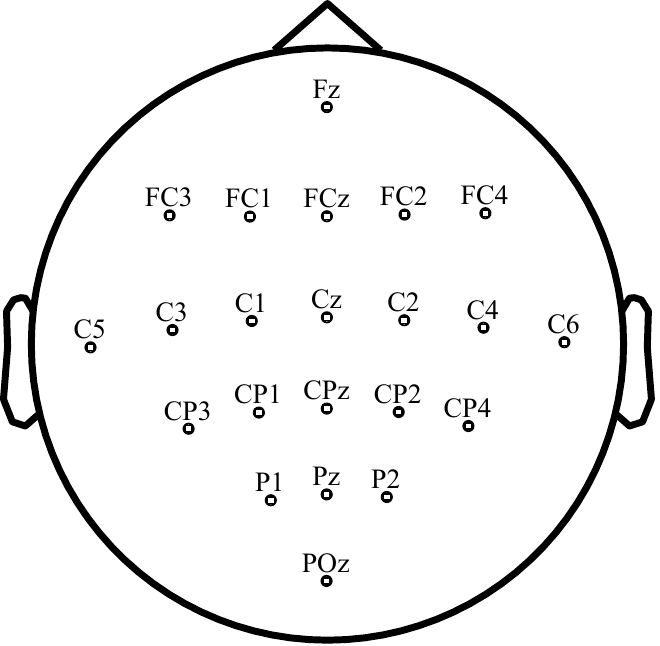}}
	\end{minipage}
	%\quad

	\begin{minipage}[ht!]{\textwidth}
		\subfloat[][]{\includegraphics[width = 0.74\textwidth]{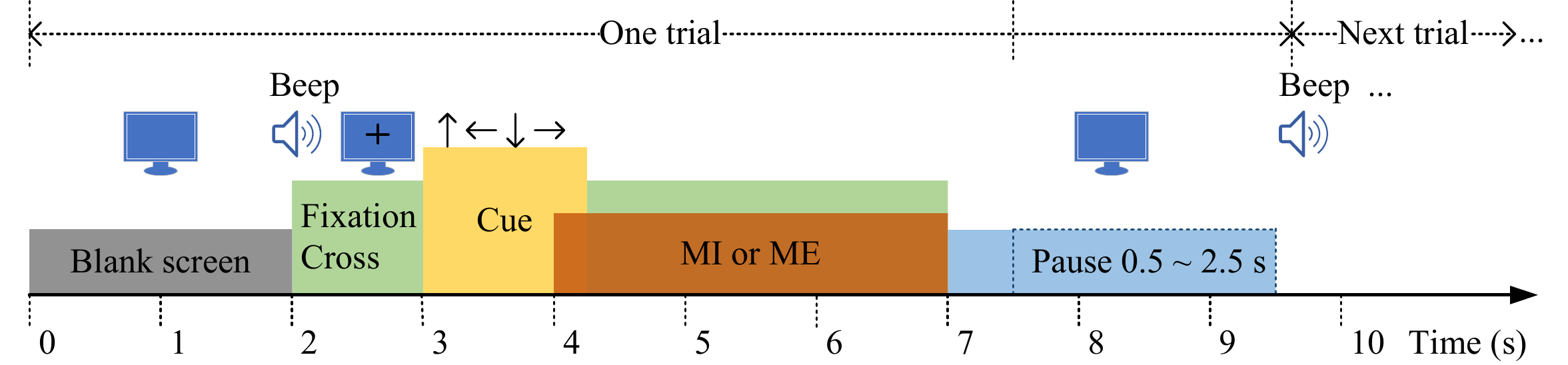}\label{subfig:timeline-ds2}}
		\hfill
		\subfloat[][\label{subfig:layout-ds2}]{\includegraphics[width = 0.25\textwidth]{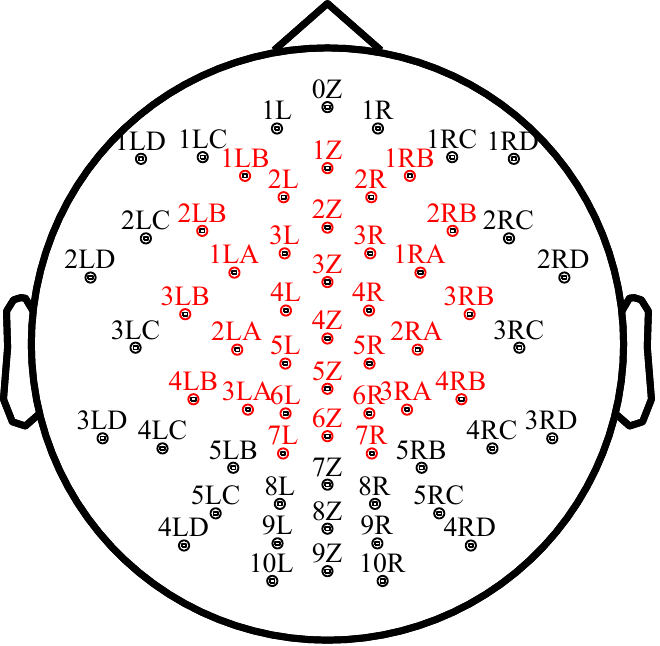}}
	\end{minipage}
	\caption{The trial timelines of the cued MI paradigm and electrode layouts used in DS1 (a and b) and in DS2 (c and d).}
	\label{fig:timeline-and-layout}
\end{figure}

Dataset DS1 consists of EEG recordings from 22 Ag/AgCl electrodes and nine healthy subjects (subjects A01-A09) performing four cue-based MI tasks: left hand, right hand, tongue, and foot (L, R, T, F for short, respectively). For each subject, two sessions (training session and testing session) on different days were recorded\deleted{ for each subject}, each of which is comprised of 72 trials per class, yielding a total of 288 trials per session. \cref{subfig:timeline-ds1} \mreplaced{is depicted}{depicts} the timeline of a trial in the cue-based MI-BCI paradigm. The electrode layout corresponding to the international 10-20 system is shown in Figure \cref{subfig:layout-ds1}. For more detail about the dataset, we refer to \cite{tangermann2012ReviewBCICompetition} and the BCI competition website (\url{https://www.bbci.de/competition/iv/}).

%\subref{subfig:timeline-ds2}
In DS2, the EEG recordings were performed at the Institute of Biomedical Engineering and Informatics of Technische Universit\"{a}t Ilmenau. This study complied with the ethical standards outlined in the Declaration of Helsinki and was approved by the local Ethics Committee. All volunteers \deleted{were }provided written informed consent before they participated in the study. The dataset is comprised of two sessions \added{recorded on different days}: a motor execution (ME) session followed by a MI session. The ME session is used to help subjects to perform the MI tasks. Each session includes 6 runs separated by \added{a} short break of \added{a} few minutes according to the subject\added{'s needs}. Each run includes 40 trials (10 trials per MI/ME task), yielding a total of 240 trials for each subject (60 trials per task). EEG data were collected from 10 healthy subjects (subjects B01-B10) performing four different MI/ME tasks (L, R, T, and F) following the cue-based paradigm (as shown in \cref{subfig:timeline-ds2}). The subjects were sitting \added{in} a comfortable armchair in front of a computer screen. At the beginning of each trial, a blank screen was displayed for 2 s. Then\added{,} a short acoustic stimulus indicated the beginning of the trial ($t = 2$ s) and a fixation cross was presented at the center of the screen for the next 5 s, which helped the subjects to focus on the screen. At $t = 3$ s, a cue in the form of an arrow pointing either to the left, right, up\added{,} or down (corresponding to one of the four MI tasks: L, R, T\added{,} or F, respectively) was displayed for 1.25 s. This prompted the subjects to perform the desired MI tasks until the fixation cross disappeared from the screen at $t = 7$ s. The order of cues was random. A short break of 0.5--2.5 s followed\added{,} where the screen turn\added{ed} to blank again allowing the subjects to relax. The brain activities were acquired using a portable 64-channel dry EEG cap with multipin Ag-AgCl electrodes (waveguard\textsuperscript{TM} touch, ANT neuro) corresponding to the equidistant layout, which is shown in \cref{subfig:layout-ds2}, with an eego\textsuperscript{TM} amplifier (ANT neuro) at a sampling frequency of 1024 Hz (24-bit resolution). Data were referenced to the right mastoid while \added{the} left \deleted{ear }mastoid \added{was} grounded. During the experiments, \mreplaced{electrode impedances}{the impedances of reference and ground electrode} were kept to be lower than 5 k$\Omega$. The raw signals were bandpass filtered between 0.5 and 128 Hz with an additional 50 Hz notch filter to remove line noise (Chebyshev Type II filters of order 10 with stopband ripple 50 dB) and then down-sampled to 256 Hz. In this study, only 32 channels fully covering the sensorimotor area, \deleted{which are }indicated by \added{the} red color in \cref{subfig:layout-ds2}, were used for the MI decoding. Only the MI dataset\added{s} were used, while the first four runs were grouped as the training set and the last two as \added{the} testing set. \added{An overview of the analyzed data sets DS1 and DS2 \mreplaced{are listed}{is given} in \cref{tab:dataset-info}.}

\begin{table}[ht!]
	%\scriptsize
	\smaller
	\centering
	\caption{\added{Overview of the investigated data sets.}}
	\label{tab:dataset-info}
	\begin{tabular}{*{10}{c}} % 9 columns
		\toprule
		&\multirow{2}{*}{\thead{Sub-\\jects}} &\multirow{2}{*}{\thead{Elec-\\trode}} &\multirow{2}{*}{\thead{Channels}} &\multirow{2}{*}{\thead{MI\\ tasks}} &\multirow{2}{*}{\thead{Sampling\\ frequency}} &\multicolumn{2}{c}{\thead{Trial number\\ in training set}} &\multicolumn{2}{c}{\thead{Trial number\\ in testing set}} \\
		\cline{7-10} &&&&& &\thead{Binary\\ problem} &\thead{Multi-\\class} &\thead{Binary\\ problem} &\thead{Multi-\\class} \\
		\midrule
		DS1 &\thead{A01-\\A09} &Wet &22 EEG &\thead{4 (L, R,\\ T, F)} &100 Hz &144 &288 	&144 &288 \\
		DS2 &\thead{B01-\\B10} &Dry &32 EEG &\thead{4 (L, R,\\ T, F)} &256 Hz &80  &160 	&40	 &80 \\ 
		\bottomrule
	\end{tabular}
\end{table}

\deleted{The challenges in DS2 are that the dataset is a relatively small one and the signal \deleted{quality }from dry electrodes is in general known to be \deleted{worse}somewhat more prone to artifacts (\mreplaced{in terms}{i.e.} of lower SNR\deleted{, for example}) compared to \deleted{that with }wet electrodes. Moreover, all the subjects in DS2 \mreplaced{do}{did} not have any experience on MI-BCIs (some of them even \mreplaced{know}{knew} nothing about MI-BCIs), i.e., they are all \deleted{the }so-called BCI-\mreplaced{n\"aive}{na\"ive} subjects. However, noting that practical MI-BCI applications with conventional wet electrodes still suffer some drawbacks such as the complexity \mreplaced{to}{of} install\added{ation}, MI-BCI using dry electrodes, especially with \mreplaced{n\"aive}{na\"ive} subjects, has the potential to be usable out of the lab.}

\subsection{\added{Experimental setup}}

As for preprocessing, the general settings for frequency band and time segment, which are the most commonly used in literature \citep{lotte2011RegularizingCommonSpatial, samek2012StationaryCommonSpatial, ang2012FilterBankCommon}, are applied to all the datasets. The time interval located from 0.5 s to 2.5 s after the cue is used, i.e., $2.5 \text{ s} \leq t \leq 4.5$ s for DS1 and $3.5 \text{ s} \leq t \leq 5.5$ s for DS2. EEG signals are bandpass-filtered between 7 Hz and 31 Hz covering both $\mu$-rhythm (8-14 Hz) and $\beta$-rhythm (14-30 Hz) with a fifth-order Butterworth filter. Six spatial filters ($m = 3$) selected according to the eigenvalues are used for feature extraction, as recommended in \citep{blankertz2008OptimizingSpatialFilters}. In case when scaCSP\mreplaced{-extraSub}{$_{(extraSub)}$} is applied, six extra spatial filters from subspaces are used. In the regularized CSP algorithms (TRCSP, sCSP, and sTRCSP), the regularization parameters $\alpha$ and $\beta$ are selected from the set of $\{0, 10^{-6}, 10^{-5}, \cdots, 10^0\}$ by 10-fold cross-validation during the training phase where the ones yielding the maximum averaged accuracy on the training set are chosen. It should be noted that we do not apply any rejection of trials or electrodes neither manually nor automatically in this work.

In the experiments, all the analyzed methods are trained in a subject-specific way. For all the datasets, the performance of the methods is evaluated by measuring the ratio of trials correctly classified to the total number of test trials for each subject using a\added{n} LDA classifier. The classification accuracy in the training phase is computed with a 10-fold cross-validation and that in testing phase is then obtained with the trained classifier.

%In the regularized CSP algorithms (TRCSP, sCSP, and sTRCSP), the regularization parameters $\alpha$ and $\beta$ are required to be determined. In order to set a reasonable range for these hyperparameters, all the covariance matrices are normalized by dividing them by their traces. We select the optimal subject-specific hyperparameter values from the set of $\{0, 10^{-6}, 10^{-5}, \cdots, 10^0\}$ by 10-fold cross-validation during the training phase where the ones yielding the maximum averaged accuracy on the training set are chosen. Extracted features are fed into the binary or multi-class classifier.

%In the experiments, both the proposed method and the competing methods are trained in a subject-specific way. More precisely, spatial filters are learned from the training sets for each subject individually. For all the datasets, the performance of the method is evaluated by measuring the ratio of trials correctly classified to the total number of test trials for each subject using a LDA classifier. 
%%==============================================================================
%%

%%==============================================================================
%%
\section{Results}

\subsection{Empirical evaluation of informative subspaces} \label{sec:empirical}

\begin{figure}[ht!]
	\centering
	\begin{minipage}[ht!]{0.90\textwidth}
		\centering
		\begin{subfigure}[ht]{0.48\textwidth}
			\centering
			\includegraphics[width = \textwidth]{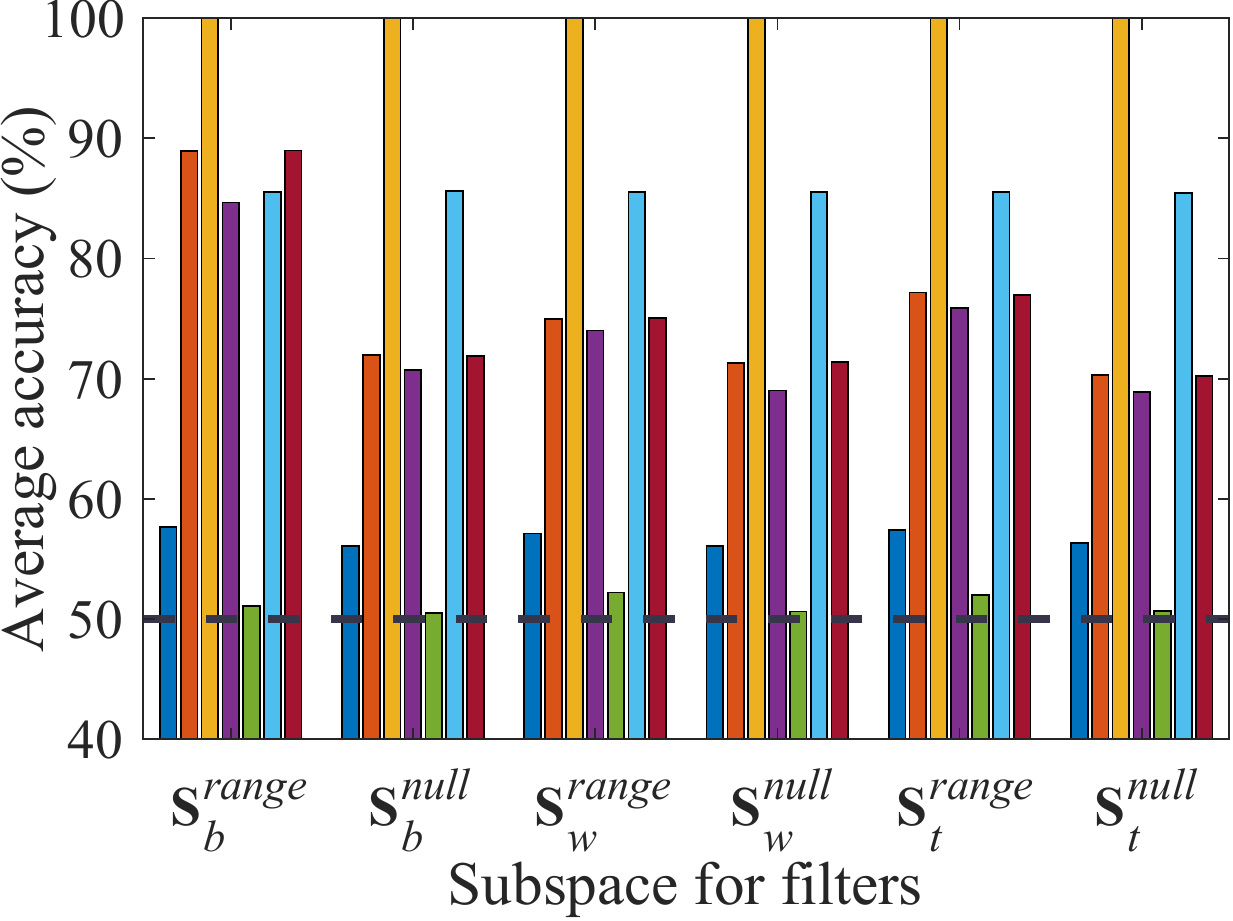}
			\caption{Training session in DS1}
			\label{subfig:empirical-binaryT-ds1}
		\end{subfigure}
		\hfill
		\begin{subfigure}[ht]{0.48\textwidth}
			\centering
			\includegraphics[width = \textwidth]{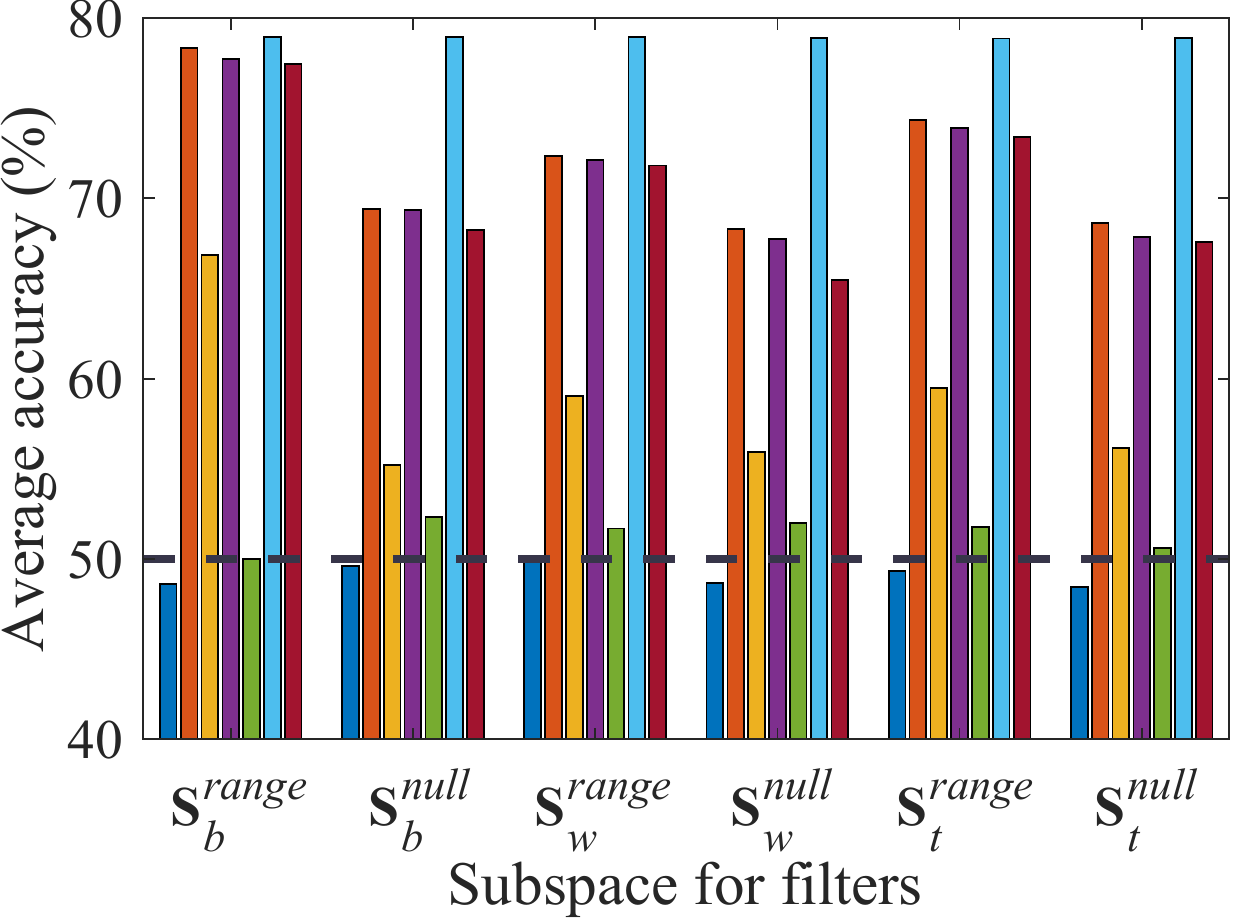}
			\caption{Testing session in DS1}
			\label{subfig:empirical-binaryE-ds1}
		\end{subfigure}
		\quad
		\begin{subfigure}[ht]{0.48\textwidth}
			\centering
			\includegraphics[width = \textwidth]{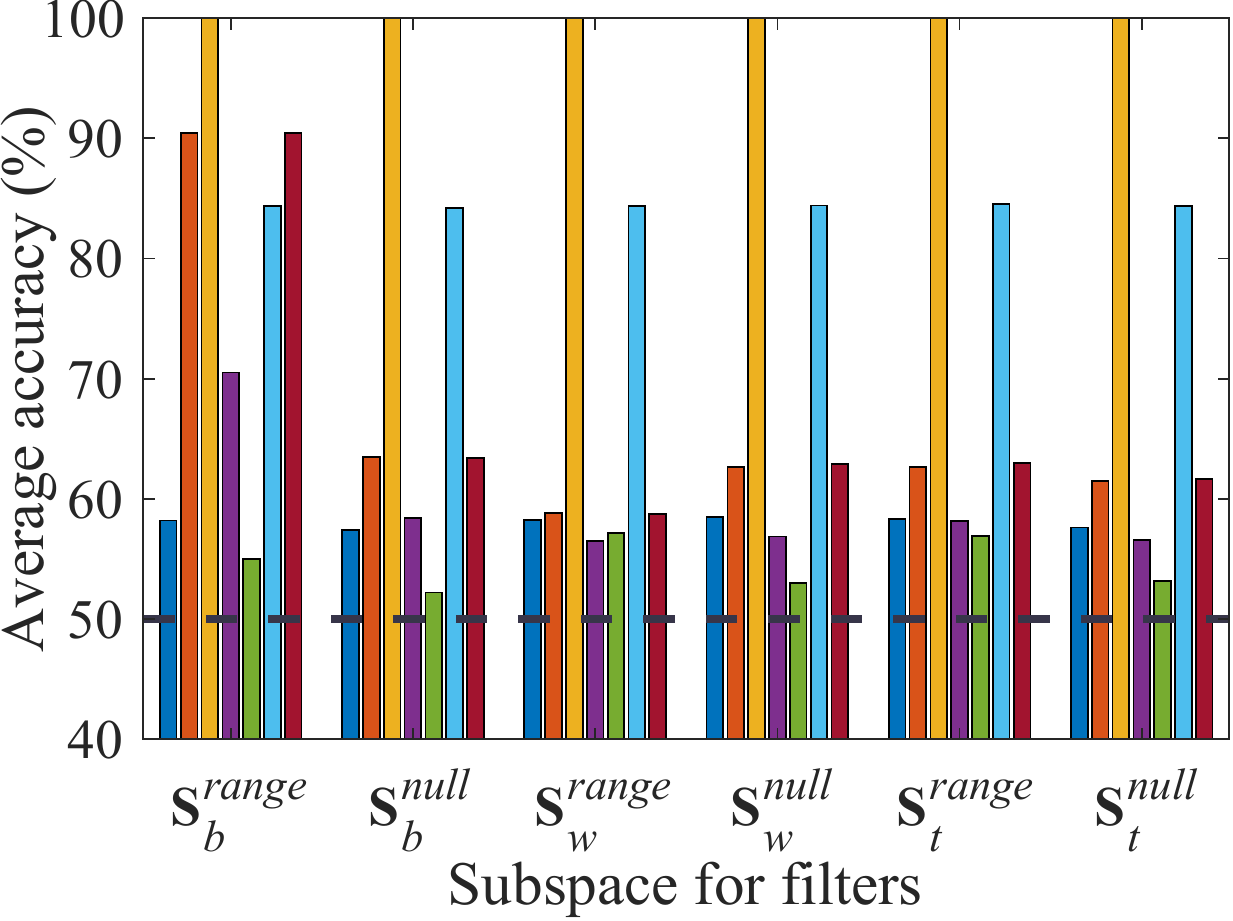}
			\caption{Training session in DS2}
			\label{subfig:empirical-binaryT-ds2}
		\end{subfigure}
		\hfill
		\begin{subfigure}[ht]{0.48\textwidth}
			\centering
			\includegraphics[width = \textwidth]{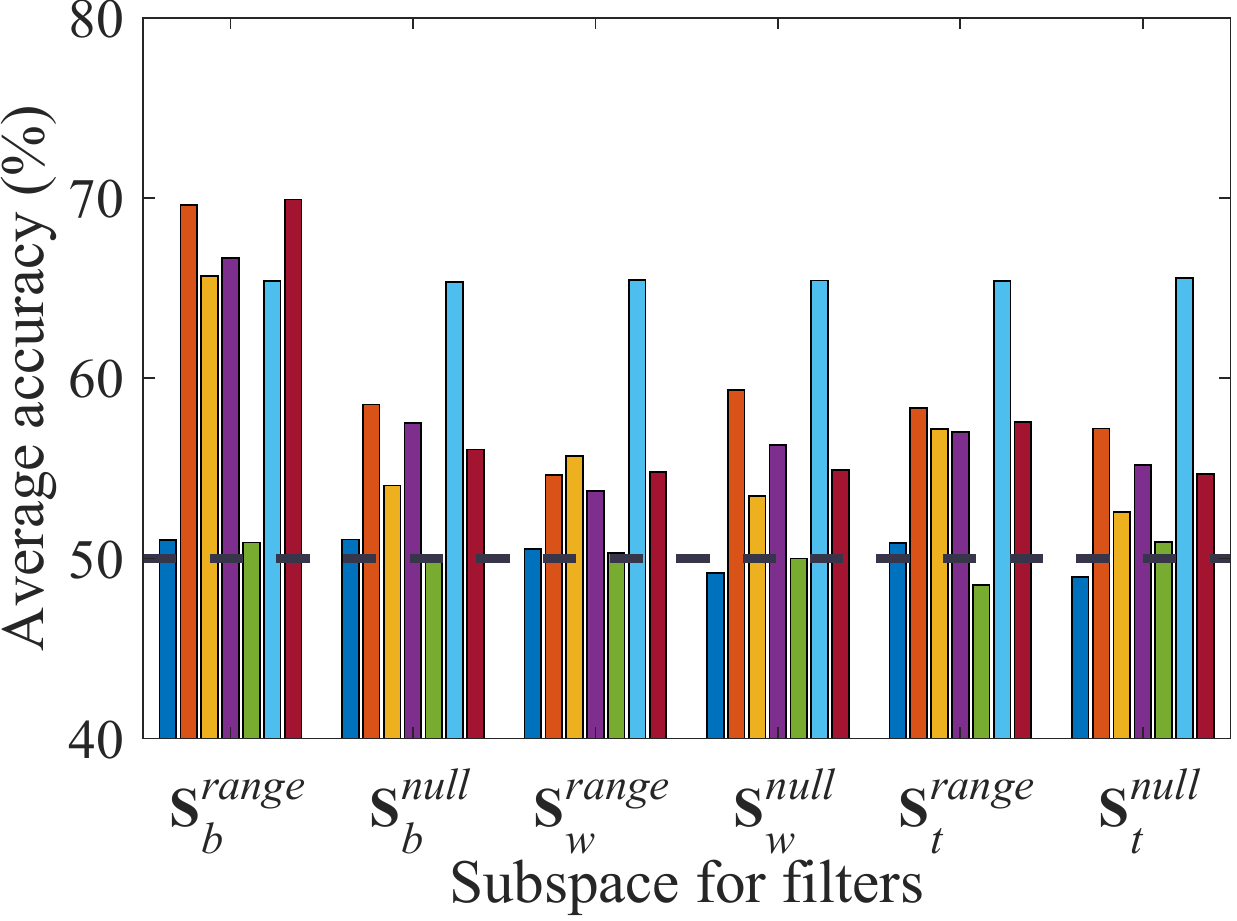}
			\caption{Testing session in DS2}
			\label{subfig:empirical-binaryE-ds2}
		\end{subfigure}
	\end{minipage}
	\hfill
	\begin{minipage}[ht!]{0.08\textwidth}
		\centering
		\begin{subfigure}[ht!]{\textwidth}
			\centering
			\includegraphics[width = \textwidth]{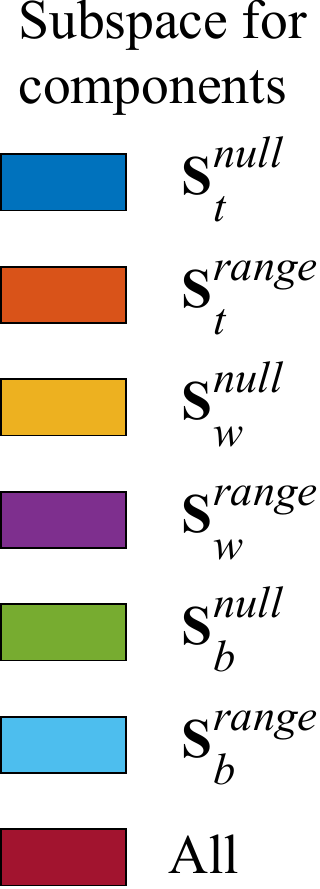}
			%\vspace{14pt}
		\end{subfigure}
	\end{minipage}
	\caption{Average classification accuracy of binary pairs for DS1 and DS2 using filters extracted from spaces $\mmat{S}_b^{range}$, $\mmat{S}_b^{null}$, $\mmat{S}_w^{range}$, $\mmat{S}_w^{null}$, $\mmat{S}_t^{range}$, and $\mmat{S}_t^{null}$ with data components projected onto them or without component projecting. The average was taken over all the six binary \mreplaced{task-pairs}{task pairs} for all the subjects in each dataset. The horizontal dashed line represents the chance level from a random guess, i.e., 50\% for the binary classification. \added{Please observe that the y-axis is differently scaled for training and testing sessions.}}
	%The training session is used to train a LDA classifier which is then used to classify the testing session. 
	\label{fig:empirical-binary}
\end{figure}

\begin{figure}[ht!]
	\centering
	\begin{minipage}[ht!]{0.90\textwidth}
		\centering
		\begin{subfigure}[ht]{0.48\textwidth}
			\centering
			\includegraphics[width = \textwidth]{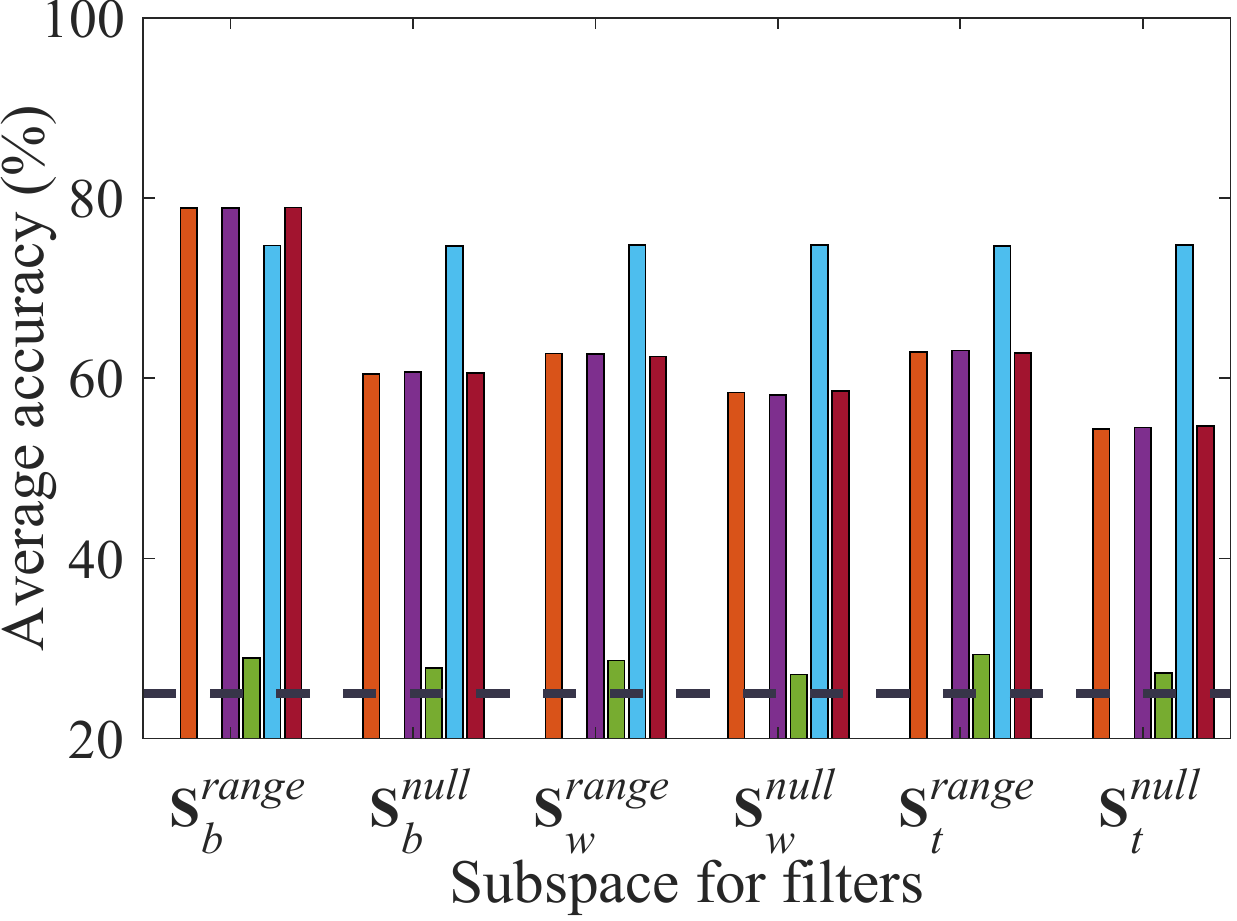}
			\caption{Training session in DS1}
			\label{subfig:empirical-multiT-ds1}
		\end{subfigure}
		\hfill
		\begin{subfigure}[ht]{0.48\textwidth}
			\centering
			\includegraphics[width = \textwidth]{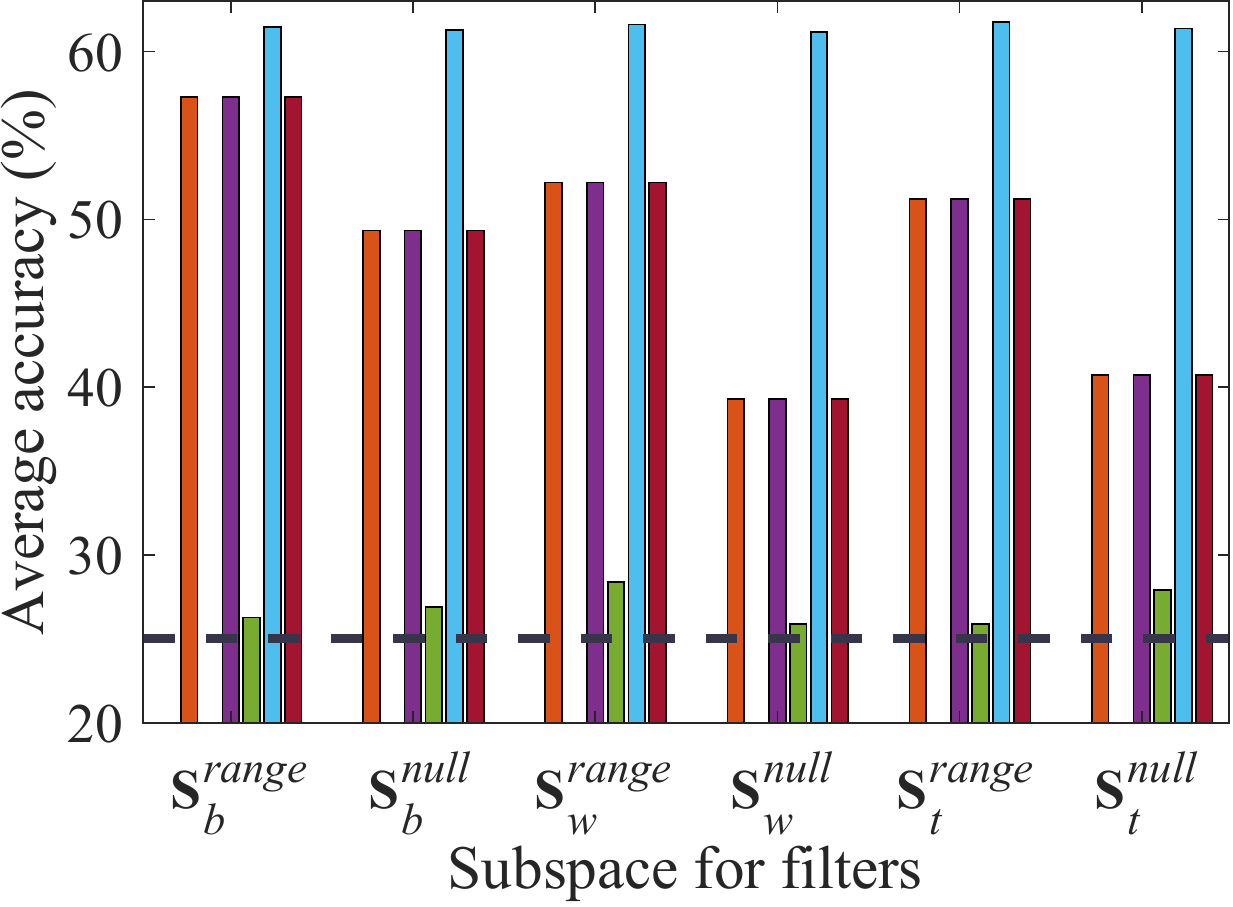}
			\caption{Testing session in DS1}
			\label{subfig:empirical-multiE-ds1}
		\end{subfigure}
		\quad
		\begin{subfigure}[ht]{0.48\textwidth}
			\centering
			\includegraphics[width = \textwidth]{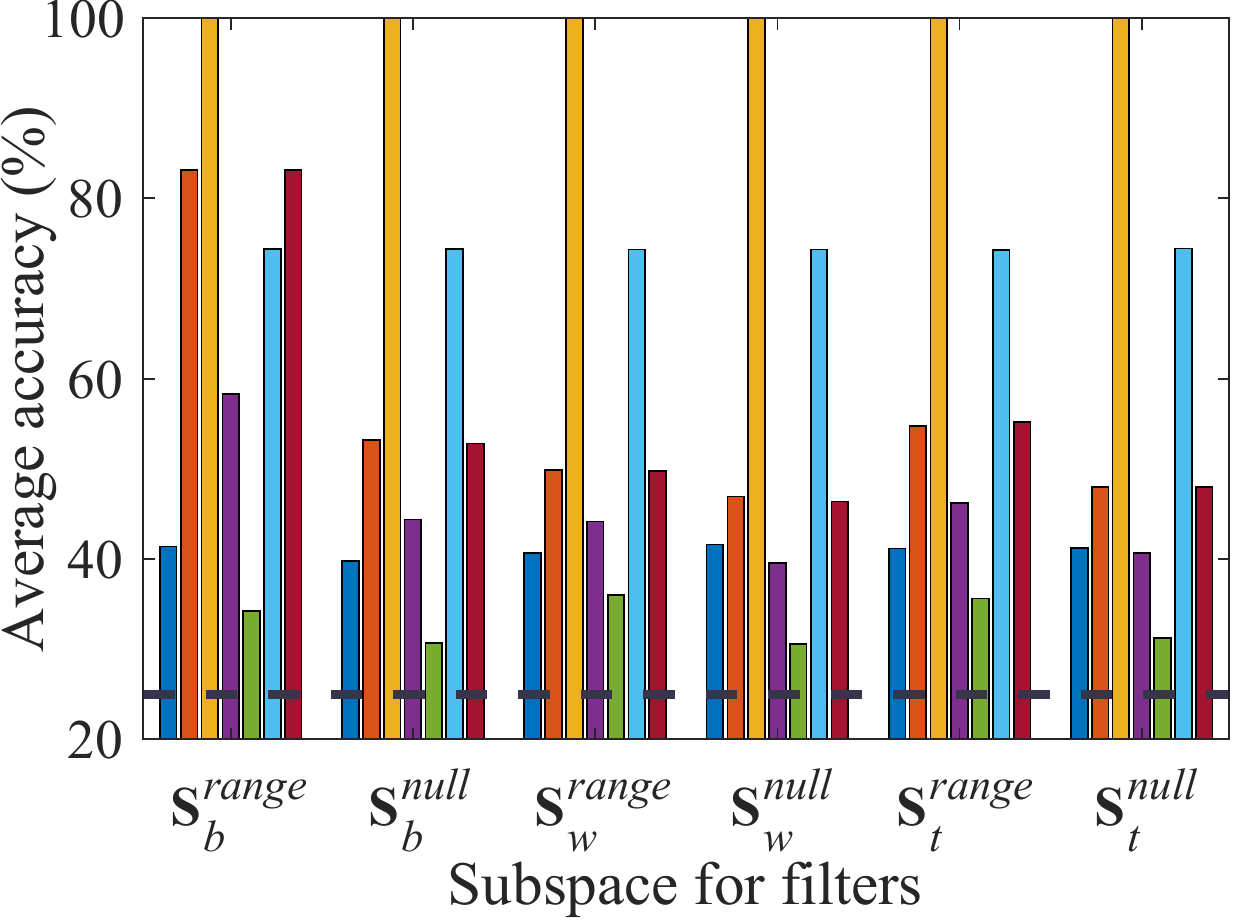}
			\caption{Training session in DS2}
			\label{subfig:empirical-multiT-ds2}
		\end{subfigure}
		\hfill
		\begin{subfigure}[ht]{0.48\textwidth}
			\centering
			\includegraphics[width = \textwidth]{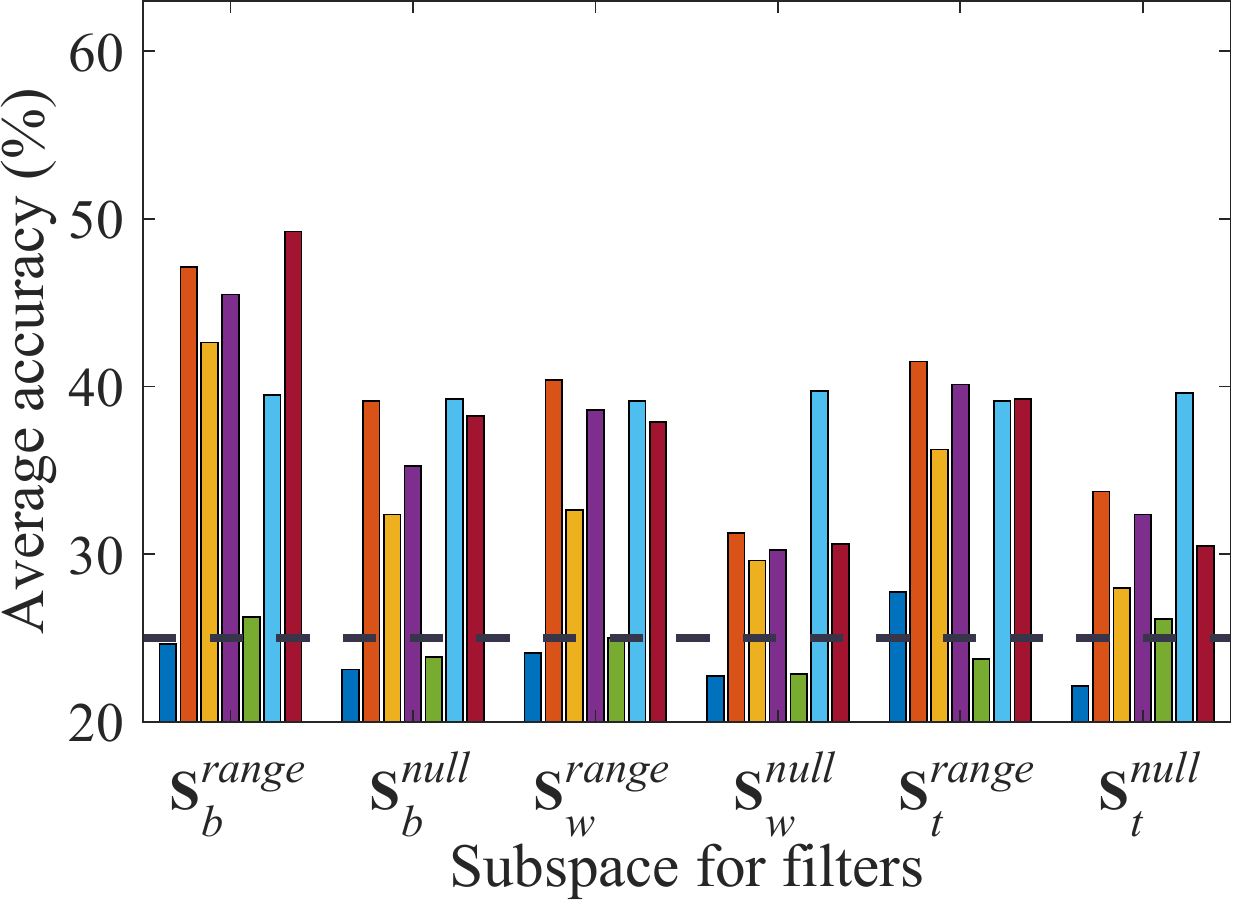}
			\caption{Testing session in DS2}
			\label{subfig:empirical-multiE-ds2}
		\end{subfigure}
	\end{minipage}
	\hfill
	\begin{minipage}[ht!]{0.08\textwidth}
		\centering
		\begin{subfigure}[htbp!]{\textwidth}
			\centering
			\includegraphics[width = \textwidth]{empirical_subfig_legend.pdf}
			%\vspace{14pt}
		\end{subfigure}
	\end{minipage}
	\caption{Average multi-class classification accuracy for DS1 and DS2 using spaces $\mmat{S}_b^{range}$, $\mmat{S}_b^{null}$, $\mmat{S}_w^{range}$, $\mmat{S}_w^{null}$, $\mmat{S}_t^{range}$, and $\mmat{S}_t^{null}$ with data components projected onto them or without component projecting. The average was taken over all the subjects in each dataset. The horizontal dashed line represents the chance level from a random guess, i.e., 25\% for the 4-class classification.}
	\label{fig:empirical-multi}
\end{figure}

An empirical analysis is carried out to show that we can extract useful spatial filters for classification from all the subspaces namely, $\mmat{S}_b^{range}$, $\mmat{S}_b^{null}$, $\mmat{S}_w^{range}$, $\mmat{S}_w^{null}$, $\mmat{S}_t^{range}$, and $\mmat{S}_t^{null}$. In order to demonstrate this, we compute spatial filters from these subspaces using \cref{alg:scaCSP-extraSub}, where $2m(N_\Omega - 1)$ filters are selected according to $|\mvec{d}|$ from each subspace individually and no combination of subspaces are considered. More precisely, we choose $m = 3$, i.e., in the binary classification 6 filters are used while 18 filters in the multi-class problem. Furthermore, data components in these subspaces could make different contributions to the classification performance. To analyze their effects on the performance, data components in each subspace are also  used individually for classification. Similar to the steps introduced in scaCSP-NSR, the components are computed by projecting the original covariance samples onto each corresponding subspace. In this way, the informative subspaces are considered in two different scenarios, i.e., subspaces for \added{the} spatial filter learning scenario and subspaces for \added{the} component projecting scenario. We evaluate the performance in terms of classification accuracy with all the combinations of subspaces for filters and for components.

\cref{fig:empirical-binary} and \cref{fig:empirical-multi} show the average classification accuracy for the binary case and the multi-class case\added{,} respectively. The grand average is computed over all the subjects in each dataset. Moreover, for the binary classification, the average is also taken over all the binary \mreplaced{task-pairs}{task pairs}\added{,} namely, L/R, L/F, L/T, R/F, R/T, and F/T. It should be noted that in terms of subspace for data component projecting, "\mreplaced{None}{All}" indicates no projecting is applied and the original data is then used for classification. Furthermore, when $\mmat{S}_b^{range}$ is used as the subspace for extracting filters, it is equivalent to applying the binary-scaCSP (resp., multi-scaCSP) using the \deleted{original data or }data components in $\mmat{S}_b^{range}$, $\mmat{S}_b^{null}$, $\mmat{S}_w^{range}$, $\mmat{S}_w^{null}$, $\mmat{S}_t^{range}$, \deleted{and }$\mmat{S}_t^{null}$\added{, and "All" subspaces, i.e., the original data} for the binary (resp., multi-class) problem. In terms of the $N_\Omega$-class classification on DS1 with $N_\Omega = 4$, there are $|\Omega| = 288$ covariance samples in the training session and $N_c = 22$ EEG channels. From \eqref{eq:rank-of-scatter-matrix}, we have $\text{rank}(\mmat{S}_w) = \text{rank}(\mmat{S}_t) = 253$. In this work, a\added{n} $N_c^2 \times N_c^2$ scatter matrix of the covariance vectorization defined in \eqref{eq:scatter-of-covariance} is called \added{a} semi-full rank matrix if its rank is $\frac{N_c(N_c + 1)}{2}$ noting that this value is the maximal rank the scatter matrix could reach because of the symmetric property of covariance. Therefore, in DS1, both $\mmat{S}_w$ and $\mmat{S}_t$ are semi-full rank matrices. In this case, $\mmat{S}_w^{null}$ and $\mmat{S}_t^{null}$ are nonempty but purely contain data components coming from vectorized non-symmetrical matrices. Note that the samples considered in the scatter-based framework are the vectorization of covariance matrices. Therefore, $\mmat{S}_w^{null}$ and $\mmat{S}_t^{null}$ contain no information about the samples. In this case, they are termed \deleted{as }semi-empty. Data components projected onto $\mmat{S}_w^{null}$ and $\mmat{S}_t^{null}$ in DS1 are always zero. So \added{the} classification of data components in $\mmat{S}_w^{null}$ and $\mmat{S}_t^{null}$ are omitted in the empirical analysis, as shown in \cref{subfig:empirical-multiT-ds1} and \cref{subfig:empirical-multiE-ds1}.

In terms of \added{the} data component projecting scenario, it can be observed from both \cref{fig:empirical-binary} and \cref{fig:empirical-multi} that range spaces ($\mmat{S}_b^{range}$, $\mmat{S}_w^{range}$, and $\mmat{S}_t^{range}$) contain significant information in both binary and multi-class classification since it yields much higher accuracy by using data components in these individual spaces. Among \added{the} three range spaces, $\mmat{S}_b^{range}$ contains discriminant information that is very robust to spatial filters for both binary and multi-class problems since the classification accuracy keeps to \deleted{be }almost the same no matter the filters are extracted from which subspace. Whilst, null spaces ($\mmat{S}_b^{null}$ and $\mmat{S}_t^{null}$) make almost no contribution to classification noting that the accuracy is around the level of a random guess when using data components in $\mmat{S}_b^{null}$ and $\mmat{S}_t^{null}$. Though $\mmat{S}_w^{null}$ is less effective, it still contains some discriminant information useful for classification. Especially, in the training phase the accuracy using $\mmat{S}_w^{null}$ components is always 100\% since in this case the within-class scatters of the extracted features are always zero regardless of the spatial filters. More precisely, let $\mmat{S}_w^{null}$ be spanned by columns of $\mmat{Q}$. The data components in $\mmat{S}_w^{null}$ are given by $\mmat{Q} \mmat{Q}\transpose \mvec{r}_i$. Then given matrix $\mmat{V}$ whose columns are the learned spatial filters, the within-class scatters of the extracted features are computed by $\mmat{V}\transpose \mmat{Q} \mmat{Q}\transpose \mmat{S}_w \mmat{Q} \mmat{Q}\transpose \mmat{V}$ \added{resulting in a zero matrix $\mmat{0}$ all of whose entries are zero} \deleted{which are a matrix of zeros $\mmat{0}$ }noting that $\mmat{Q}\transpose \mmat{S}_w \mmat{Q} \equiv \mmat{0}$. However, \deleted{by }using $\mmat{S}_w^{null}$ components \deleted{it }is very sensitive to the non-stationarity between sessions and may lead to unsatisfactory classification performance in the testing phase. Due to the between-session non-stationary property and even other noises, the null spaces of the between-class scatter matrices for the training data and testing data are not equal. When using the $\mmat{S}_w^{null}$ components to learn spatial filters and to train the classifier, only discriminative characteristics in this space could be learned and the generalization would be poor.

In terms of \added{the} spatial filter learning scenario, the filters extracted from $\mmat{S}_b^{range}$ are the most effective yielding the highest average classification accuracy in both binary and multi-class classification, as can be seen \mreplaced{from}{in} \cref{fig:empirical-binary} and \cref{fig:empirical-multi}. Though spatial filters extracted from other subspaces ($\mmat{S}_b^{null}$, $\mmat{S}_w^{range}$, $\mmat{S}_w^{null}$, $\mmat{S}_t^{range}$, and $\mmat{S}_t^{null}$) are less effective, they still provide considerable accuracy higher than the random guess level, especially when data components in $\mmat{S}_b^{range}$ are used.

\subsection{Classification results}

In this section, we present the effectiveness of the proposed scatter-based framework by comparing its classification results with those of \deleted{the }other \deleted{competing }methods, namely, the conventional CSP, TRCSP, sCSP, and sTRCSP for the binary classification and CSP-PW and CSP-OVR for multi-class classification. For the binary problems, we extract spatial filters for each binary \mreplaced{task-pair}{task pair} (L/R, L/F, L/T, R/F, R/T, and F/T) and the average classification accuracy is computed for each subject. Since we have 288 trials and 22 channels for the multi-class problem in DS1, both $\mmat{S}_t^{null}$ and $\mmat{S}_w^{null}$ are semi-empty. The common NSR process can not be applied in this case noting that components in $\mmat{S}_t^{null}$ are all zeros. Therefore, scaCSP-BNSR and scaCSP-BNSR\mreplaced{-extraSub}{$_{(extraSub)}$} are used. Whereas, for all the binary problems in DS1 and DS2 and the multi-class problem in DS2, $\mmat{S}_t^{null}$ are not semi-empty and thus scaCSP-CNSR and scaCSP-CNSR\mreplaced{-extraSub}{$_{(extraSub)}$} are applied. Extra spatial filters from range spaces $\mmat{S}_w^{range}$, $\mmat{S}_t^{range}$, and both of them are used in the binary classifications\deleted{, which are termed as scaCSP-extraSub ($\mmat{S}_w^{range}$), scaCSP-extraSub ($\mmat{S}_t^{range}$), and scaCSP-extraSub ($\mmat{S}_w^{range} + \mmat{S}_t^{range}$) for the scaCSP-extraSub group and scaCSP-CNSR-extraSub ($\mmat{S}_w^{range}$), scaCSP-CNSR-extraSub ($\mmat{S}_t^{range}$), and scaCSP-CNSR-extraSub ($\mmat{S}_w^{range} + \mmat{S}_t^{range}$) for the scaCSP-CNSR-extraSub group, respectively}. Regarding the multi-class problem, we use $\mmat{S}_b^{null}$ and $\mmat{S}_w^{range}$ as the candidate subspaces to extract extra filters for DS1 and $\mmat{S}_w^{range}$ and $\mmat{S}_w^{null}$ for DS2 \added{, respectively}\deleted{namely, scaCSP-extraSub ($\mmat{S}_b^{null}$), scaCSP-extraSub ($\mmat{S}_w^{range}$), scaCSP-extraSub ($\mmat{S}_b^{null} + \mmat{S}_w^{range}$), scaCSP-BNSR-extraSub ($\mmat{S}_b^{null}$), scaCSP-BNSR-extraSub ($\mmat{S}_w^{range}$), and scaCSP-BNSR-extraSub ($\mmat{S}_b^{null} + \mmat{S}_w^{range}$) for DS1 whilst scaCSP-extraSub ($\mmat{S}_w^{range}$), scaCSP-extraSub ($\mmat{S}_w^{null}$), scaCSP-extraSub ($\mmat{S}_w^{range} + \mmat{S}_w^{null}$), scaCSP-CNSR-extraSub ($\mmat{S}_w^{range}$), scaCSP-CNSR-extraSub ($\mmat{S}_w^{null}$), and scaCSP-CNSR-extraSub ($\mmat{S}_w^{range} + \mmat{S}_w^{null}$) for DS2}.

\begin{figure}[ht]
	\centering
	\includegraphics[width = 0.85\textwidth]{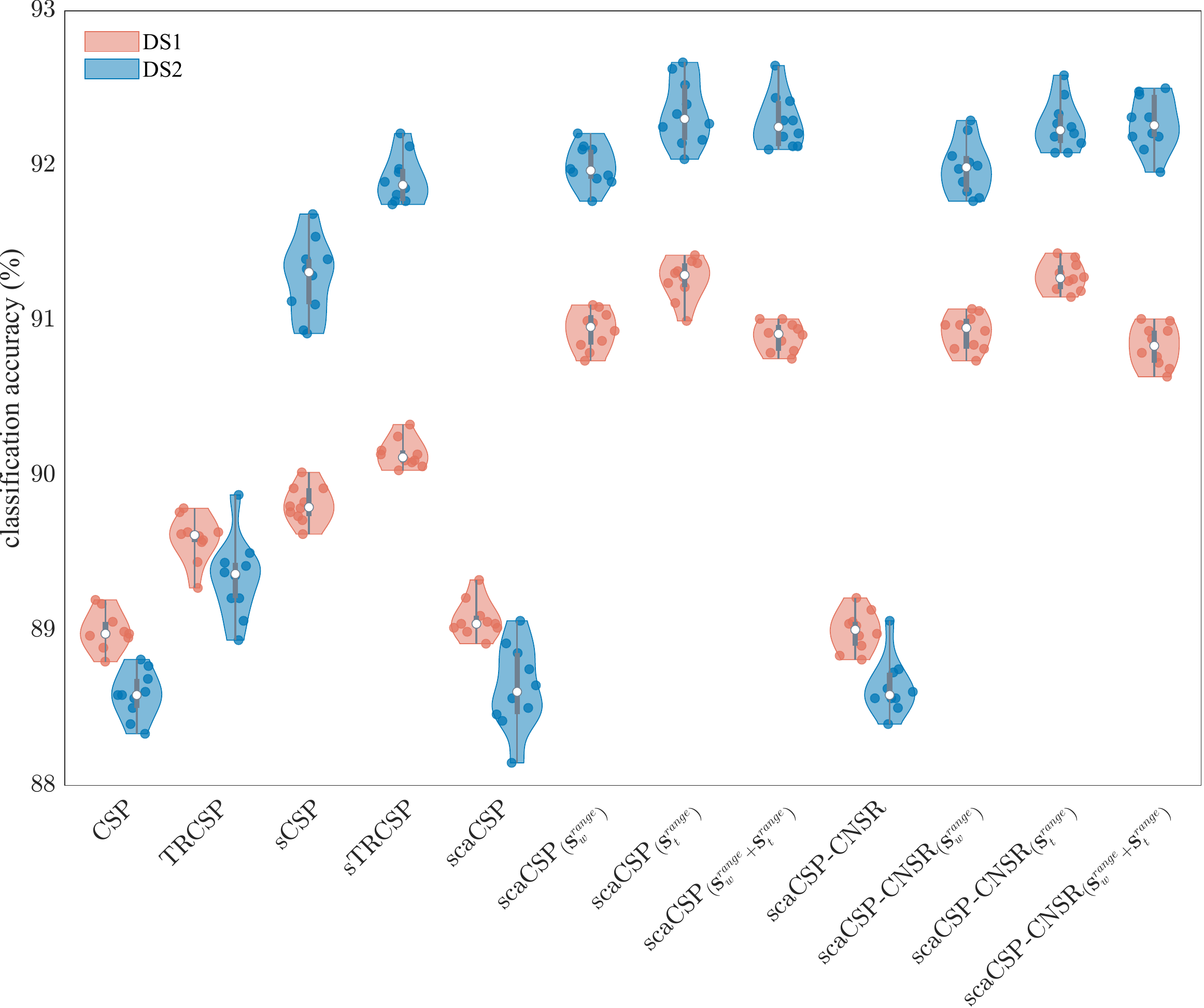}
	\caption{\added{Violin plots showing the 10-fold cross-validation results of averaged classification accuracy of binary problems in the training phase for both DS1 and DS2. The lower and upper sides of the \added{thick} gray \mreplaced{boxes}{bars} denote the 25\% and 75\% percentiles, respectively, and the white circles inside represent the median classification performance. The average was taken across all the subjects and binary task pairs in each dataset}.}
	\label{fig:results-training-binary-violin}
\end{figure}

\begin{figure}[ht]
	\centering
	\includegraphics[width = 0.85\textwidth]{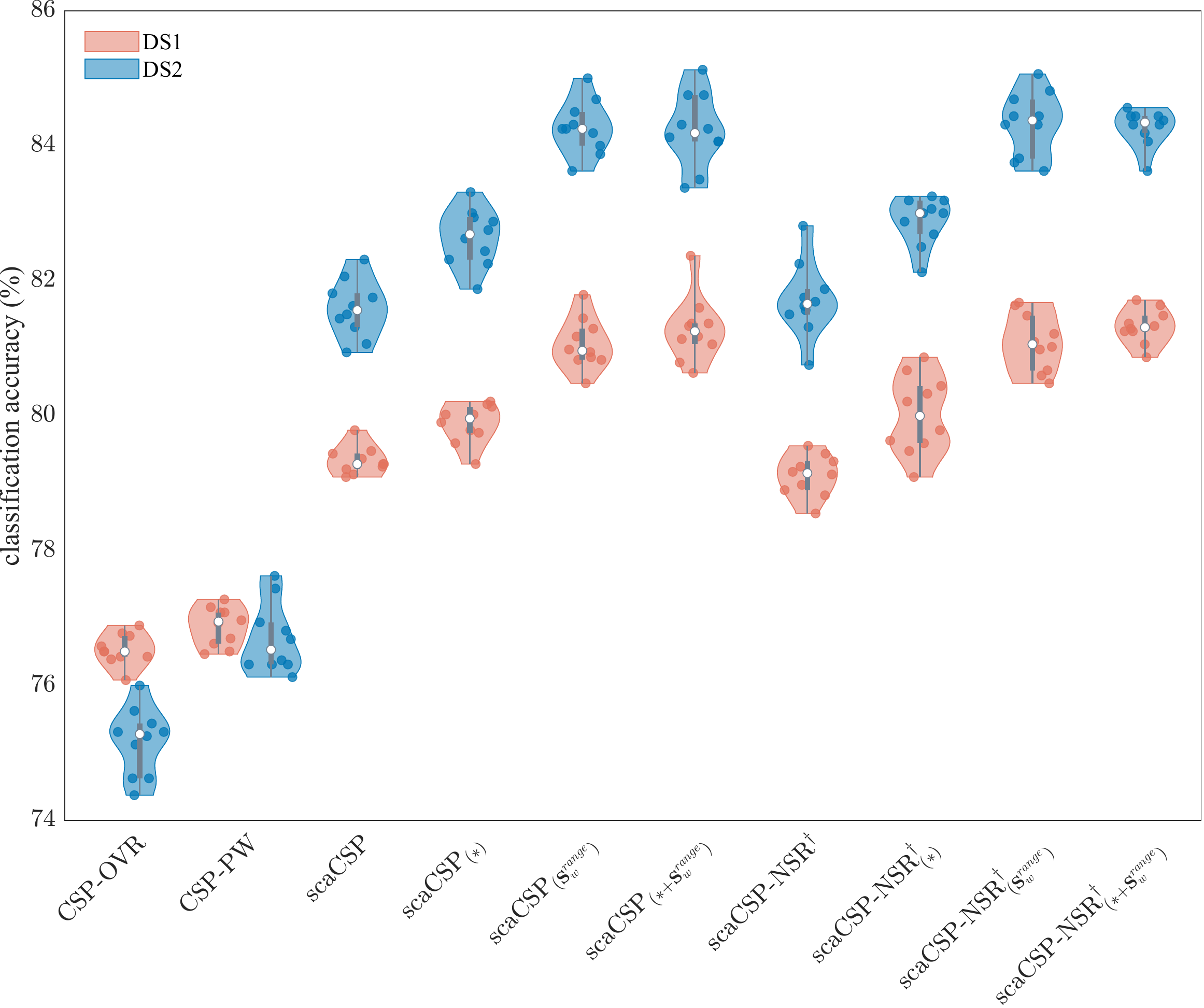}
	\caption{\added{Violin plots showing the 10-fold cross-validation results of averaged classification accuracy of multi-class problems in the training phase for both DS1 and DS2. The lower and upper sides of the \added{thick} gray \mreplaced{boxes}{bars} denote the 25\% and 75\% percentiles, respectively, and the white circles inside represent the median classification performance. The average was taken across all the subjects in each data set. The extra null subspace \(\mmat{S}_b^{null}\) and $\mmat{S}_w^{null}$ are used for DS1 and DS2 respectively, which are indicated by $*$ in order to keep the labels in horizontal axis more compact. Similarly, the CNSR and BNSR processes are used in the scaCSP-NSR algorithms for DS1 and DS2, respectively, which are indicated by $\dagger$.}}
	\label{fig:results-training-multi-violin}
\end{figure}

\cref{fig:results-training-binary-violin} and \cref{fig:results-training-multi-violin} depict the 10-fold cross-validation results of averaged classification accuracy in the training phase for binary and multi-class problems, respectively, for DS1 and DS2. \added{In the violin plots, a normal distribution is used as the kernel function for data smoothing, where the bandwidths of the kernel function are 0.11 and 0.24 for binary problems (\cref{fig:results-training-binary-violin}) and multi-class problems (\cref{fig:results-training-multi-violin}), respectively.} Regarding the binary problems, it can be observed from \cref{fig:results-training-binary-violin} that the median classification performance of CSP and scaCSP are quite similar as can be expected from \cref{theorem1} since they are mathematically equivalent. The difference mainly comes from the cross-validation process when the training datasets are randomly partitioned into 10 subsets. Though \mreplaced{significant}{considerable} improvements compared to the original CSP algorithm can be obtained by using either the regularized CSP algorithms (TRCSP, sCSP, and sTRCSP) or the proposed subspace enhanced scaCSP algorithms (scaCSP\mreplaced{-extraSub}{$_{(extraSub)}$} and scaCSP-CNSR\mreplaced{-extraSub}{$_{(extraSub)}$} except scaCSP-CNSR), scaCSP\mreplaced{-extraSub}{$_{(extraSub)}$} and scaCSP-CNSR\mreplaced{-extraSub}{$_{(extraSub)}$} algorithms yield much higher averaged classification accuracy in both DS1 and DS2. In terms of the multi-class problems, we can see from \cref{fig:results-training-multi-violin} that the proposed multi-class scaCSP outperforms both CSP-OVR and CSP-PW since it gives much higher averaged classification accuracy in both DS1 and DS2. Moreover, the classification performance of scaCSP can be further improved with the subspaces enhancement process by using scaCSP\mreplaced{-extraSub}{$_{(extraSub)}$} and scaCSP-NSR\mreplaced{-extraSub}{$_{(extraSub)}$} algorithms. It should be noted that although the NSR process does not help too much with scaCSP for improving the performance, it still gives much higher accuracy than CSP-OVR and CSP-PW.

%\begin{figure}[ht!]
%	\centering
%	\begin{subfigure}[ht]{0.48\textwidth}
%		\centering
%		\includegraphics[width = \textwidth]{results_multiT_ds1_subfig_violin.pdf}
%		\caption{Multi-class classification on DS1}
%		\label{subfig:results-multiT-ds1-violin}
%	\end{subfigure}
%	\hfill
%	\begin{subfigure}[ht]{0.48\textwidth}
%		\centering
%		\includegraphics[width = \textwidth]{results_multiT_ds2_subfig_violin.pdf}
%		\caption{Multi-class classification on DS2}
%		\label{subfig:results-multiT-ds2-violin}
%	\end{subfigure}
%	\caption{\added{Violin plots showing the 10-fold cross-validation results of averaged classification accuracy of multi-class problems in the training phase for both DS1 (a) and DS2 (b). The lower and upper sides of the \added{thick} gray \mreplaced{boxes}{bars} denote the 25\% and 75\% percentiles, respectively, and the white circles inside represent the median classification performance. The average was taken across all the subjects in each data set.}} 
%	\label{fig:results-training-multi-violin}
%\end{figure}

\cref{tab:Results-binaryE-scaCSP} shows the averaged classification accuracies of the binary problems in the testing phase for each subject of DS1 and DS2. From the results, we see that on average all the regularized CSP algorithms and the proposed subspace enhanced scaCSP algorithms perform better than CSP in both DS1 and DS2. Particularly, scaCSP-CNSR\mreplaced{-extraSub($\mmat{S}_w^{range}$)}{$_{(\mmat{S}_w^{range})}$} outperforms all the other considered approaches yielding the highest accuracy in DS1 (78.69\%) while scaCSP-CNSR\mreplaced{-extraSub($\mmat{S}_t^{range}$)}{$_{(\mmat{S}_t^{range})}$} gives the highest accuracy in DS2 (71.29\%) as well as the overall (74.47\%), showing an \mreplaced{significant}{considerable} improvement comparing to CSP (77.44\%, 68.58\%, and 72.78\% in DS1, DS2 and overall, respectively). As can be expected, the binary scaCSP performs totally the same as CSP. Moreover, all the proposed subspace enhanced scaCSP approaches (scaCSP\mreplaced{-extraSub}{$_{(extraSub)}$}, scaCSP-CNSR, scaCSP-CNSR\mreplaced{-extraSub}{$_{(extraSub)}$}) perform better than CSP and scaCSP, showing the effectiveness of the subspace enhancement process. In addition, regarding the competing methods considered in this paper, sTRCSP outperforms the other CSP-like approaches (CSP, TRCSP, sCSP) in both datasets, where similar results can also be found in \cite{samek2012StationaryCommonSpatial}. However, all the \mreplaced{propposed}{proposed} subspace enhanced scaCSP algorithms except scaCSP\mreplaced{-extraSub ($\mmat{S}_t^{range}$)}{$_{(\mmat{S}_t^{range})}$} perform better than sTRCSP in DS1 and the approaches scaCSP-CNSR\mreplaced{-extraSub ($\mmat{S}_t^{range}$)}{$_{(\mmat{S}_t^{range})}$} and scaCSP-CNSR\mreplaced{-extraSub ($\mmat{S}_w^{range} + \mmat{S}_t^{range}$)}{$_{(\mmat{S}_w^{range} + \mmat{S}_t^{range})}$} perform better than sTRCSP in DS2.

\cref{tab:Results-multi-scaCSP} depicts the multi-class classification accuracies in the testing phase for each subject of DS1 and DS2. It can be observed from the results that all the subspace enhanced scaSCP approaches perform better on average than both CSP-OVR and CSP-PW. Specifically in DS1, scaCSP-BNSR\mreplaced{-extraSub ($\mmat{S}_b^{null} + \mmat{S}_w^{range}$)}{$_{(\mmat{S}_b^{null} + \mmat{S}_w^{range})}$} yields the highest classification accuracy (61.57\%), which is \mreplaced{significantly}{considerably} higher than CSP-OVR (57.45\%) and CSP-PW (56.48\%). In DS2, the highest classification accuracy (46.25\%) is obtained by scaCSP-CNSR\mreplaced{-extraSub ($\mmat{S}_w^{range}$)}{$_{(\mmat{S}_w^{range})}$}, which is also much higher than both CSP-OVR (41.88\%) and CSP-PW (43.00\%). Regarding the scaCSP algorithm without subspace enhancement, it is superior to both CSP-OVR and CSP-PW in terms of the overall grand mean value. Moreover, the accuracy by scaCSP (57.29\%) is only slightly smaller than CSP-OVR (57.45\%) in DS1 whilst larger than CSP-PW in both DS1 (56.48\%) and DS2 (43.00\%) as well as CSP-OVR in DS2 (41.88\%).

In summary, the experimental results demonstrated the effectiveness of the proposed scatter-based framework for spatial filter learning and feature extraction. From the results in \cref{fig:results-training-binary-violin}, \cref{fig:results-training-multi-violin}, \cref{tab:Results-binaryE-scaCSP} and \cref{tab:Results-multi-scaCSP}, the averaged classification accuracies demonstrate the superiority of the proposed scaCSP algorithms in both binary and multi-class MI classification tasks in both training phase and testing phase.

\begin{landscape}
	\begin{table}[ht!]
		%\scriptsize
		\small
		\centering
		\caption{Average classification accuracy (\%) of binary \mreplaced{task-pairs}{task pairs} using competing methods (CSP, RTCSP, sCSP, and sTRCSP) and proposed scaCSP algorithms for each subject in the testing phase. The highest accuracy is marked in boldface for each subject. }
		\label{tab:Results-binaryE-scaCSP}
		\begin{tabular}{*{13}{c}} % 13 columns
			\toprule
			\multirow{2}{*}{Subject} &\multirow{2}{*}{CSP}  &\multirow{2}{*}{TRCSP} &\multirow{2}{*}{sCSP} &\multirow{2}{*}{sTRCSP} &\multirow{2}{*}{scaCSP} &\multicolumn{3}{c}{scaCSP\mreplaced{-extraSub}{$_{(extraSub)}$}} &scaCSP &\multicolumn{3}{c}{scaCSP-CNSR\mreplaced{-extraSub}{$_{(extraSub)}$}} \\
			\cline{7-9} \cline{11-13} &&&&& &$\mmat{S}_w^{range}$ &$\mmat{S}_t^{range}$ &$\mmat{S}_w^{range} + \mmat{S}_t^{range}$ &-CNSR &$\mmat{S}_w^{range}$ &$\mmat{S}_t^{range}$ &$\mmat{S}_w^{range} + \mmat{S}_t^{range}$ \\
			\midrule
			A01	&89.70	&90.86	&89.81	&90.74	&89.70	&89.81	&90.05	&90.39	&91.32	&\textbf{92.01}	&91.44	&90.63	\\
			A02	&65.39	&66.67	&65.51	&65.39	&65.39	&64.35	&64.24	&64.93	&67.01	&65.39	&65.05	&\textbf{67.25}	\\
			A03	&86.69	&87.04	&86.34	&86.69	&86.69	&87.38	&85.65	&\textbf{87.96}	&87.73	&87.04	&86.46	&86.81	\\
			A04	&77.78	&78.82	&\textbf{80.67}	&78.94	&77.78	&78.13	&78.94	&78.94	&79.40	&79.98	&80.56	&80.44	\\
			A05	&57.06	&57.41	&57.06	&57.52	&57.06	&\textbf{60.30}	&59.61	&59.37	&59.03	&58.56	&58.22	&56.71	\\
			A06	&64.70	&64.35	&64.00	&\textbf{65.05}	&64.70	&64.24	&64.24	&64.93	&63.77	&64.93	&64.81	&64.81	\\
			A07	&90.51	&89.58	&89.81	&89.70	&90.51	&90.74	&90.05	&89.70	&\textbf{91.20}	&90.51	&90.28	&90.16	\\
			A08	&80.21	&83.10	&81.83	&\textbf{83.91}	&80.21	&82.18	&82.18	&82.75	&81.02	&82.75	&82.29	&83.22	\\
			A09	&84.95	&85.65	&84.72	&86.57	&84.95	&87.96	&\textbf{88.77}	&88.31	&84.61	&87.04	&87.96	&87.62	\\
			\hline
			Mean &77.44	&78.16	&77.75	&78.28	&77.44	&78.34	&78.19	&78.59	&78.34	&\textbf{78.69}	&78.56	&78.63 \\
			$\pm$std    &11.52	&11.57	&11.57	&11.69	&11.52	&11.52	&11.55	&11.57	&11.47	&11.77	&11.82	&11.78 \\
			\hline
			B01	&65.00	&64.58	&62.08	&62.92	&65.00	&58.75	&59.58	&59.17	&65.83	&63.75	&\textbf{66.25}	&63.33 \\
			B02	&65.83	&65.42	&65.83	&65.83	&65.83	&\textbf{70.83}	&69.58	&70.00	&66.67	&70.42	&70.00	&\textbf{70.83} \\
			B03	&\textbf{58.75}	&56.67	&58.33	&57.50	&\textbf{58.75}	&56.25	&55.42	&55.83	&56.25	&54.58	&53.75	&54.58 \\
			B04	&67.08	&68.75	&65.42	&66.25	&67.08	&66.67	&68.75	&64.58	&70.00	&69.58	&69.17	&\textbf{71.67} \\
			B05	&61.25	&60.83	&\textbf{79.17}	&\textbf{79.17}	&61.25	&70.83	&68.33	&74.58	&62.92	&75.83	&77.08	&73.33 \\
			B06	&52.08	&54.58	&54.58	&\textbf{56.25}	&52.08	&52.50	&50.42	&53.33	&52.50	&50.83	&53.33	&50.42 \\
			B07	&62.92	&63.75	&65.42	&63.75	&62.92	&62.92	&61.25	&62.50	&64.58	&66.67	&\textbf{67.50}	&65.83 \\
			B08	&81.67	&82.08	&80.42	&82.08	&81.67	&77.92	&78.33	&77.50	&79.58	&81.25	&81.25	&\textbf{83.75} \\
			B09	&80.00	&81.67	&83.33	&83.33	&80.00	&80.00	&80.83	&81.67	&81.25	&82.08	&80.83	&\textbf{83.75} \\
			B10	&91.25	&91.67	&93.33	&93.33	&91.25	&93.75	&93.75	&93.33	&92.92	&94.17	&93.75	&\textbf{94.17} \\
			\hline
			Mean &68.58	&69.00	&70.79	&71.04	&68.58	&69.04	&68.63	&69.25	&69.25	&70.92	&\textbf{71.29}	&71.17 \\
			$\pm$std	&11.37	&11.54	&11.85	&11.86	&11.37	&11.84	&12.32	&11.97	&11.58	&12.40	&11.85	&12.83 \\
			
			\midrule
			Grand mean &72.78	&73.34	&74.09	&74.47	&72.78	&73.45	&73.16	&73.67	&73.56	&74.60	&\textbf{74.74}	&74.70 \\
			$\pm$std   &12.6	&12.77	&12.56	&12.66	&12.6	&12.92	&13.23	&13.02	&12.73	&13.06	&12.72	&13.25 \\
			\bottomrule
		\end{tabular}
	\end{table}
\end{landscape}

\begin{landscape}
	\begin{table}[ht!]
		%\scriptsize
		\small
		\centering
		\caption{Multi-class classification accuracy (\%) using competing multi-class extensions of CSP (CSP-PW and CSP-OVR) and proposed scaCSP algorithms for each subject in the testing phase. The highest accuracy is marked in boldface for each subject.}
		\label{tab:Results-multi-scaCSP}
		\begin{tabular}{c cc c ccc c ccc} % 11 columns
			\toprule
			\multirow{2}{*}{Subject} &\multirow{2}{*}{CSP-OVR}  &\multirow{2}{*}{CSP-PW} &\multirow{2}{*}{scaCSP} &\multicolumn{3}{c}{scaCSP\mreplaced{-extraSub}{$_{(extraSub)}$}} &scaCSP &\multicolumn{3}{c}{scaCSP-BNSR\mreplaced{-extraSub}{$_{(extraSub)}$}} \\
			\cline{5-7} \cline{9-11} &&& &$\mmat{S}_b^{null}$ &$\mmat{S}_w^{range}$ &$\mmat{S}_b^{null} + \mmat{S}_w^{range}$ &-BNSR &$\mmat{S}_b^{null}$ &$\mmat{S}_w^{range}$ &$\mmat{S}_b^{null} + \mmat{S}_w^{range}$\\
			\midrule
			A01	&67.71	&72.57	&75.35	&73.96	&76.74	&73.96	&77.78	&\textbf{78.82}	&77.43	&78.13 \\
			A02	&42.01	&39.58	&40.97	&39.58	&39.93	&39.24	&43.40	&\textbf{45.49}	&45.14	&45.14 \\
			A03	&67.01	&\textbf{71.88}	&67.01	&68.06	&69.44	&68.06	&68.40	&68.40	&67.36	&67.36 \\
			A04	&\textbf{60.76}	&54.17	&58.68	&56.94	&55.21	&58.33	&\textbf{60.76}	&\textbf{60.76}	&59.72	&59.72 \\
			A05	&30.21	&26.74	&28.47	&31.25	&32.29	&33.33	&42.01	&41.32	&41.67	&\textbf{43.40} \\
			A06	&41.32	&42.01	&40.63	&40.63	&40.63	&41.32	&43.75	&43.75	&\textbf{44.44}	&\textbf{44.44} \\
			A07	&\textbf{74.31}	&73.96	&68.40	&68.75	&64.93	&68.75	&71.18	&69.79	&68.75	&71.18 \\
			A08	&66.67	&59.03	&65.63	&69.79	&69.44	&\textbf{70.14}	&65.63	&63.89	&65.28	&66.32 \\
			A09	&67.01	&68.40	&70.49	&70.83	&72.22	&70.83	&77.43	&\textbf{78.82}	&78.13	&78.47 \\
			\hline
			Mean &57.45	&56.48	&57.29	&57.75	&57.87	&58.22	&61.15	&61.23	&60.88	&\textbf{61.57} \\
			$\pm$std    &14.56	&16.11	&15.51	&15.39	&15.49	&14.99	&13.74	&13.75	&13.26	&13.35 \\
			%\bottomrule
			\midrule
			\multirow{2}{*}{Subject} &\multirow{2}{*}{CSP-OVR}  &\multirow{2}{*}{CSP-PW} &\multirow{2}{*}{scaCSP} &\multicolumn{3}{c}{scaCSP\mreplaced{-extraSub}{$_{(extraSub)}$}} &scaCSP &\multicolumn{3}{c}{scaCSP-CNSR\mreplaced{-extraSub}{$_{(extraSub)}$}} \\
			\cline{5-7} \cline{9-11} &&& &$\mmat{S}_w^{range}$ &$\mmat{S}_w^{null}$ &$\mmat{S}_w^{range} + \mmat{S}_w^{null}$ &-CNSR &$\mmat{S}_w^{range}$ &$\mmat{S}_w^{null}$ &$\mmat{S}_w^{range} + \mmat{S}_w^{null}$\\
			\midrule
			B01	&33.75	&\textbf{41.25}	&\textbf{41.25}	&40.00	&40.00	&40.00	&31.25	&31.25	&32.50	&31.25 \\
			B02	&33.75	&33.75	&51.25	&53.75	&47.50	&53.75	&\textbf{58.75}	&\textbf{58.75}	&57.50	&\textbf{58.75} \\
			B03	&23.75	&\textbf{32.50}	&30.00	&30.00	&26.25	&30.00	&28.75	&30.00	&27.50	&30.00 \\
			B04	&\textbf{42.50}	&41.25	&36.25	&36.25	&35.00	&36.25	&38.75	&38.75	&37.50	&38.75 \\
			B05	&32.50	&38.75	&53.75	&\textbf{57.50}	&56.25	&\textbf{57.50}	&56.25	&\textbf{57.50}	&52.50	&\textbf{57.50} \\
			B06	&25.00	&28.75	&26.25	&\textbf{30.00}	&21.25	&\textbf{30.00}	&26.25	&27.50	&25.00	&27.50 \\
			B07	&35.00	&36.25	&33.75	&35.00	&28.75	&36.25	&33.75	&\textbf{38.75}	&35.00	&37.50 \\
			B08	&51.25	&51.25	&\textbf{52.50}	&46.25	&48.75	&46.25	&46.25	&50.00	&48.75	&50.00 \\
			B09	&\textbf{61.25}	&51.25	&51.25	&53.75	&53.75	&53.75	&48.75	&53.75	&50.00	&53.75 \\
			B10	&\textbf{80.00}	&75.00	&78.75	&72.50	&72.50	&72.50	&\textbf{80.00}	&76.25	&77.50	&76.25 \\
			\hline
			Mean &41.88	&43.00	&45.50	&45.50	&43.00	&45.63	&44.88	&\textbf{46.25}	&44.38	&46.13 \\
			$\pm$std	&16.73	&12.76	&14.63	&13.10	&14.98	&13.01	&15.89	&14.85	&15.24	&14.92 \\
			\midrule
			Grand mean &49.25	&49.39	&51.09	&51.3	&50.04	&51.59	&52.58	&53.34	&52.19	&\textbf{53.44} \\
			$\pm$std   &18.04	&16.37	&16.61	&15.92	&17.40	&15.75	&17.45	&16.62	&16.99	&16.60 \\
			\bottomrule
		\end{tabular}
	\end{table}
\end{landscape}

\subsection{\added{Computational efficiency}}

\added{
We also investigated the computational efficiency of the proposed algorithms and compared \added{them} with the competing methods. The results are summarized in \cref{tab:Runtime-binary} and \cref{tab:Runtime-multi} for the binary problems and multi-class problems, respectively. All the computations are conducted on a personal computer with an Intel Core i5-2400 CPU at 3.1 GHz and with 16 GB memory, running \deleted{in} Windows 10 and MATLAB R2021a. The run\deleted{ning }time is the period of computational stages including learning spatial filters, feature extraction and training classifiers in the training phase, whilst the period of feature extraction and classification in the testing phase. The time consumed in the preprocessing stage including the bandpass-filtering and cutting signals into trials is excluded since these steps are \deleted{all} the same for all the evaluated algorithms. Moreover, the running time in the testing phase is the total time for all the trials in the testing sets, i.e., 144 trials of 22-channel (resp. 20 trials of 32-channel) EEG for the binary problems and 288 trials of 22-channel (resp. 80 trials of 32-channel) EEG for the multi-class problems in DS1 (resp. DS2). As summarized in \cref{tab:dataset-info}, in the training phase there are 144 trials of 22-channel EEG in DS1 (resp. 40 trials of 32-channel EEG in DS2) for binary problems and 288 trials of 22-channel EEG in DS1 (resp. 160 trials of 32-channel EEG in DS2) for multi-class problems. }

\begin{table}[ht!]
	%\scriptsize
	\small
	\centering
	\caption{Running time (second) comparison between the competing and proposed algorithms for binary problems. Standard deviation is reported after the $\pm$. The lowest running time is marked in boldface for each dataset.}
	\label{tab:Runtime-binary}
	\begin{tabular}{*{5}{c}} % 5 columns
		\toprule
		\multirow{2}{*}{Algorithms} &\multicolumn{2}{c}{Training time ($\pm$ std)} &\multicolumn{2}{c}{Testing time ($\pm$ std)} \\
		\cline{2-3} \cline{4-5} &DS1 &DS2 &DS1 &DS2 \\
		\midrule
		CSP 	
		&\textbf{64.04} (1.39) 	&\textbf{63.48} (2.50)	&3.35 (0.12)			&2.24 (0.15)	\\
		TRCSP	
		&477.23 (3.87)			&469.42 (14.69)			&3.34 (0.07)			&2.23 (0.15)	\\
		sCSP	
		&489.55 (3.55)			&476.92 (19.79)			&3.34 (0.09)			&2.23 (0.17)	\\
		sTRCSP	
		&3798.78 (24.32)		&3748.62 (144.59)		&3.33 (0.07)			&2.23 (0.15)	\\
		scaCSP	
		&103.50 (4.16)			&91.49 (6.67)			&\textbf{2.26} (0.07)	&2.02 (0.19)	\\
		scaCSP$_{(\mmat{S}_w^{range})}$ 
		&159.25 (6.54)			&126.36 (10.95)			&2.40 (0.05)			&2.09 (0.15)	\\
		scaCSP$_{(\mmat{S}_t^{range})}$ 
		&158.99 (5.49)			&125.91 (10.79)			&2.40 (0.07)			&2.09 (0.13)	\\
		scaCSP$_{(\mmat{S}_w^{range} + \mmat{S}_t^{range})}$ 	
		&193.68 (8.55)			&143.84 (11.40)			&2.55 (0.09)			&2.20 (0.31)	\\
		scaCSP-CNSR 
		&152.72 (6.13)			&123.30 (10.22)			&\textbf{2.26} (0.05)	&\textbf{2.01} (0.13)	\\
		scaCSP-CNSR$_{(\mmat{S}_w^{range})}$ 
		&184.94 (3.34)			&142.53 (11.66)			&2.39 (0.05)			&2.11 (0.16)	\\
		scaCSP-CNSR$_{(\mmat{S}_t^{range})}$ 
		&168.23 (4.65)			&131.14 (9.78)			&2.39 (0.04)			&2.09 (0.08)	\\
		scaCSP-CNSR$_{(\mmat{S}_w^{range} + \mmat{S}_t^{range})}$ 
		&203.50 (10.68)			&148.43 (8.52)			&2.55 (0.11)			&2.20 (0.13)	\\
		\bottomrule
	\end{tabular}
\end{table}

\begin{table}[ht!]
	%\scriptsize
	\small
	\centering
	\caption{Running time (second) comparison between the competing and proposed algorithms for multi-class problems. The lowest running time is marked in boldface for each dataset. The extra null subspace \(\mmat{S}_b^{null}\) and $\mmat{S}_w^{null}$ are used for DS1 and DS2 respectively, which are indicated by $*$ in order to keep the table more compact. Similarly, the CNSR and BNSR processes are used in the scaCSP-NSR algorithms for DS1 and DS2, respectively, which are indicated by $\dagger$.}
	\label{tab:Runtime-multi}
	\begin{tabular}{*{5}{c}} % 5 columns
		\toprule
		\multirow{2}{*}{Algorithms} &\multicolumn{2}{c}{Training time ($\pm$ std)} &\multicolumn{2}{c}{Testing time ($\pm$ std)} \\
		\cline{2-3} \cline{4-5} &DS1 &DS2 &DS1 &DS2 \\
		\midrule
		CSP-OVR 
		&259.83 (10.09) 		&244.25 (1.45)			&20.99 (0.86)			&7.54 (0.15)	\\
		CSP-PW	
		&382.39 (6.59) 			&364.88 (0.83)			&31.51 (1.18)			&11.37 (0.18)	\\
		scaCSP	
		&\textbf{121.95} (4.28) &\textbf{103.16} (6.33)	&3.84 (0.07) 			&2.45 (0.09)	\\
		scaCSP$_{(*)}$ 
		&176.76 (3.70)			&146.83 (3.57			&4.06 (0.09)			&2.59 (0.07)	\\
		scaCSP$_{(\mmat{S}_w^{range})}$ 
		&195.26 (0.81)			&154.32 (3.54)			&4.01 (0.07)			&2.59 (0.09)	\\
		scaCSP$_{(* + \mmat{S}_w^{range})}$ 
		&251.05 (1.02)			&171.47 (4.40)			&4.26 (0.07)			&2.77 (0.11)	\\
		scaCSP-NSR$^\dagger$ 
		&133.90 (4.71)			&135.87 (3.53)			&\textbf{3.71} (0.10) 	&\textbf{2.44} (0.04)	\\
		scaCSP-NSR$^\dagger_{(*)}$ 
		&188.12 (4.81)			&177.78 (4.54)			&3.91 (0.14)			&2.58 (0.07)	\\
		scaCSP-NSR$^\dagger_{(\mmat{S}_w^{range})}$ 
		&209.22 (9.16)			&171.19 (1.91)			&3.88 (0.09)			&2.59 (0.05)	\\
		scaCSP-NSR$^\dagger_{(* + \mmat{S}_w^{range})}$ 
		&262.54 (9.12)			&187.76 (2.43)			&4.16 (0.18)			&2.76 (0.13)	\\
		\bottomrule
	\end{tabular}
\end{table}

\added{\cref{tab:Runtime-binary} indicates that the training time of the standard CSP is the least for the binary problems in both DS1 and DS2, whereas the training time of all the proposed scaCSP algorithms is also similarly comparable. The regularized CSP algorithms (TRCSP, sCSP, and sTRCSP) require much longer training time where most of the time \deleted{costed} is spent on the inner loop cross-validation for the optimization of the regularization parameters. The results indicate that all the algorithms (except sTRCSP) can be implemented efficiently \mreplaced{in a few minutes in the training phase}{to finish the training phase in a few minutes}. In the testing phase, scaCSP-CNSR takes the lowest testing time for both DS1 and DS2, but the testing time of all the other evaluated algorithms are also similarly comparable. It is also noteworthy that the testing time of all the proposed scaCSP algorithms is smaller than that of all the competing CSP algorithms.}

\added{Regarding the multi-class problems, it can be seen from \cref{tab:Runtime-multi} that both \added{the} training time and testing time of the proposed scaCSP algorithms are smaller than that of the competing CSP algorithms (CSP-OVR and CSP-PW) in both DS1 and DS2. The lowest training time is \mreplaced{costed by}{observed for} scaCSP in DS1 and DS2. The testing time of scaCSP-BNSR (resp. scaCSP-CNSR) is the least in DS1 (resp. DS2). The longer training time and testing time of multi-class CSP algorithms are due to the process \mreplaced{by}{of} splitting the multi-class problem into several binary problems and then learning spatial filters and training classifiers separately for each binary pair. }

\added{The results show the advantages of the proposed scaCSP approaches in both computational efficiency and \deleted{the} classification performance, especially for the multi-class problems. Moreover, although the training takes a \deleted{little} bit longer time, it can be conducted offline. Therefore, noting the small testing time, the proposed algorithms are also suitable for \mreplaced{real}{practical} BCI applications which are usually operated in real-time environments. }
%%==============================================================================
%%

%%==============================================================================
%%
\section{Discussion}

%General settings for time window and frequency band. Regularization. Feature selection for scaCSP in the subject specific way or even task-pair specific way. More extra subspaces or subspace combinations. Semi-empty or not.

Spatial filtering is an essential processing step to extract effective features for decoding brain states in MI-BCI applications. In this paper, we propose a novel spatial filtering framework called scaCSP based on the scatter matrices of the vectorized covariance samples including the between-class scatter matrix $\mmat{S}_b$, within-class scatter matrix $\mmat{S}_w$, and total scatter matrix $\mmat{S}_t$. The scaCSP algorithm computes spatial filters by the eigendecomposition of basis vectors of the null space of $\mmat{S}_b$ namely $\mmat{S}_b^{range}$, where the resulting eigenvectors are the filters and the corresponding eigenvalues act as an intrinsic criterion for filter selection. We have proved that specifically in the binary case, scsCSP is mathematically equivalent to CSP, which is the most commonly used and successful spatial filtering algorithm in MI-BCIs. However, the conventional CSP algorithm is limited to binary problems. On the opposite, the proposed scaCSP algorithm is suitable for not only binary classification problems but also multi-class problems. By revisiting CSP within our scatter-based framework, we see that CSP takes into account only the information in $\mmat{S}_b^{range}$. With an empirical analysis, we show that other subspaces ($\mmat{S}_w^{range}$, $\mmat{S}_w^{null}$, and $\mmat{S}_t^{range}$) also contain significant discriminant information on the one hand and we can extract extra spatial filters from all the other subspaces ($\mmat{S}_b^{null}$, $\mmat{S}_w^{range}$, $\mmat{S}_w^{null}$, $\mmat{S}_t^{range}$, and $\mmat{S}_t^{null}$) that are useful for classification on the other hand. To this end, the subspace enhanced scaCSP algorithms are also introduced by incorporating the discriminant information in these subspaces in two scenarios namely, \added{the} extra spatial filter learning scenario (scaCSP\mreplaced{-extraSub}{$_{(extraSub)}$}) and \added{the} subspace components reduction scenario (scaCSP-NSR) or both of them (scaCSP-NSR\mreplaced{-extraSub}{$_{(extraSub)}$}).

\added{We evaluated the proposed algorithms on a public data set DS1 and an in-house collected data set DS2 and compared \added{them} to the state-of-the-art CSP algorithms for both binary and multi-class problems. The experimental results show that the proposed scatter-based framework yields spatial filters outperforming the competing CSP algorithms (CSP, TRCSP, sCSP, and sTRCSP for binary problems and CSP-OVR and CSP-PW for multi-class problems) for MI classifications. Moreover, the challenges in DS2 are that the data set is a relatively small one and the signal \deleted{quality }from dry electrodes is in general known to be \deleted{worse}somewhat more prone to artifacts (\mreplaced{in terms}{i.e.} of lower SNR\deleted{, for example}) compared to \deleted{that with }wet electrodes. \mreplaced{Meanwhile}{Additionally}, all the subjects in DS2 \mreplaced{do}{did} not have any experience \mreplaced{on}{with} MI-BCIs (some of them even \mreplaced{know}{knew} nothing about MI-BCIs), i.e., they are all \deleted{the }so-called BCI-\mreplaced{n\"aive}{na\"ive} subjects. Noting that practical MI-BCI applications with conventional wet electrodes still suffer some drawbacks such as the complexity \mreplaced{to}{of} install\added{ation}, MI-BCI using dry electrodes, especially with \mreplaced{n\"aive}{na\"ive} subjects, has the potential to be usable out of the lab} \citep{popescu2007SingleTrialClassification, chi2012DryNoncontactEEG, grozea2011BristlesensorsLowcostFlexible, fiedler2015NovelMultipinElectrode, fiedler2021HighDensity256, lin2016DevelopmentWearableMotorImageryBased}. \added{The high classification performance and the low computational costs of the proposed scaCSP approach indicate its superiority in more practical BCI applications. }

Most significantly, it should be noted that the proposed scaCSP algorithms can be used as a straightforward replacement for CSP algorithms. Therefore, the regularization techniques introduced in \cite{lotte2011RegularizingCommonSpatial} and \cite{samek2012StationaryCommonSpatial} can be conveniently incorporated into our scaCSP algorithms. From \cref{sec:csp} and \cref{sec:rcsp} we see that the conventional CSP and RCSP algorithms with penalty terms in terms of quadratic forms can be solved within two steps: a whitening process and a PCA process. The regularization modifies the whitening process by regularizing the composite covariance in the denominator of \eqref{eq:J-rcsp}. Similarly as can be see\added{n} \mreplaced{from}{in} \cref{sec:scacsp-unwhitening}, the scaCSP approaches can be solved with the whitening process, which is totally the same as that in CSP algorithms, followed by two eigendecomposition processes. Therefore, the penalty terms in quadratic forms $P(\mvec{w}) = \mvec{w}\transpose \mmat{K} \mvec{w}$ used in RCSP can be directly added into the objective function of scaCSP in \eqref{eq:Qv-SB} yielding 
\begin{equation}
	Q^*_{reg}(\mvec{v}) = \dfrac{\mvec{v}\transpose \mmat{S}_B \mvec{v}}{\mvec{v}\transpose \left((\tilde{\mmat{C}} + \alpha \mmat{K}) \otimes (\tilde{\mmat{C}} + \alpha \mmat{K})\right) \mvec{v}} 
\end{equation}
with the regularization parameter $\alpha > 0$. Finally, the effects of matrix $\mmat{K}$ that encodes \textit{a priori} information go\deleted{es} into the whitening transform matrix $\mmat{P}$ in \mreplaced{\eqref{eq:P-appendix}}{\cref{sec:scacsp-unwhitening}}. Alternatively, scatter matrices used in scaCSP algorithms ($\mmat{S}_b$, $\mmat{S}_w$\added{,} and $\mmat{S}_t$) can also be regularized directly with some penalty terms incorporating the prior information. \mreplaced{While t}{T}he choice of the penalty terms and the regularization parameter is a complex and time\added{-}consuming problem, which goes \mreplaced{out of our main}{beyond the} scope \mreplaced{in}{of} this paper.

\begin{table}[ht!]
	%\scriptsize
	\smaller
	\centering
	\caption{\added{Number of bad channels and artifact-contaminated trials for each subject and session in DS1 and DS2.}}
	\label{tab:badchan-info}
	\begin{tabular}{*{12}{c}} % 9 columns
		\toprule
		\multicolumn{2}{c}{DS1} 
		&A01 	&A02 	&A03 	&A04 	&A05 	&A06 	&A07 	&A08 	&A09 	&	\\
		\midrule
		\multirow{2}{*}{\thead{Training\\session}} &bad channels
		&0 		&0 		&0 		&0 		&0 		&0 		&0 		&0 		&0 		&	\\
		&artifact trials 
		&6     	&1     	&5     	&6    	&12    	&21    	&21    	&15    	&17 	&	\\
		\hline
		\multirow{2}{*}{\thead{Testing\\session}} &bad channels 
		&0 		&0 		&0 		&0 		&0 		&0 		&0 		&0 		&0 		&	\\
		&artifact trials 
		&10     &8     	&6    	&10     &8    	&22    	&17    	&21     &7 		&	\\
		\midrule
		\multicolumn{2}{c}{DS2} 
		&B01 	&B02 	&B03 	&B04 	&B05 	&B06 	&B07 	&B08 	&B09 	&B10\\
		\midrule
		\multirow{2}{*}{\thead{Training\\session}} &bad channels 
		&0 		&0 		&0 		&3 		&0 		&0 		&1 		&3 		&0 		&2\\
		&artifact trials 
		&0    	&19    	&15    	&10    	&68     &5     	&6    	&33    	&13    	&12\\
		\hline
		\multirow{2}{*}{\thead{Testing\\session}} &bad channels 
		&0 		&0 		&0 		&2 		&0 		&1 		&1 		&2 		&0 		&2\\
		&artifact trials 
		&17    	&11     &5    	&11    	&45     &1     	&3     	&6     	&8     	&0\\
		\bottomrule
	\end{tabular}
\end{table}

The goal of this paper is to investigate the effectiveness of the proposed spatial filtering framework compared to the state-of-the-art competing CSP algorithms. Therefore, general settings for the time window, frequency band, and classifiers are used in the experiments in this work. Moreover, the experiments in this work do not apply any rejections of signal trial\added{s} and electrodes, which is also an important aspect to improve \deleted{the }signal quality and \deleted{reduce} SNR in BCI applications. \added{\textcolor{black}{In fact, a simple variance-based artifact detection is applied to mark artifact trials and bad channels with evident amplitude abnormalities in the used time interval between 0.5 s and 2.5 s after the cue. In particular, if the variance (across time) of the bandpass filtered signals exceeds a z-score (across electrodes) threshold of 4 or the standard deviation is smaller than 1, these electrodes are declared as bad channels for a given session. These bad channels are excluded from the subsequent artifact trial\deleted{s} detection steps. Then, if the bandpass filtered trial from any remaining electrodes has a magnitude greater than $\pm$ 125 $\mu$V or its variance (across time) exceeds a z-score (across trials) threshold of 4, this trial is declared as \added{an} artifact trial\deleted{s}. The detected bad channels and artifact trials are summarized in \cref{tab:badchan-info}. The average ratio\added{s} of artifact trials over all subjects are 4\% and 4.2\% in DS1, while 11.3\% and 13.4\% in DS2 for training and testing sessions, respectively. Additionally, there are \deleted{also} bad channels in DS2 \mreplaced{whereas no bad channels}{both not} in DS1. This makes it more challenging to extract robust spatial filters in DS2. In the training phase for binary problems (\cref{fig:results-training-binary-violin}), the stationary CSP variants (sCSP and sTRCSP) and the extra spatial filter enhanced scaCSP algorithms (scaCSP$_{(extraSub)}$) and scaCSP-NSR$_{(extraSub)}$) give higher classification accuracy in DS2 than that in DS1, indicating their robustness against artifacts. Similarly in the training phase for multi-class problems (\cref{fig:results-training-multi-violin}), all the proposed algorithms are more robust against artifacts than conventional multi-class extensions of CSP (CSP-OVR and CSP-PW). Similar results can also be found in the testing phase from \cref{tab:Results-binaryE-scaCSP} and \cref{tab:Results-multi-scaCSP}.}} This study preliminarily verifies the feasibility of using subspaces of \added{the} scatter matrices of covariances for brain signal analysis. The classification performance may be further enhanced by incorporating optimal spectral and temporal parameters\deleted{ using} such as the filter bank techniques \citep{ang2012FilterBankCommon,park2018FilterBankRegularized} and sliding window optimizations \citep{gaur2021SlidingWindowCommon,miao2021LearningCommonTimeFrequencySpatial}, exploiting other advanced classifiers \citep{lotte2018ReviewClassificationAlgorithms}, and incorporating further trial rejection and task-related electrodes selection algorithms \added{\textcolor{black}{such as independent component analysis (ICA)}} \citep{vigario2000IndependentComponentApproach}.

In future works, we will explore the application and effects of regularization techniques and spectral and temporal parameter optimizations on the scaCSP approaches more extensively. Moreover, the proposed approach will be implemented into MNE\mreplaced{-}{ }Scan \citep{esch2018MNEScanSoftware}, an open-source and cross-platform application allowing for the real-time processing of EEG/MEG signals. In combination with MagCPP \citep{oppermann2020MagCPPToolboxCombining}, which provides the external control of Magstim TMS device\added{s}, a\added{n} MI-based closed-loop TMS-EEG paradigm will be \deleted{future} evaluated \added{in future}.
%%==============================================================================
%%

%%==============================================================================
%%
\section{Conclusions}

 This work presents a novel spatial filtering framework (scaCSP) using the null spaces and range spaces of the scatter matrices of the vectorized covariance samples. The most important advantage of scaCSP is that it is a unified theoretical spatial filtering framework for both binary and multi-class classification problems. Instead of the conventional CSP algorithms which only take\deleted{s} into account information in $\mmat{S}_b^{range}$ within the spatial filter learning scenario, scaCSP makes use of the information contained in the covariances more extensively by incorporating all the subspaces into the spatial filtering framework in both the NSR process scenario and the spatial filter learning scenario. We evaluate the proposed algorithms on two EEG MI datasets, one from the public BCI competition dataset (9 subjects) for a baseline evaluation and a more challenging in-house dataset recorded with dry EEG electrodes from 10 BCI-\mreplaced{n\"aive}{na\"ive} subjects. Experimental results show that the proposed scaCSP algorithms outperform the state-of-the-art competing methods in terms of average classification accuracy in both binary and multi-class MI classification problems. \added{Moreover, the testing time of the proposed framework is \deleted{also} \mreplaced{less}{smaller} than that required by the competing methods for both binary and multi-class problems in all the evaluated data sets, indicating its superiority in both classification performance and computational efficiency.} The proposed framework \mreplaced{is}{has} shown to be a promising approach able to provide reliable EEG signal decoding results for MI-BCIs. 
%%==============================================================================
%%

%%==============================================================================
%%
\section*{Acknowledgments}

\added{This work was supported by the Deutsche Forschungsgemeinschaft (DFG) under Grant Ha2899/26-1 and by the Austrian \mreplaced{Wissenschaftsfonds}{Science Fund} (FWF) under Grant I3790-B27 and Grant P35949-B. This project received funding from the Free State of Thuringia (2018 IZN 004), co-financed by the European Union under the European Regional Development Fund (ERDF).}
%%==============================================================================
%%

%%==============================================================================
%%
\appendix
\section{Appendix} 

\subsection{Proof of Theorem 1} \label{sec:proof-theom1}

In this section we give the proof of \autoref{theorem1} as follows.

\mreplaced{Putting}{Inserting} \eqref{eq:eigen-tilde-R1} and \eqref{eq:eigen-tilde-R2} into \eqref{eq:tilde-rk}, we have 
\begin{subequations} \label{eq:eigen-tilde-rk}
	\begin{align}
		\tilde{\mvec{r}}_1 &= \sum_{j=1}^{N_c} \tilde{\lambda}_j \tilde{\mvec{u}}_j \otimes \tilde{\mvec{u}}_j
		= \sum_{j=1}^{N_c} \tilde{\lambda}_j \tilde{\mvec{v}}_j 
		= \tilde{\mmat{V}}_1 \tilde{\mvec{d}}_1 \label{eq:eigen-tilde-r1} \\
		\tilde{\mvec{r}}_2 &= \sum_{j=1}^{N_c} (1 - \tilde{\lambda}_j) \tilde{\mvec{u}}_j \otimes \tilde{\mvec{u}}_j
		= \sum_{j=1}^{N_c} (1 - \tilde{\lambda}_j) \tilde{\mvec{v}}_j 
		= \tilde{\mmat{V}}_1 (\mvec{1}^{N_c} - \tilde{\mvec{d}}_1) \label{eq:eigen-tilde-r2}
	\end{align}
\end{subequations}
where $\mathbb{R}^{N_c^2 \times N_c} \ni \tilde{\mmat{V}}_1 = [\tilde{\mvec{v}}_1, \cdots, \tilde{\mvec{v}}_{N_c}]$ whose elements $\tilde{\mvec{v}}_j = \tilde{\mvec{u}}_j \otimes \tilde{\mvec{u}}_j$ are the vectorizations of the eigenvectors and $\mvec{1}^{N_c}$ is a $N_c$-dimensional column vector filled with one.

By inserting \eqref{eq:eigen-tilde-rk} and \eqref{eq:tilde-r} into \eqref{eq:subeq-Sb}, $\mmat{S}_b$ can be expressed in terms of vectorized class-mean covariance matrices by
\begin{equation}
	\begin{split}
		\mmat{S}_b &= \dfrac{|\Omega_1| |\Omega_2|}{|\Omega|} \tilde{\mmat{V}}_1 \left( 2\tilde{\mvec{d}}_1 - \mvec{1}^{N_c} \right) \left( 2\tilde{\mvec{d}}_1 - \mvec{1}^{N_c} \right)\transpose \tilde{\mmat{V}}_1\transpose \\ 
		&= \lambda_1 \mvec{v}_1 \mvec{v}_1\transpose
	\end{split}
\end{equation}
where $\mvec{v}_1 \in \mathbb{R}^{N_c^2}$ is the eigenvector corresponding to the only nonzero eigenvalue $\lambda_1$ of the rank-1 matrix $\mmat{S}_b$. Moreover, note that 
\begin{equation}
	\begin{split}
		\tilde{\mvec{v}}_i\transpose \tilde{\mvec{v}}_j &= (\tilde{\mvec{u}}_i \otimes \tilde{\mvec{u}}_i) \transpose (\tilde{\mvec{u}}_j \otimes \tilde{\mvec{u}}_j) \\
		&= (\tilde{\mvec{u}}_i\transpose \tilde{\mvec{u}}_i) \otimes (\tilde{\mvec{u}}_j\transpose \tilde{\mvec{u}}_j) 
		= (\tilde{\mvec{u}}_i\transpose \tilde{\mvec{u}}_j)^2 
		= \begin{cases}
			1, i = j \\
			0, i \neq j
		\end{cases}
	\end{split}
\end{equation}
indicating that 
\begin{equation}
	\tilde{\mmat{V}}_1\transpose \tilde{\mmat{V}}_1 = \mmat{I}^{N_c}
\end{equation}
where $\mmat{I}^{N_c}$ is the $N_c \times N_c$ identity matrix. Then we have 
\begin{equation} 
	\begin{split}
		\mmat{S}_b \mvec{v}_1 &= \lambda_1 \mvec{v}_1 \\
		\text{with }
		\mvec{v}_1 &= \dfrac{\tilde{\mmat{V}}_1 (2\tilde{\mvec{d}}_1 - \mvec{1}^{N_c})}{\| 2\tilde{\mvec{d}}_1 - \mvec{1}^{N_c} \|_2} \\
		\lambda_1 &= \dfrac{|\Omega_1| |\Omega_2|}{|\Omega|} \| 2\tilde{\mvec{d}}_1 - \mvec{1}^{N_c} \|_2^2
	\end{split}
\end{equation}

Let $\mvec{v}_1 = \text{vec}(\mmat{A})$ be the vectorization of a $N_c \times N_c$ matrix $\mmat{A}$ whose eigendecomposition is given by \eqref{eq:eigen-A}. Then we have 
\begin{equation} 
	\text{vec}(\mmat{A}) = \dfrac{\tilde{\mmat{V}}_1 (2\tilde{\mvec{d}}_1 - \mvec{1}^{N_c})}{\| 2\tilde{\mvec{d}}_1 - \mvec{1}^{N_c} \|_2} = \sum_{j=1}^{N_c} \dfrac{2\tilde{\lambda}_j - 1}{\| 2\tilde{\mvec{d}}_1 - \mvec{1}^{N_c} \|_2} \text{vec}(\tilde{\mvec{u}}_j \tilde{\mvec{u}}_j\transpose)
\end{equation}
which, by noting that the operator vec($\cdot$) is a linear transformation, indicates that $\mmat{A}$ also has the eigendecomposition
\begin{equation} 
	\mmat{A} = \sum_{j=1}^{N_c} \dfrac{2\tilde{\lambda}_j - 1}{\| 2\tilde{\mvec{d}}_1 - \mvec{1}^{N_c} \|_2} \tilde{\mvec{u}}_j \tilde{\mvec{u}}_j\transpose = \dfrac{1}{\| 2\tilde{\mvec{d}}_1 - \mvec{1}^{N_c} \|_2} \tilde{\mmat{U}}_1 (2\tilde{\mmat{\Lambda}}_1 - \mmat{I}^{N_c}) \tilde{\mmat{U}}_1\transpose
\end{equation}

Comparing to \eqref{eq:eigen-A}, we finally have 
\begin{equation} 
	\begin{split}
		\mmat{U}_a &= \tilde{\mmat{U}}_1 \\
		\mvec{d}_a &= \dfrac{2\tilde{\mvec{d}}_1 - \mvec{1}^{N_c}}{\| 2\tilde{\mvec{d}}_1 - \mvec{1}^{N_c} \|_2}
	\end{split}
\end{equation}
leading to identical spatial filters $\mmat{W} = \mmat{W}_a$ (sorted by $\tilde{\lambda}_i$ and $\lambda_{ai}$, respectively) since we are using the same whitening projection $\mmat{P}_c$. The proof is thus completed.

\subsection{scaCSP with un-whitened covariance} \label{sec:scacsp-unwhitening}

In this section, we introduce the scaCSP formulated with the un-whitened covariance matrices as the conventional CSP does in order to keep the scatter-based framework consistent and complete.

Let $\mmat{X}_i = [\mvec{x}_{i1}, \cdots, \mvec{x}_{iN_t}]$ be the EEG data matrix of $i$-th trial and $\mmat{C}_i = \frac{1}{N_t - 1} \mmat{X}_i \mmat{X}_i\transpose$ its covariance matrix. The vectorization of $\mmat{C}_i$ is 
\begin{equation}
	\mvec{c}_i = \text{vec}(\mmat{C}_i) = \frac{1}{N_t - 1} \sum_{j=1}^{N_t} \mvec{x}_{ij} \otimes \mvec{x}_{ij} 
\end{equation}
The class mean and total mean vectors are then given by 
\begin{subequations}
	\begin{align}
		&\tilde{\mvec{c}}_k = \text{vec}(\tilde{\mmat{C}}_k) = \frac{1}{|\Omega_k|} \sum_{i \in \Omega_k} \mvec{c}_i, k = 1, \cdots, N_\Omega \label{eq:tilde-ck} \\
		&\tilde{\mvec{c}} = \frac{1}{|\Omega|} \sum_{i \in \Omega} \mvec{c}_i 
		= \dfrac{1}{|\Omega|} \sum_{k=1}^{N_\Omega} |\Omega_k| \tilde{\mvec{c}}_k \label{eq:tilde-c}
	\end{align}
\end{subequations}
The scatter matrices are defined as 
\begin{subequations} \label{eq:scatter-of-covariance-unwhitened}
	\begin{align}
		&\mmat{S}_W = \sum_{k=1}^{N_\Omega} \sum_{i \in \Omega_k} \left( \mvec{c}_i - \tilde{\mvec{c}}_k \right) \left( \mvec{c}_i - \tilde{\mvec{c}}_k \right)\transpose \label{eq:subeq-SW} \\
		&\mmat{S}_B = \sum_{k=1}^{N_\Omega} |\Omega_k| \left( \tilde{\mvec{c}}_k - \tilde{\mvec{c}} \right) \left( \tilde{\mvec{c}}_k - \tilde{\mvec{c}} \right)\transpose \label{eq:subeq-SB} \\
		&\mmat{S}_T = \sum_{i \in \Omega} (\mvec{c}_i - \tilde{\mvec{c}}) (\mvec{c}_i - \tilde{\mvec{c}})\transpose \label{eq:subeq-ST}
	\end{align}
\end{subequations}
with $\mmat{S}_T = \mmat{S}_W + \mmat{S}_B$. The objective function in \eqref{eq:Qv-Sb} is modified as 
\begin{equation} \label{eq:Qv-SB}
	Q^*(\mvec{v}) = \dfrac{\mvec{v}\transpose \mmat{S}_B \mvec{v}}{\mvec{v}\transpose (\tilde{\mmat{C}} \otimes \tilde{\mmat{C}}) \mvec{v}}
\end{equation}
where $\tilde{\mmat{C}}$ is the composite covariance matrix given by $\tilde{\mmat{C}} = \sum_{k=1}^{N_\Omega} \tilde{\mmat{C}}_k$.

Similar to the conventional CSP algorithm, the solution to maximizing the Rayleigh quotient in \eqref{eq:Qv-SB} can be given by the simultaneous diagonalization approach. Noting $\tilde{\mmat{C}}$ has the eigendecomposition $\tilde{\mmat{C}} = \tilde{\mmat{U}}_c \tilde{\mmat{\Lambda}}_c \tilde{\mmat{U}}_c\transpose$, the eigendecomposition of $\tilde{\mmat{C}} \otimes \tilde{\mmat{C}}$ can be given by \citep{golub2013MatrixComputations}
\begin{equation} 
	\tilde{\mmat{C}} \otimes \tilde{\mmat{C}} = (\tilde{\mmat{U}}_c \otimes \tilde{\mmat{U}}_c) (\tilde{\mmat{\Lambda}}_c \otimes \tilde{\mmat{\Lambda}}_c) (\tilde{\mmat{U}}_c \otimes \tilde{\mmat{U}}_c)\transpose
\end{equation}
It should be noted that the eigenvectors are not unique considering the structure property of the Kronecker product. Eigendecompositions computed directly from $\tilde{\mmat{C}} \otimes \tilde{\mmat{C}}$ (e.g., using the function eig($\cdot$) in MATLAB) may not lead\deleted{s} to the proper results. We use the whitening transform with 
\begin{equation}
	\mmat{P} = (\tilde{\mmat{U}}_c \otimes \tilde{\mmat{U}}_c) (\tilde{\mmat{\Lambda}}_c \otimes \tilde{\mmat{\Lambda}}_c)^{(-1/2)}
\end{equation}
which yields 
\begin{equation} \label{eq:P-appendix}
	\mmat{P}\transpose (\tilde{\mmat{C}} \otimes \tilde{\mmat{C}}) \mmat{P} = \mmat{I}
\end{equation}
and 
\begin{equation} 
	\mmat{P}\transpose \mmat{S}_B \mmat{P} = \mmat{S}_b
\end{equation}
where $\mmat{S}_b$ is the between-scatter matrix defined in \eqref{eq:subeq-Sb}. Then maximizing \eqref{eq:Qv-SB} is equivalent to maximizing \eqref{eq:Qv-Sb} in addition to a whitening transform with $\mmat{P}$. 
%%==============================================================================
%%

%%==============================================================================
%%
% Bibliography
\bibliography{Biblio_exportedZotero.bib}

\begin{thebibliography}{65}
\expandafter\ifx\csname natexlab\endcsname\relax\def\natexlab#1{#1}\fi
\providecommand{\url}[1]{\texttt{#1}}
\providecommand{\href}[2]{#2}
\providecommand{\path}[1]{#1}
\providecommand{\DOIprefix}{doi:}
\providecommand{\ArXivprefix}{arXiv:}
\providecommand{\URLprefix}{URL: }
\providecommand{\Pubmedprefix}{pmid:}
\providecommand{\doi}[1]{\href{http://dx.doi.org/#1}{\path{#1}}}
\providecommand{\Pubmed}[1]{\href{pmid:#1}{\path{#1}}}
\providecommand{\bibinfo}[2]{#2}
\ifx\xfnm\relax \def\xfnm[#1]{\unskip,\space#1}\fi
%Type = Article
\bibitem[{Ang et~al.(2012)Ang, Chin, Wang, Guan \&
  Zhang}]{ang2012FilterBankCommon}
\bibinfo{author}{Ang, K.~K.}, \bibinfo{author}{Chin, Z.~Y.},
  \bibinfo{author}{Wang, C.}, \bibinfo{author}{Guan, C.}, \&
  \bibinfo{author}{Zhang, H.} (\bibinfo{year}{2012}).
\newblock \bibinfo{title}{Filter {{Bank Common Spatial Pattern Algorithm}} on
  {{BCI Competition IV Datasets}} 2a and 2b}.
\newblock {\it \bibinfo{journal}{Frontiers in Neuroscience}\/},  {\it
  \bibinfo{volume}{6}\/}. \DOIprefix\doi{10.3389/fnins.2012.00039}.
%Type = Article
\bibitem[{Arpaia et~al.(2022)Arpaia, Esposito, Natalizio \&
  Parvis}]{arpaia2022HowSuccessfullyClassify}
\bibinfo{author}{Arpaia, P.}, \bibinfo{author}{Esposito, A.},
  \bibinfo{author}{Natalizio, A.}, \& \bibinfo{author}{Parvis, M.}
  (\bibinfo{year}{2022}).
\newblock \bibinfo{title}{How to successfully classify {{EEG}} in motor imagery
  {{BCI}}: A metrological analysis of the state of the art}.
\newblock {\it \bibinfo{journal}{Journal of Neural Engineering}\/},  {\it
  \bibinfo{volume}{19}\/}, \bibinfo{pages}{031002}.
  \DOIprefix\doi{10.1088/1741-2552/ac74e0}.
%Type = Article
\bibitem[{Arvaneh et~al.(2013)Arvaneh, {Cuntai Guan}, {Kai Keng Ang} \& {Chai
  Quek}}]{arvaneh2013OptimizingSpatialFilters}
\bibinfo{author}{Arvaneh, M.}, \bibinfo{author}{{Cuntai Guan}},
  \bibinfo{author}{{Kai Keng Ang}}, \& \bibinfo{author}{{Chai Quek}}
  (\bibinfo{year}{2013}).
\newblock \bibinfo{title}{Optimizing {{Spatial Filters}} by {{Minimizing
  Within-Class Dissimilarities}} in {{Electroencephalogram-Based
  Brain}}\textendash{{Computer Interface}}}.
\newblock {\it \bibinfo{journal}{IEEE Transactions on Neural Networks and
  Learning Systems}\/},  {\it \bibinfo{volume}{24}\/},
  \bibinfo{pages}{610--619}. \DOIprefix\doi{10.1109/TNNLS.2013.2239310}.
%Type = Article
\bibitem[{Azab et~al.(2020)Azab, Ahmadi, Mihaylova \&
  Arvaneh}]{azab2020DynamicTimeWarpingbased}
\bibinfo{author}{Azab, A.~M.}, \bibinfo{author}{Ahmadi, H.},
  \bibinfo{author}{Mihaylova, L.}, \& \bibinfo{author}{Arvaneh, M.}
  (\bibinfo{year}{2020}).
\newblock \bibinfo{title}{Dynamic time warping-based transfer learning for
  improving common spatial patterns in brain\textendash computer interface}.
\newblock {\it \bibinfo{journal}{Journal of Neural Engineering}\/},  {\it
  \bibinfo{volume}{17}\/}, \bibinfo{pages}{016061}.
  \DOIprefix\doi{10.1088/1741-2552/ab64a0}.
%Type = Article
\bibitem[{Balzi et~al.(2015)Balzi, Yger \&
  Sugiyama}]{balzi2015ImportanceweightedCovarianceEstimation}
\bibinfo{author}{Balzi, A.}, \bibinfo{author}{Yger, F.}, \&
  \bibinfo{author}{Sugiyama, M.} (\bibinfo{year}{2015}).
\newblock \bibinfo{title}{Importance-weighted covariance estimation for robust
  common spatial pattern}.
\newblock {\it \bibinfo{journal}{Pattern Recognition Letters}\/},  {\it
  \bibinfo{volume}{68}\/}, \bibinfo{pages}{139--145}.
  \DOIprefix\doi{10.1016/j.patrec.2015.09.003}.
%Type = Article
\bibitem[{Bennett et~al.(2021)Bennett, John, Grayden \&
  Burkitt}]{bennett2021NeurophysiologicalApproachSpatial}
\bibinfo{author}{Bennett, J.~D.}, \bibinfo{author}{John, S.~E.},
  \bibinfo{author}{Grayden, D.~B.}, \& \bibinfo{author}{Burkitt, A.~N.}
  (\bibinfo{year}{2021}).
\newblock \bibinfo{title}{A neurophysiological approach to spatial filter
  selection for adaptive brain\textendash computer interfaces}.
\newblock {\it \bibinfo{journal}{Journal of Neural Engineering}\/},  {\it
  \bibinfo{volume}{18}\/}, \bibinfo{pages}{026017}.
  \DOIprefix\doi{10.1088/1741-2552/abd51f}.
%Type = Book
\bibitem[{Bishop(2006)}]{bishop2006PatternRecognitionMachine}
\bibinfo{author}{Bishop, C.~M.} (\bibinfo{year}{2006}).
\newblock {\it \bibinfo{title}{Pattern Recognition and Machine Learning}\/}.
\newblock Information Science and Statistics.
\newblock \bibinfo{address}{{New York}}: \bibinfo{publisher}{{Springer}}.
%Type = Article
\bibitem[{Blankertz et~al.(2007)Blankertz, Dornhege, Krauledat, M{\"u}ller \&
  Curio}]{blankertz2007NoninvasiveBerlinBrain}
\bibinfo{author}{Blankertz, B.}, \bibinfo{author}{Dornhege, G.},
  \bibinfo{author}{Krauledat, M.}, \bibinfo{author}{M{\"u}ller, K.-R.}, \&
  \bibinfo{author}{Curio, G.} (\bibinfo{year}{2007}).
\newblock \bibinfo{title}{The non-invasive {{Berlin
  Brain}}\textendash{{Computer Interface}}: {{Fast}} acquisition of effective
  performance in untrained subjects}.
\newblock {\it \bibinfo{journal}{NeuroImage}\/},  {\it \bibinfo{volume}{37}\/},
  \bibinfo{pages}{539--550}. \DOIprefix\doi{10.1016/j.neuroimage.2007.01.051}.
%Type = Article
\bibitem[{Blankertz et~al.(2004)Blankertz, Muller, Curio, Vaughan, Schalk,
  Wolpaw, Schlogl, Neuper, Pfurtscheller, Hinterberger, Schroder \&
  Birbaumer}]{blankertz2004BCICompetition2003}
\bibinfo{author}{Blankertz, B.}, \bibinfo{author}{Muller, K.-R.},
  \bibinfo{author}{Curio, G.}, \bibinfo{author}{Vaughan, T.},
  \bibinfo{author}{Schalk, G.}, \bibinfo{author}{Wolpaw, J.},
  \bibinfo{author}{Schlogl, A.}, \bibinfo{author}{Neuper, C.},
  \bibinfo{author}{Pfurtscheller, G.}, \bibinfo{author}{Hinterberger, T.},
  \bibinfo{author}{Schroder, M.}, \& \bibinfo{author}{Birbaumer, N.}
  (\bibinfo{year}{2004}).
\newblock \bibinfo{title}{The {{BCI}} competition 2003: Progress and
  perspectives in detection and discrimination of {{EEG}} single trials}.
\newblock {\it \bibinfo{journal}{IEEE Transactions on Biomedical
  Engineering}\/},  {\it \bibinfo{volume}{51}\/}, \bibinfo{pages}{1044--1051}.
  \DOIprefix\doi{10.1109/TBME.2004.826692}.
%Type = Article
\bibitem[{Blankertz et~al.(2006)Blankertz, Muller, Krusienski, Schalk, Wolpaw,
  Schlogl, Pfurtscheller, Millan, Schroder \&
  Birbaumer}]{blankertz2006BCICompetitionIII}
\bibinfo{author}{Blankertz, B.}, \bibinfo{author}{Muller, K.-R.},
  \bibinfo{author}{Krusienski, D.}, \bibinfo{author}{Schalk, G.},
  \bibinfo{author}{Wolpaw, J.}, \bibinfo{author}{Schlogl, A.},
  \bibinfo{author}{Pfurtscheller, G.}, \bibinfo{author}{Millan, J.},
  \bibinfo{author}{Schroder, M.}, \& \bibinfo{author}{Birbaumer, N.}
  (\bibinfo{year}{2006}).
\newblock \bibinfo{title}{The {{BCI}} competition {{III}}: Validating
  alternative approaches to actual {{BCI}} problems}.
\newblock {\it \bibinfo{journal}{IEEE Transactions on Neural Systems and
  Rehabilitation Engineering}\/},  {\it \bibinfo{volume}{14}\/},
  \bibinfo{pages}{153--159}. \DOIprefix\doi{10.1109/TNSRE.2006.875642}.
%Type = Article
\bibitem[{Blankertz et~al.(2008)Blankertz, Tomioka, Lemm, Kawanabe \&
  Muller}]{blankertz2008OptimizingSpatialFilters}
\bibinfo{author}{Blankertz, B.}, \bibinfo{author}{Tomioka, R.},
  \bibinfo{author}{Lemm, S.}, \bibinfo{author}{Kawanabe, M.}, \&
  \bibinfo{author}{Muller, K.-r.} (\bibinfo{year}{2008}).
\newblock \bibinfo{title}{Optimizing {{Spatial}} filters for {{Robust EEG
  Single-Trial Analysis}}}.
\newblock {\it \bibinfo{journal}{IEEE Signal Processing Magazine}\/},  {\it
  \bibinfo{volume}{25}\/}, \bibinfo{pages}{41--56}.
  \DOIprefix\doi{10.1109/MSP.2008.4408441}.
%Type = Article
\bibitem[{Brunner et~al.(2015)Brunner, Birbaumer, Blankertz, Guger, K{\"u}bler,
  Mattia, Mill{\'a}n, Miralles, Nijholt, Opisso, Ramsey, Salomon \&
  {M{\"u}ller-Putz}}]{brunner2015BNCIHorizon2020}
\bibinfo{author}{Brunner, C.}, \bibinfo{author}{Birbaumer, N.},
  \bibinfo{author}{Blankertz, B.}, \bibinfo{author}{Guger, C.},
  \bibinfo{author}{K{\"u}bler, A.}, \bibinfo{author}{Mattia, D.},
  \bibinfo{author}{Mill{\'a}n, J. d.~R.}, \bibinfo{author}{Miralles, F.},
  \bibinfo{author}{Nijholt, A.}, \bibinfo{author}{Opisso, E.},
  \bibinfo{author}{Ramsey, N.}, \bibinfo{author}{Salomon, P.}, \&
  \bibinfo{author}{{M{\"u}ller-Putz}, G.~R.} (\bibinfo{year}{2015}).
\newblock \bibinfo{title}{{{BNCI Horizon}} 2020: Towards a roadmap for the
  {{BCI}} community}.
\newblock {\it \bibinfo{journal}{Brain-Computer Interfaces}\/},  {\it
  \bibinfo{volume}{2}\/}, \bibinfo{pages}{1--10}.
  \DOIprefix\doi{10.1080/2326263X.2015.1008956}.
%Type = Article
\bibitem[{Cai et~al.(2021)Cai, Gong, Deng \&
  Wang}]{cai2021SingleTrialEEGClassification}
\bibinfo{author}{Cai, Q.}, \bibinfo{author}{Gong, W.}, \bibinfo{author}{Deng,
  Y.}, \& \bibinfo{author}{Wang, H.} (\bibinfo{year}{2021}).
\newblock \bibinfo{title}{Single-{{Trial EEG Classification}} via {{Common
  Spatial Patterns}} with {{Mixed Lp-}} and {{Lq-Norms}}}.
\newblock {\it \bibinfo{journal}{Mathematical Problems in Engineering}\/},
  {\it \bibinfo{volume}{2021}\/}, \bibinfo{pages}{1--13}.
  \DOIprefix\doi{10.1155/2021/6645322}.
%Type = Article
\bibitem[{Chakraborty et~al.(2020)Chakraborty, Ghosh \&
  Konar}]{chakraborty2020DesigningPhaseSensitiveCommon}
\bibinfo{author}{Chakraborty, B.}, \bibinfo{author}{Ghosh, L.}, \&
  \bibinfo{author}{Konar, A.} (\bibinfo{year}{2020}).
\newblock \bibinfo{title}{Designing {{Phase-Sensitive Common Spatial Pattern
  Filter}} to {{Improve Brain-Computer Interfacing}}}.
\newblock {\it \bibinfo{journal}{IEEE Transactions on Biomedical
  Engineering}\/},  (pp. \bibinfo{pages}{1--1}).
  \DOIprefix\doi{10.1109/TBME.2019.2954470}.
%Type = Article
\bibitem[{Chaudhary et~al.(2016)Chaudhary, Birbaumer \&
  {Ramos-Murguialday}}]{chaudhary2016BrainComputerInterfaces}
\bibinfo{author}{Chaudhary, U.}, \bibinfo{author}{Birbaumer, N.}, \&
  \bibinfo{author}{{Ramos-Murguialday}, A.} (\bibinfo{year}{2016}).
\newblock \bibinfo{title}{Brain\textendash computer interfaces for
  communication and rehabilitation}.
\newblock {\it \bibinfo{journal}{Nature Reviews Neurology}\/},  {\it
  \bibinfo{volume}{12}\/}, \bibinfo{pages}{513--525}.
  \DOIprefix\doi{10.1038/nrneurol.2016.113}.
%Type = Article
\bibitem[{Chi et~al.(2012)Chi, Wang, Wang, Maier, Jung \&
  Cauwenberghs}]{chi2012DryNoncontactEEG}
\bibinfo{author}{Chi, Y.~M.}, \bibinfo{author}{Wang, Y.-T.},
  \bibinfo{author}{Wang, Y.}, \bibinfo{author}{Maier, C.},
  \bibinfo{author}{Jung, T.-P.}, \& \bibinfo{author}{Cauwenberghs, G.}
  (\bibinfo{year}{2012}).
\newblock \bibinfo{title}{Dry and {{Noncontact EEG Sensors}} for {{Mobile
  Brain}}\textendash{{Computer Interfaces}}}.
\newblock {\it \bibinfo{journal}{IEEE Transactions on Neural Systems and
  Rehabilitation Engineering}\/},  {\it \bibinfo{volume}{20}\/},
  \bibinfo{pages}{228--235}. \DOIprefix\doi{10.1109/TNSRE.2011.2174652}.
%Type = Book
\bibitem[{Dornhege(2007)}]{dornhege2007BraincomputerInterfacing}
\bibinfo{editor}{Dornhege, G.} (Ed.) (\bibinfo{year}{2007}).
\newblock {\it \bibinfo{title}{Toward Brain-Computer Interfacing}\/}.
\newblock Neural Information Processing Series.
\newblock \bibinfo{address}{{Cambridge, Mass}}: \bibinfo{publisher}{{MIT
  Press}}.
%Type = Article
\bibitem[{Dornhege et~al.(2006)Dornhege, Blankertz, Krauledat, Losch, Curio \&
  Muller}]{dornhege2006CombinedOptimizationSpatial}
\bibinfo{author}{Dornhege, G.}, \bibinfo{author}{Blankertz, B.},
  \bibinfo{author}{Krauledat, M.}, \bibinfo{author}{Losch, F.},
  \bibinfo{author}{Curio, G.}, \& \bibinfo{author}{Muller, K.-R.}
  (\bibinfo{year}{2006}).
\newblock \bibinfo{title}{Combined {{Optimization}} of {{Spatial}} and
  {{Temporal Filters}} for {{Improving Brain-Computer Interfacing}}}.
\newblock {\it \bibinfo{journal}{IEEE Transactions on Biomedical
  Engineering}\/},  {\it \bibinfo{volume}{53}\/}, \bibinfo{pages}{2274--2281}.
  \DOIprefix\doi{10.1109/TBME.2006.883649}.
%Type = Article
\bibitem[{Esch et~al.(2018)Esch, Sun, Kl{\"u}ber, Lew, Baumgarten, Grant,
  Okada, Haueisen, H{\"a}m{\"a}l{\"a}inen \& Dinh}]{esch2018MNEScanSoftware}
\bibinfo{author}{Esch, L.}, \bibinfo{author}{Sun, L.},
  \bibinfo{author}{Kl{\"u}ber, V.}, \bibinfo{author}{Lew, S.},
  \bibinfo{author}{Baumgarten, D.}, \bibinfo{author}{Grant, P.~E.},
  \bibinfo{author}{Okada, Y.}, \bibinfo{author}{Haueisen, J.},
  \bibinfo{author}{H{\"a}m{\"a}l{\"a}inen, M.~S.}, \& \bibinfo{author}{Dinh,
  C.} (\bibinfo{year}{2018}).
\newblock \bibinfo{title}{{{MNE Scan}}: {{Software}} for real-time processing
  of electrophysiological data}.
\newblock {\it \bibinfo{journal}{Journal of Neuroscience Methods}\/},  {\it
  \bibinfo{volume}{303}\/}, \bibinfo{pages}{55--67}.
  \DOIprefix\doi{10.1016/j.jneumeth.2018.03.020}.
%Type = Article
\bibitem[{Feng et~al.(2018)Feng, Yin, Jin, Saab, Daly, Wang, Hu \&
  Cichocki}]{feng2018CorrelationbasedTimeWindow}
\bibinfo{author}{Feng, J.}, \bibinfo{author}{Yin, E.}, \bibinfo{author}{Jin,
  J.}, \bibinfo{author}{Saab, R.}, \bibinfo{author}{Daly, I.},
  \bibinfo{author}{Wang, X.}, \bibinfo{author}{Hu, D.}, \&
  \bibinfo{author}{Cichocki, A.} (\bibinfo{year}{2018}).
\newblock \bibinfo{title}{Towards correlation-based time window selection
  method for motor imagery {{BCIs}}}.
\newblock {\it \bibinfo{journal}{Neural Networks}\/},  {\it
  \bibinfo{volume}{102}\/}, \bibinfo{pages}{87--95}.
  \DOIprefix\doi{10.1016/j.neunet.2018.02.011}.
%Type = Article
\bibitem[{Fiedler et~al.(2021)Fiedler, Fonseca, Supriyanto, Zanow \&
  Haueisen}]{fiedler2021HighDensity256}
\bibinfo{author}{Fiedler, P.}, \bibinfo{author}{Fonseca, C.},
  \bibinfo{author}{Supriyanto, E.}, \bibinfo{author}{Zanow, F.}, \&
  \bibinfo{author}{Haueisen, J.} (\bibinfo{year}{2021}).
\newblock \bibinfo{title}{A high-density 256-channel cap for dry
  electroencephalography}.
\newblock {\it \bibinfo{journal}{Human Brain Mapping}\/},  (p.
  \bibinfo{pages}{hbm.25721}). \DOIprefix\doi{10.1002/hbm.25721}.
%Type = Article
\bibitem[{Fiedler et~al.(2015)Fiedler, Pedrosa, Griebel, Fonseca, Vaz,
  Supriyanto, Zanow \& Haueisen}]{fiedler2015NovelMultipinElectrode}
\bibinfo{author}{Fiedler, P.}, \bibinfo{author}{Pedrosa, P.},
  \bibinfo{author}{Griebel, S.}, \bibinfo{author}{Fonseca, C.},
  \bibinfo{author}{Vaz, F.}, \bibinfo{author}{Supriyanto, E.},
  \bibinfo{author}{Zanow, F.}, \& \bibinfo{author}{Haueisen, J.}
  (\bibinfo{year}{2015}).
\newblock \bibinfo{title}{Novel {{Multipin Electrode Cap System}} for {{Dry
  Electroencephalography}}}.
\newblock {\it \bibinfo{journal}{Brain Topography}\/},  {\it
  \bibinfo{volume}{28}\/}, \bibinfo{pages}{647--656}.
  \DOIprefix\doi{10.1007/s10548-015-0435-5}.
%Type = Book
\bibitem[{Fukunaga(1990)}]{fukunaga1990IntroductionStatisticalPattern}
\bibinfo{author}{Fukunaga, K.} (\bibinfo{year}{1990}).
\newblock {\it \bibinfo{title}{Introduction to Statistical Pattern
  Recognition}\/}.
\newblock Computer Science and Scientific Computing (\bibinfo{edition}{2nd}
  ed.).
\newblock \bibinfo{address}{{Boston}}: \bibinfo{publisher}{{Academic Press}}.
%Type = Article
\bibitem[{Gaur et~al.(2021)Gaur, Gupta, Chowdhury, McCreadie, Pachori \&
  Wang}]{gaur2021SlidingWindowCommon}
\bibinfo{author}{Gaur, P.}, \bibinfo{author}{Gupta, H.},
  \bibinfo{author}{Chowdhury, A.}, \bibinfo{author}{McCreadie, K.},
  \bibinfo{author}{Pachori, R.~B.}, \& \bibinfo{author}{Wang, H.}
  (\bibinfo{year}{2021}).
\newblock \bibinfo{title}{A {{Sliding Window Common Spatial Pattern}} for
  {{Enhancing Motor Imagery Classification}} in {{EEG-BCI}}}.
\newblock {\it \bibinfo{journal}{IEEE Transactions on Instrumentation and
  Measurement}\/},  (pp. \bibinfo{pages}{1--1}).
  \DOIprefix\doi{10.1109/TIM.2021.3051996}.
%Type = Article
\bibitem[{Ghanbar et~al.(2021)Ghanbar, Rezaii, Farzamnia \&
  Saad}]{ghanbar2021CorrelationbasedCommonSpatial}
\bibinfo{author}{Ghanbar, K.~D.}, \bibinfo{author}{Rezaii, T.~Y.},
  \bibinfo{author}{Farzamnia, A.}, \& \bibinfo{author}{Saad, I.}
  (\bibinfo{year}{2021}).
\newblock \bibinfo{title}{Correlation-based common spatial pattern ({{CCSP}}):
  {{A}} novel extension of {{CSP}} for classification of motor imagery signal}.
\newblock {\it \bibinfo{journal}{PLOS ONE}\/},  {\it \bibinfo{volume}{16}\/},
  \bibinfo{pages}{e0248511}. \DOIprefix\doi{10.1371/journal.pone.0248511}.
%Type = Article
\bibitem[{Ghojogh et~al.(2019)Ghojogh, Karray \&
  Crowley}]{ghojogh2019eigenvalue}
\bibinfo{author}{Ghojogh, B.}, \bibinfo{author}{Karray, F.}, \&
  \bibinfo{author}{Crowley, M.} (\bibinfo{year}{2019}).
\newblock \bibinfo{title}{Eigenvalue and generalized eigenvalue problems:
  Tutorial}.
\newblock {\it \bibinfo{journal}{arXiv preprint arXiv:1903.11240}\/}, .
%Type = Book
\bibitem[{Golub \& Van~Loan(2013)}]{golub2013MatrixComputations}
\bibinfo{author}{Golub, G.~H.}, \& \bibinfo{author}{Van~Loan, C.~F.}
  (\bibinfo{year}{2013}).
\newblock {\it \bibinfo{title}{Matrix Computations}\/}.
\newblock Johns {{Hopkins}} Studies in the Mathematical Sciences
  (\bibinfo{edition}{fourth edition} ed.).
\newblock \bibinfo{address}{{Baltimore}}: \bibinfo{publisher}{{The Johns
  Hopkins University Press}}.
%Type = Article
\bibitem[{{Gouy-Pailler} et~al.(2010){Gouy-Pailler}, Congedo, Brunner, Jutten
  \& Pfurtscheller}]{gouy-pailler2010NonstationaryBrainSource}
\bibinfo{author}{{Gouy-Pailler}, C.}, \bibinfo{author}{Congedo, M.},
  \bibinfo{author}{Brunner, C.}, \bibinfo{author}{Jutten, C.}, \&
  \bibinfo{author}{Pfurtscheller, G.} (\bibinfo{year}{2010}).
\newblock \bibinfo{title}{Nonstationary {{Brain Source Separation}} for
  {{Multiclass Motor Imagery}}}.
\newblock {\it \bibinfo{journal}{IEEE Transactions on Biomedical
  Engineering}\/},  {\it \bibinfo{volume}{57}\/}, \bibinfo{pages}{469--478}.
  \DOIprefix\doi{10.1109/TBME.2009.2032162}.
%Type = Article
\bibitem[{{Grosse-Wentrup} \&
  Buss(2008)}]{grosse-wentrup2008MulticlassCommonSpatial}
\bibinfo{author}{{Grosse-Wentrup}, M.}, \& \bibinfo{author}{Buss, M.}
  (\bibinfo{year}{2008}).
\newblock \bibinfo{title}{Multiclass {{Common Spatial Patterns}} and
  {{Information Theoretic Feature Extraction}}}.
\newblock {\it \bibinfo{journal}{IEEE Transactions on Biomedical
  Engineering}\/},  {\it \bibinfo{volume}{55}\/}, \bibinfo{pages}{1991--2000}.
  \DOIprefix\doi{10.1109/TBME.2008.921154}.
%Type = Article
\bibitem[{Grozea et~al.(2011)Grozea, Voinescu \&
  Fazli}]{grozea2011BristlesensorsLowcostFlexible}
\bibinfo{author}{Grozea, C.}, \bibinfo{author}{Voinescu, C.~D.}, \&
  \bibinfo{author}{Fazli, S.} (\bibinfo{year}{2011}).
\newblock \bibinfo{title}{Bristle-sensors\textemdash low-cost flexible passive
  dry {{EEG}} electrodes for neurofeedback and {{BCI}} applications}.
\newblock {\it \bibinfo{journal}{Journal of Neural Engineering}\/},  {\it
  \bibinfo{volume}{8}\/}, \bibinfo{pages}{025008}.
  \DOIprefix\doi{10.1088/1741-2560/8/2/025008}.
%Type = Article
\bibitem[{Gu et~al.(2021)Gu, Wei, Guo \& Wang}]{gu2021CommonSpatialPattern}
\bibinfo{author}{Gu, J.}, \bibinfo{author}{Wei, M.}, \bibinfo{author}{Guo, Y.},
  \& \bibinfo{author}{Wang, H.} (\bibinfo{year}{2021}).
\newblock \bibinfo{title}{Common {{Spatial Pattern}} with {{L21-Norm}}}.
\newblock {\it \bibinfo{journal}{Neural Processing Letters}\/},  {\it
  \bibinfo{volume}{53}\/}, \bibinfo{pages}{3619--3638}.
  \DOIprefix\doi{10.1007/s11063-021-10567-x}.
%Type = Article
\bibitem[{He et~al.(2015)He, Baxter, Edelman, Cline \&
  Ye}]{he2015NoninvasiveBrainComputerInterfaces}
\bibinfo{author}{He, B.}, \bibinfo{author}{Baxter, B.},
  \bibinfo{author}{Edelman, B.~J.}, \bibinfo{author}{Cline, C.~C.}, \&
  \bibinfo{author}{Ye, W.~W.} (\bibinfo{year}{2015}).
\newblock \bibinfo{title}{Noninvasive {{Brain-Computer Interfaces Based}} on
  {{Sensorimotor Rhythms}}}.
\newblock {\it \bibinfo{journal}{Proceedings of the IEEE}\/},  {\it
  \bibinfo{volume}{103}\/}, \bibinfo{pages}{907--925}.
  \DOIprefix\doi{10.1109/JPROC.2015.2407272}.
%Type = Article
\bibitem[{Higashi \& Tanaka(2013)}]{higashi2013SimultaneousDesignFIR}
\bibinfo{author}{Higashi, H.}, \& \bibinfo{author}{Tanaka, T.}
  (\bibinfo{year}{2013}).
\newblock \bibinfo{title}{Simultaneous {{Design}} of {{FIR Filter Banks}} and
  {{Spatial Patterns}} for {{EEG Signal Classification}}}.
\newblock {\it \bibinfo{journal}{IEEE Transactions on Biomedical
  Engineering}\/},  {\it \bibinfo{volume}{60}\/}, \bibinfo{pages}{1100--1110}.
  \DOIprefix\doi{10.1109/TBME.2012.2215960}.
%Type = Inproceedings
\bibitem[{Huang et~al.(2002)Huang, Liu, Lu \& Ma}]{huang2002SolvingSmallSample}
\bibinfo{author}{Huang, R.}, \bibinfo{author}{Liu, Q.}, \bibinfo{author}{Lu,
  H.}, \& \bibinfo{author}{Ma, S.} (\bibinfo{year}{2002}).
\newblock \bibinfo{title}{Solving the small sample size problem of {{LDA}}}.
\newblock In {\it \bibinfo{booktitle}{Proceedings of 16th {{International
  Conference}} on {{Pattern Recognition}}}\/} (pp. \bibinfo{pages}{29--32}).
\newblock \bibinfo{address}{{Quebec, Canada}}: \bibinfo{publisher}{{IEEE}}.
\newblock \DOIprefix\doi{10.1109/ICPR.2002.1047787}.
%Type = Article
\bibitem[{Jin et~al.(2021)Jin, Xiao, Daly, Miao, Wang \&
  Cichocki}]{jin2021InternalFeatureSelection}
\bibinfo{author}{Jin, J.}, \bibinfo{author}{Xiao, R.}, \bibinfo{author}{Daly,
  I.}, \bibinfo{author}{Miao, Y.}, \bibinfo{author}{Wang, X.}, \&
  \bibinfo{author}{Cichocki, A.} (\bibinfo{year}{2021}).
\newblock \bibinfo{title}{Internal {{Feature Selection Method}} of {{CSP
  Based}} on {{L1-Norm}} and {{Dempster}}\textendash{{Shafer Theory}}}.
\newblock {\it \bibinfo{journal}{IEEE Transactions on Neural Networks and
  Learning Systems}\/},  {\it \bibinfo{volume}{32}\/},
  \bibinfo{pages}{4814--4825}. \DOIprefix\doi{10.1109/TNNLS.2020.3015505}.
%Type = Article
\bibitem[{Lebedev \& Nicolelis(2017)}]{lebedev2017BrainMachineInterfacesBasic}
\bibinfo{author}{Lebedev, M.~A.}, \& \bibinfo{author}{Nicolelis, M. A.~L.}
  (\bibinfo{year}{2017}).
\newblock \bibinfo{title}{Brain-{{Machine Interfaces}}: {{From Basic Science}}
  to {{Neuroprostheses}} and {{Neurorehabilitation}}}.
\newblock {\it \bibinfo{journal}{Physiological Reviews}\/},  {\it
  \bibinfo{volume}{97}\/}, \bibinfo{pages}{767--837}.
  \DOIprefix\doi{10.1152/physrev.00027.2016}.
%Type = Article
\bibitem[{Lemm et~al.(2005)Lemm, Blankertz, Curio \&
  Muller}]{lemm2005SpatioSpectralFiltersImproving}
\bibinfo{author}{Lemm, S.}, \bibinfo{author}{Blankertz, B.},
  \bibinfo{author}{Curio, G.}, \& \bibinfo{author}{Muller, K.-R.}
  (\bibinfo{year}{2005}).
\newblock \bibinfo{title}{Spatio-{{Spectral Filters}} for {{Improving}} the
  {{Classification}} of {{Single Trial EEG}}}.
\newblock {\it \bibinfo{journal}{IEEE Transactions on Biomedical
  Engineering}\/},  {\it \bibinfo{volume}{52}\/}, \bibinfo{pages}{1541--1548}.
  \DOIprefix\doi{10.1109/TBME.2005.851521}.
%Type = Article
\bibitem[{Li et~al.(2019)Li, Fan, Wang \& Wang}]{li2019CommonSpatialPatterns}
\bibinfo{author}{Li, X.}, \bibinfo{author}{Fan, H.}, \bibinfo{author}{Wang,
  H.}, \& \bibinfo{author}{Wang, L.} (\bibinfo{year}{2019}).
\newblock \bibinfo{title}{Common spatial patterns combined with phase
  synchronization information for classification of {{EEG}} signals}.
\newblock {\it \bibinfo{journal}{Biomedical Signal Processing and Control}\/},
  {\it \bibinfo{volume}{52}\/}, \bibinfo{pages}{248--256}.
  \DOIprefix\doi{10.1016/j.bspc.2019.04.034}.
%Type = Article
\bibitem[{Lin et~al.(2016)Lin, Pan, Chu \&
  Lin}]{lin2016DevelopmentWearableMotorImageryBased}
\bibinfo{author}{Lin, B.-S.}, \bibinfo{author}{Pan, J.-S.},
  \bibinfo{author}{Chu, T.-Y.}, \& \bibinfo{author}{Lin, B.-S.}
  (\bibinfo{year}{2016}).
\newblock \bibinfo{title}{Development of a {{Wearable Motor-Imagery-Based
  Brain}}\textendash{{Computer Interface}}}.
\newblock {\it \bibinfo{journal}{Journal of Medical Systems}\/},  {\it
  \bibinfo{volume}{40}\/}, \bibinfo{pages}{71}.
  \DOIprefix\doi{10.1007/s10916-015-0429-6}.
%Type = Article
\bibitem[{Liu et~al.(2021{\natexlab{a}})Liu, Jin, Xu, Li, Zuo, Sun, Wang \&
  Cichocki}]{liu2021DistinguishableSpatialspectralFeaturea}
\bibinfo{author}{Liu, C.}, \bibinfo{author}{Jin, J.}, \bibinfo{author}{Xu, R.},
  \bibinfo{author}{Li, S.}, \bibinfo{author}{Zuo, C.}, \bibinfo{author}{Sun,
  H.}, \bibinfo{author}{Wang, X.}, \& \bibinfo{author}{Cichocki, A.}
  (\bibinfo{year}{2021}{\natexlab{a}}).
\newblock \bibinfo{title}{Distinguishable spatial-spectral feature learning
  neural network framework for motor imagery-based brain\textendash computer
  interface}.
\newblock {\it \bibinfo{journal}{Journal of Neural Engineering}\/},  {\it
  \bibinfo{volume}{18}\/}, \bibinfo{pages}{0460e4}.
  \DOIprefix\doi{10.1088/1741-2552/ac1d36}.
%Type = Article
\bibitem[{Liu et~al.(2021{\natexlab{b}})Liu, Zheng, Chen, Ma \&
  Ai}]{liu2021OnlineDetectionClassimbalanced}
\bibinfo{author}{Liu, Q.}, \bibinfo{author}{Zheng, W.}, \bibinfo{author}{Chen,
  K.}, \bibinfo{author}{Ma, L.}, \& \bibinfo{author}{Ai, Q.}
  (\bibinfo{year}{2021}{\natexlab{b}}).
\newblock \bibinfo{title}{Online detection of class-imbalanced error-related
  potentials evoked by motor imagery}.
\newblock {\it \bibinfo{journal}{Journal of Neural Engineering}\/},  {\it
  \bibinfo{volume}{18}\/}, \bibinfo{pages}{046032}.
  \DOIprefix\doi{10.1088/1741-2552/abf522}.
%Type = Article
\bibitem[{Lotte et~al.(2018)Lotte, Bougrain, Cichocki, Clerc, Congedo,
  Rakotomamonjy \& Yger}]{lotte2018ReviewClassificationAlgorithms}
\bibinfo{author}{Lotte, F.}, \bibinfo{author}{Bougrain, L.},
  \bibinfo{author}{Cichocki, A.}, \bibinfo{author}{Clerc, M.},
  \bibinfo{author}{Congedo, M.}, \bibinfo{author}{Rakotomamonjy, A.}, \&
  \bibinfo{author}{Yger, F.} (\bibinfo{year}{2018}).
\newblock \bibinfo{title}{A review of classification algorithms for
  {{EEG-based}} brain\textendash computer interfaces: A 10 year update}.
\newblock {\it \bibinfo{journal}{Journal of Neural Engineering}\/},  {\it
  \bibinfo{volume}{15}\/}, \bibinfo{pages}{031005}.
  \DOIprefix\doi{10.1088/1741-2552/aab2f2}.
%Type = Article
\bibitem[{Lotte \& {Cuntai Guan}(2011)}]{lotte2011RegularizingCommonSpatial}
\bibinfo{author}{Lotte, F.}, \& \bibinfo{author}{{Cuntai Guan}}
  (\bibinfo{year}{2011}).
\newblock \bibinfo{title}{Regularizing {{Common Spatial Patterns}} to {{Improve
  BCI Designs}}: {{Unified Theory}} and {{New Algorithms}}}.
\newblock {\it \bibinfo{journal}{IEEE Transactions on Biomedical
  Engineering}\/},  {\it \bibinfo{volume}{58}\/}, \bibinfo{pages}{355--362}.
  \DOIprefix\doi{10.1109/TBME.2010.2082539}.
%Type = Article
\bibitem[{Miao et~al.(2021)Miao, Jin, Daly, Zuo, Wang, Cichocki \&
  Jung}]{miao2021LearningCommonTimeFrequencySpatial}
\bibinfo{author}{Miao, Y.}, \bibinfo{author}{Jin, J.}, \bibinfo{author}{Daly,
  I.}, \bibinfo{author}{Zuo, C.}, \bibinfo{author}{Wang, X.},
  \bibinfo{author}{Cichocki, A.}, \& \bibinfo{author}{Jung, T.-P.}
  (\bibinfo{year}{2021}).
\newblock \bibinfo{title}{Learning {{Common Time-Frequency-Spatial Patterns}}
  for {{Motor Imagery Classification}}}.
\newblock {\it \bibinfo{journal}{IEEE Transactions on Neural Systems and
  Rehabilitation Engineering}\/},  {\it \bibinfo{volume}{29}\/},
  \bibinfo{pages}{699--707}. \DOIprefix\doi{10.1109/TNSRE.2021.3071140}.
%Type = Article
\bibitem[{Olias et~al.(2019)Olias, {Martin-Clemente}, {Sarmiento-Vega} \&
  Cruces}]{olias2019EEGSignalProcessing}
\bibinfo{author}{Olias, J.}, \bibinfo{author}{{Martin-Clemente}, R.},
  \bibinfo{author}{{Sarmiento-Vega}, M.~A.}, \& \bibinfo{author}{Cruces, S.}
  (\bibinfo{year}{2019}).
\newblock \bibinfo{title}{{{EEG Signal Processing}} in {{MI-BCI Applications
  With Improved Covariance Matrix Estimators}}}.
\newblock {\it \bibinfo{journal}{IEEE Transactions on Neural Systems and
  Rehabilitation Engineering}\/},  {\it \bibinfo{volume}{27}\/},
  \bibinfo{pages}{895--904}. \DOIprefix\doi{10.1109/TNSRE.2019.2905894}.
%Type = Article
\bibitem[{Onaran et~al.(2013)Onaran, Ince \&
  Cetin}]{onaran2013SparseSpatialFilter}
\bibinfo{author}{Onaran, I.}, \bibinfo{author}{Ince, N.~F.}, \&
  \bibinfo{author}{Cetin, A.~E.} (\bibinfo{year}{2013}).
\newblock \bibinfo{title}{Sparse spatial filter via a novel objective function
  minimization with smooth {$\mathscr{l}$}1 regularization}.
\newblock {\it \bibinfo{journal}{Biomedical Signal Processing and Control}\/},
  {\it \bibinfo{volume}{8}\/}, \bibinfo{pages}{282--288}.
  \DOIprefix\doi{10.1016/j.bspc.2012.10.003}.
%Type = Article
\bibitem[{Oppermann et~al.(2020)Oppermann, Wichum, Haueisen, Klemm \&
  Esch}]{oppermann2020MagCPPToolboxCombining}
\bibinfo{author}{Oppermann, H.}, \bibinfo{author}{Wichum, F.},
  \bibinfo{author}{Haueisen, J.}, \bibinfo{author}{Klemm, M.}, \&
  \bibinfo{author}{Esch, L.} (\bibinfo{year}{2020}).
\newblock \bibinfo{title}{{{MagCPP}}: {{A C}}++ toolbox for {{Combining
  Neurofeedback}} with {{Magstim}} transcranial magnetic stimulators}.
\newblock {\it \bibinfo{journal}{Current Directions in Biomedical
  Engineering}\/},  {\it \bibinfo{volume}{6}\/}, \bibinfo{pages}{497--500}.
  \DOIprefix\doi{10.1515/cdbme-2020-3128}.
%Type = Article
\bibitem[{Park et~al.(2018)Park, Lee \& Lee}]{park2018FilterBankRegularized}
\bibinfo{author}{Park, S.-H.}, \bibinfo{author}{Lee, D.}, \&
  \bibinfo{author}{Lee, S.-G.} (\bibinfo{year}{2018}).
\newblock \bibinfo{title}{Filter {{Bank Regularized Common Spatial Pattern
  Ensemble}} for {{Small Sample Motor Imagery Classification}}}.
\newblock {\it \bibinfo{journal}{IEEE Transactions on Neural Systems and
  Rehabilitation Engineering}\/},  {\it \bibinfo{volume}{26}\/},
  \bibinfo{pages}{498--505}. \DOIprefix\doi{10.1109/TNSRE.2017.2757519}.
%Type = Article
\bibitem[{Pfurtscheller et~al.(2006)Pfurtscheller, Brunner, Schl{\"o}gl \&
  {Lopes da Silva}}]{pfurtscheller2006MuRhythmSynchronization}
\bibinfo{author}{Pfurtscheller, G.}, \bibinfo{author}{Brunner, C.},
  \bibinfo{author}{Schl{\"o}gl, A.}, \& \bibinfo{author}{{Lopes da Silva}, F.}
  (\bibinfo{year}{2006}).
\newblock \bibinfo{title}{Mu rhythm (de)synchronization and {{EEG}}
  single-trial classification of different motor imagery tasks}.
\newblock {\it \bibinfo{journal}{NeuroImage}\/},  {\it \bibinfo{volume}{31}\/},
  \bibinfo{pages}{153--159}. \DOIprefix\doi{10.1016/j.neuroimage.2005.12.003}.
%Type = Article
\bibitem[{Popescu et~al.(2007)Popescu, Fazli, Badower, Blankertz \&
  M{\"u}ller}]{popescu2007SingleTrialClassification}
\bibinfo{author}{Popescu, F.}, \bibinfo{author}{Fazli, S.},
  \bibinfo{author}{Badower, Y.}, \bibinfo{author}{Blankertz, B.}, \&
  \bibinfo{author}{M{\"u}ller, K.-R.} (\bibinfo{year}{2007}).
\newblock \bibinfo{title}{Single {{Trial Classification}} of {{Motor
  Imagination Using}} 6 {{Dry EEG Electrodes}}}.
\newblock {\it \bibinfo{journal}{PLoS ONE}\/},  {\it \bibinfo{volume}{2}\/},
  \bibinfo{pages}{e637}. \DOIprefix\doi{10.1371/journal.pone.0000637}.
%Type = Article
\bibitem[{Ramadan \& Vasilakos(2017)}]{ramadan2017BrainComputerInterface}
\bibinfo{author}{Ramadan, R.~A.}, \& \bibinfo{author}{Vasilakos, A.~V.}
  (\bibinfo{year}{2017}).
\newblock \bibinfo{title}{Brain computer interface: Control signals review}.
\newblock {\it \bibinfo{journal}{Neurocomputing}\/},  {\it
  \bibinfo{volume}{223}\/}, \bibinfo{pages}{26--44}.
  \DOIprefix\doi{10.1016/j.neucom.2016.10.024}.
%Type = Article
\bibitem[{Roijendijk et~al.(2016)Roijendijk, Gielen \&
  Farquhar}]{roijendijk2016ClassifyingRegularizedSensor}
\bibinfo{author}{Roijendijk, L.}, \bibinfo{author}{Gielen, S.}, \&
  \bibinfo{author}{Farquhar, J.} (\bibinfo{year}{2016}).
\newblock \bibinfo{title}{Classifying {{Regularized Sensor Covariance
  Matrices}}: {{An Alternative}} to {{CSP}}}.
\newblock {\it \bibinfo{journal}{IEEE Transactions on Neural Systems and
  Rehabilitation Engineering}\/},  {\it \bibinfo{volume}{24}\/},
  \bibinfo{pages}{893--900}. \DOIprefix\doi{10.1109/TNSRE.2015.2477687}.
%Type = Article
\bibitem[{Samek et~al.(2014)Samek, Kawanabe \&
  Muller}]{samek2014DivergenceBasedFrameworkCommon}
\bibinfo{author}{Samek, W.}, \bibinfo{author}{Kawanabe, M.}, \&
  \bibinfo{author}{Muller, K.-R.} (\bibinfo{year}{2014}).
\newblock \bibinfo{title}{Divergence-{{Based Framework}} for {{Common Spatial
  Patterns Algorithms}}}.
\newblock {\it \bibinfo{journal}{IEEE Reviews in Biomedical Engineering}\/},
  {\it \bibinfo{volume}{7}\/}, \bibinfo{pages}{50--72}.
  \DOIprefix\doi{10.1109/RBME.2013.2290621}.
%Type = Article
\bibitem[{Samek et~al.(2012)Samek, Vidaurre, M{\"u}ller \&
  Kawanabe}]{samek2012StationaryCommonSpatial}
\bibinfo{author}{Samek, W.}, \bibinfo{author}{Vidaurre, C.},
  \bibinfo{author}{M{\"u}ller, K.-R.}, \& \bibinfo{author}{Kawanabe, M.}
  (\bibinfo{year}{2012}).
\newblock \bibinfo{title}{Stationary common spatial patterns for
  brain\textendash computer interfacing}.
\newblock {\it \bibinfo{journal}{Journal of Neural Engineering}\/},  {\it
  \bibinfo{volume}{9}\/}, \bibinfo{pages}{026013}.
  \DOIprefix\doi{10.1088/1741-2560/9/2/026013}.
%Type = Article
\bibitem[{Tangermann et~al.(2012)Tangermann, M{\"u}ller, Aertsen, Birbaumer,
  Braun, Brunner, Leeb, Mehring, Miller, {M{\"u}ller-Putz}, Nolte,
  Pfurtscheller, Preissl, Schalk, Schl{\"o}gl, Vidaurre, Waldert \&
  Blankertz}]{tangermann2012ReviewBCICompetition}
\bibinfo{author}{Tangermann, M.}, \bibinfo{author}{M{\"u}ller, K.-R.},
  \bibinfo{author}{Aertsen, A.}, \bibinfo{author}{Birbaumer, N.},
  \bibinfo{author}{Braun, C.}, \bibinfo{author}{Brunner, C.},
  \bibinfo{author}{Leeb, R.}, \bibinfo{author}{Mehring, C.},
  \bibinfo{author}{Miller, K.~J.}, \bibinfo{author}{{M{\"u}ller-Putz}, G.~R.},
  \bibinfo{author}{Nolte, G.}, \bibinfo{author}{Pfurtscheller, G.},
  \bibinfo{author}{Preissl, H.}, \bibinfo{author}{Schalk, G.},
  \bibinfo{author}{Schl{\"o}gl, A.}, \bibinfo{author}{Vidaurre, C.},
  \bibinfo{author}{Waldert, S.}, \& \bibinfo{author}{Blankertz, B.}
  (\bibinfo{year}{2012}).
\newblock \bibinfo{title}{Review of the {{BCI Competition IV}}}.
\newblock {\it \bibinfo{journal}{Frontiers in Neuroscience}\/},  {\it
  \bibinfo{volume}{6}\/}. \DOIprefix\doi{10.3389/fnins.2012.00055}.
%Type = Article
\bibitem[{Tharwat et~al.(2017)Tharwat, Gaber, Ibrahim \&
  Hassanien}]{tharwat2017LinearDiscriminantAnalysis}
\bibinfo{author}{Tharwat, A.}, \bibinfo{author}{Gaber, T.},
  \bibinfo{author}{Ibrahim, A.}, \& \bibinfo{author}{Hassanien, A.~E.}
  (\bibinfo{year}{2017}).
\newblock \bibinfo{title}{Linear discriminant analysis: {{A}} detailed
  tutorial}.
\newblock {\it \bibinfo{journal}{AI communications}\/},  {\it
  \bibinfo{volume}{30}\/}, \bibinfo{pages}{169--190}.
  \DOIprefix\doi{10.3233/AIC-170729}.
%Type = Article
\bibitem[{Vidaurre et~al.(2021)Vidaurre, Jorajur{\'i}a, {Ramos-Murguialday},
  M{\"u}ller, G{\'o}mez \& Nikulin}]{vidaurre2021ImprovingMotorImagery}
\bibinfo{author}{Vidaurre, C.}, \bibinfo{author}{Jorajur{\'i}a, T.},
  \bibinfo{author}{{Ramos-Murguialday}, A.}, \bibinfo{author}{M{\"u}ller,
  K.-R.}, \bibinfo{author}{G{\'o}mez, M.}, \& \bibinfo{author}{Nikulin, V.~V.}
  (\bibinfo{year}{2021}).
\newblock \bibinfo{title}{Improving motor imagery classification during induced
  motor perturbations}.
\newblock {\it \bibinfo{journal}{Journal of Neural Engineering}\/},  {\it
  \bibinfo{volume}{18}\/}, \bibinfo{pages}{0460b1}.
  \DOIprefix\doi{10.1088/1741-2552/ac123f}.
%Type = Article
\bibitem[{Vigario et~al.(2000)Vigario, Sarela, Jousmiki, Hamalainen \&
  Oja}]{vigario2000IndependentComponentApproach}
\bibinfo{author}{Vigario, R.}, \bibinfo{author}{Sarela, J.},
  \bibinfo{author}{Jousmiki, V.}, \bibinfo{author}{Hamalainen, M.}, \&
  \bibinfo{author}{Oja, E.} (\bibinfo{year}{2000}).
\newblock \bibinfo{title}{Independent component approach to the analysis of
  {{EEG}} and {{MEG}} recordings}.
\newblock {\it \bibinfo{journal}{IEEE Transactions on Biomedical
  Engineering}\/},  {\it \bibinfo{volume}{47}\/}, \bibinfo{pages}{589--593}.
  \DOIprefix\doi{10.1109/10.841330}.
%Type = Article
\bibitem[{Wang et~al.(2021)Wang, Wong, Kang, Liu, Shui, Wan \&
  Chen}]{wang2021CommonSpatialPattern}
\bibinfo{author}{Wang, B.}, \bibinfo{author}{Wong, C.~M.},
  \bibinfo{author}{Kang, Z.}, \bibinfo{author}{Liu, F.}, \bibinfo{author}{Shui,
  C.}, \bibinfo{author}{Wan, F.}, \& \bibinfo{author}{Chen, C. L.~P.}
  (\bibinfo{year}{2021}).
\newblock \bibinfo{title}{Common {{Spatial Pattern Reformulated}} for
  {{Regularizations}} in {{Brain}}\textendash{{Computer Interfaces}}}.
\newblock {\it \bibinfo{journal}{IEEE Transactions on Cybernetics}\/},  {\it
  \bibinfo{volume}{51}\/}, \bibinfo{pages}{5008--5020}.
  \DOIprefix\doi{10.1109/TCYB.2020.2982901}.
%Type = Article
\bibitem[{Wang(2012)}]{wang2012HarmonicMeanKullback}
\bibinfo{author}{Wang, H.} (\bibinfo{year}{2012}).
\newblock \bibinfo{title}{Harmonic {{Mean}} of {{Kullback}}\textendash{{Leibler
  Divergences}} for {{Optimizing Multi-Class EEG Spatio-Temporal Filters}}}.
\newblock {\it \bibinfo{journal}{Neural Processing Letters}\/},  {\it
  \bibinfo{volume}{36}\/}, \bibinfo{pages}{161--171}.
  \DOIprefix\doi{10.1007/s11063-012-9228-y}.
%Type = Article
\bibitem[{Wang \& Li(2016)}]{wang2016RegularizedFiltersL1NormBased}
\bibinfo{author}{Wang, H.}, \& \bibinfo{author}{Li, X.} (\bibinfo{year}{2016}).
\newblock \bibinfo{title}{Regularized {{Filters}} for {{L1-Norm-Based Common
  Spatial Patterns}}}.
\newblock {\it \bibinfo{journal}{IEEE Transactions on Neural Systems and
  Rehabilitation Engineering}\/},  {\it \bibinfo{volume}{24}\/},
  \bibinfo{pages}{201--211}. \DOIprefix\doi{10.1109/TNSRE.2015.2474141}.
%Type = Article
\bibitem[{Wang et~al.(2012)Wang, Tang \&
  Zheng}]{wang2012L1NormBasedCommonSpatial}
\bibinfo{author}{Wang, H.}, \bibinfo{author}{Tang, Q.}, \&
  \bibinfo{author}{Zheng, W.} (\bibinfo{year}{2012}).
\newblock \bibinfo{title}{L1-{{Norm-Based Common Spatial Patterns}}}.
\newblock {\it \bibinfo{journal}{IEEE Transactions on Biomedical
  Engineering}\/},  {\it \bibinfo{volume}{59}\/}, \bibinfo{pages}{653--662}.
  \DOIprefix\doi{10.1109/TBME.2011.2177523}.
%Type = Book
\bibitem[{Wolpaw \& Wolpaw(2012)}]{wolpaw2012BraincomputerInterfacesPrinciples}
\bibinfo{editor}{Wolpaw, J.~R.}, \& \bibinfo{editor}{Wolpaw, E.~W.} (Eds.)
  (\bibinfo{year}{2012}).
\newblock {\it \bibinfo{title}{Brain-Computer Interfaces: Principles and
  Practice}\/}.
\newblock \bibinfo{address}{{Oxford ; New York}}: \bibinfo{publisher}{{Oxford
  University Press}}.
%Type = Article
\bibitem[{Wu et~al.(2008)Wu, Gao, Hong \&
  Gao}]{wu2008ClassifyingSingleTrialEEG}
\bibinfo{author}{Wu, W.}, \bibinfo{author}{Gao, X.}, \bibinfo{author}{Hong,
  B.}, \& \bibinfo{author}{Gao, S.} (\bibinfo{year}{2008}).
\newblock \bibinfo{title}{Classifying {{Single-Trial EEG During Motor Imagery}}
  by {{Iterative Spatio-Spectral Patterns Learning}} ({{ISSPL}})}.
\newblock {\it \bibinfo{journal}{IEEE Transactions on Biomedical
  Engineering}\/},  {\it \bibinfo{volume}{55}\/}, \bibinfo{pages}{1733--1743}.
  \DOIprefix\doi{10.1109/TBME.2008.919125}.
%Type = Article
\bibitem[{Zuo et~al.(2021)Zuo, Jin, Xu, Wu, Liu, Miao \&
  Wang}]{zuo2021ClusterDecomposingMultiobjective}
\bibinfo{author}{Zuo, C.}, \bibinfo{author}{Jin, J.}, \bibinfo{author}{Xu, R.},
  \bibinfo{author}{Wu, L.}, \bibinfo{author}{Liu, C.}, \bibinfo{author}{Miao,
  Y.}, \& \bibinfo{author}{Wang, X.} (\bibinfo{year}{2021}).
\newblock \bibinfo{title}{Cluster decomposing and multi-objective optimization
  based-ensemble learning framework for motor imagery-based brain\textendash
  computer interfaces}.
\newblock {\it \bibinfo{journal}{Journal of Neural Engineering}\/},  {\it
  \bibinfo{volume}{18}\/}, \bibinfo{pages}{026018}.
  \DOIprefix\doi{10.1088/1741-2552/abe20f}.

\end{thebibliography}
%%==============================================================================
%%

\end{document}